\documentclass[superscriptaddress,a4paper,nofootinbib,amsmath,amssymb,floatfix,prd]{revtex4-2}
\usepackage{graphicx}
\usepackage{dcolumn}
\usepackage{bm}
\usepackage{hyperref}
\usepackage[mathlines]{lineno}
\usepackage{multirow}
\usepackage{url}
\usepackage{float}
\usepackage{comment}

\hypersetup{pdfstartview=FitV,colorlinks=true,linkcolor=blue,citecolor=red,filecolor=black,urlcolor=blue}
\newcommand{\met}{$\cancel E_T$}

\def\issue(#1,#2,#3){{\bf #1}, #2 (#3)}

\catcode`\@=11
\def\lsim{\mathrel{\mathpalette\@versim<}}
\def\gsim{\mathrel{\mathpalette\@versim>}}
\def\@versim#1#2{\vcenter{\offinterlineskip
\ialign{$\m@th#1\hfil##\hfil$\crcr#2\crcr\sim\crcr } }}
\catcode`\@=12

\catcode`@=12

\newcommand{\newc}{\newcommand}
\newc{\wt}{\widetilde}
\newc{\ra}{\rightarrow}
\def\beq {\begin{equation}}
\def\eeq {\end{equation}}
\def\bi {\begin{itemize}}
\def\ei {\end{itemize}}
\def\bea {\begin{eqnarray}}
\def\eea {\end{eqnarray}}

\def \met{\rm E{\!\!\!/}_T}

\newcommand{\br}{\begin{eqnarray}}
\newcommand{\er}{\end{eqnarray}}
\newcommand{\be}{\begin{equation}}
\newcommand{\ee}{\end{equation}}

\newcommand{\ch}{\widetilde \chi^\pm}

\newcommand{\stau}{\widetilde\tau}

\def \ch2p {{\wt\chi_2^+}}

\def \ch2m {{\wt\chi_2^-}}

\def \chonepm{{\wt\chi_1}^{\pm}}

\newc{\dmchi}{\Delta m_{\wt\chi}}

\def \chtwopm{{\wt\chi_2}^{\pm}}

\def \lspone{\wt\chi_1^0}

\def \lsptwo{\wt\chi_2^0}

\def \lspthree{\wt\chi_3^0}

\def \lspfour{\wt\chi_4^0}

\usepackage{array}
\newcolumntype{L}[1]{>{\raggedright\let\newline\\\arraybackslash\hspace{0pt}}m{#1}}
\newcolumntype{C}[1]{>{\centering\let\newline\\\arraybackslash\hspace{0pt}}m{#1}}
\newcolumntype{R}[1]{>{\raggedleft\let\newline\\\arraybackslash\hspace{0pt}}m{#1}}

\def\issue(#1,#2,#3){{\bf #1}, #2 (#3)}

\hypersetup{
   colorlinks=true,       
   linkcolor=blue,        
   citecolor=red,         
   filecolor=magenta,      
   }

\begin{document}

\begin{flushright}
BONN-TH-2024-03
\end{flushright}

\vspace{1cm}

%

\title{Current status of the light neutralino thermal dark matter in the phenomenological MSSM}

\author{Rahool Kumar Barman}
\email{rahool.barman@ipmu.jp}
\affiliation{Kavli IPMU (WPI), UTIAS, The University of Tokyo, Kashiwa, Chiba 277-8583, Japan}

\author{Genevieve B\'elanger}
\email{belanger@lapth.cnrs.fr}
\affiliation{LAPTh, Universit\'e Savoie Mont Blanc, CNRS, B.P. 110,  
F-74941 Annecy Cedex, France}

\author{\\Biplob Bhattacherjee}
\email{biplob@iisc.ac.in}
\affiliation{Centre for High Energy Physics, 
Indian Institute of Science, Bangalore 560012, India}

\addtocounter{footnote}{-1}

\author{Rohini Godbole\,\footnotemark[1]$\dagger$\footnotetext[1]{$\dagger$ Deceased. {\it 
Sadly, our distinguished colleague Rohini Godbole passed away while this work was
being finalized. She will be missed.}}}
\email{rohini@iisc.ac.in}
\affiliation{Centre for High Energy Physics, 
Indian Institute of Science, Bangalore 560012, India}

\author{Rhitaja Sengupta}
\email{rsengupt@uni-bonn.de}
\affiliation{Bethe Center for Theoretical Physics and Physikalisches Institut der Universit\"at Bonn, \\
Nu{\ss}allee 12, 53115 Bonn, Germany}






%
\begin{abstract}
In a previous publication, we studied the parameter space of the phenomenological Minimal Supersymmetric Standard Model (pMSSM) with a light neutralino thermal dark matter ($M_{\tilde{\chi}_1^0} \leq M_h/2$) and observed that the recent results from the dark matter and collider experiments put strong constraints on this scenario. In this work, we present in detail the arguments behind the robustness of this result against scanning over the large number of parameters in pMSSM. The Run-3 of LHC will be crucial in probing the surviving regions of the parameter space. 
We further investigate the impact of light staus on our parameter space and also provide benchmarks which can be interesting for Run-3 of LHC. We analyse these benchmarks at the LHC using the machine learning framework of \texttt{XGBOOST}. Finally, we also discuss the effect of non-standard cosmology on the parameter space.
\end{abstract}





\maketitle


\addtocounter{footnote}{+1}

\section{Introduction}

Since the discovery of the 125 GeV Higgs boson~($h$) by the ATLAS and CMS collaborations in 2012, the precise quantification of its non-standard couplings has been a major cornerstone of the new physics search program at the LHC. Until now, clear evidence of physics beyond the Standard Model~(SM) is yet to be observed at the LHC.
Within experimental uncertainties, measurements indicate that the observed properties of $h$ are consistent with the expectations for the SM Higgs boson.
While the couplings of the Higgs boson with the third-generation charged fermions and gauge bosons have been measured with considerable precision, uncertainties in the second-generation Yukawa coupling measurements are gradually reducing with improved statistics. 
Nonetheless, the current data still allow non-standard decays for the discovered Higgs boson.
An exciting aspect of non-standard interactions of Higgs bosons is their decay into an invisible final state. Recent analyses of the LHC Run-II data by the ATLAS and CMS collaborations have constrained the branching fraction for Higgs invisible decay to 11\%\,\cite{ATLAS:2024lyh} and 15\%\,\cite{CMS:2023sdw} at $95\%$ CL, respectively. 
The invisible final states can be dark matter (DM) candidates, as long as they are stable or have a lifetime larger than the age of the Universe,
thus transmuting Higgs invisible searches at the LHC to potential probes for the elusive Dark Matter. 

The R-parity conserving~(RPC) Minimal Supersymmetric Standard Model~(MSSM), with no explicit terms for baryon number and lepton number violation, provides a stable lightest supersymmetric particle~(LSP), typically the neutralino $\tilde{\chi}_{1}^{0}$, which can be a WIMP~(weakly interacting massive particle) DM candidate. Charged under the $SU(2)_{L} \times U(1)_{Y}$ gauge group, the LSP neutralino can typically generate the correct relic abundances at a mass of $\mathcal{O}(100)~$GeV, making it one of the most favorable and widely studied cold DM candidates. It is also worth noting that the MSSM can address the ``naturalness" problem~\cite{Weinberg:1975gm,Gildener:1976ih, PhysRevD.20.2619} while also accommodating a scalar boson consistent with the $h$ measurements, thus remaining one of the most attractive frameworks to pursue new physics beyond the scope of the Standard Model.

In this paper, we focus on the case of a light neutralino DM with mass $M_{\tilde{\chi}_{1}^{0}} \leq M_{h}/2$ such that the Higgs boson can decay invisibly through $h \to \tilde{\chi}_{1}^{0}\tilde{\chi}_{1}^{0}$. Since charged Higgsinos and Winos are constrained to have masses above $\gtrsim 100~$GeV by LEP searches\,\cite{L3:1999onh,OPAL:2003wxm}, the $\tilde{\chi}_{1}^{0}$ with mass below $\lesssim M_{h}/2$ is left with the sole possibility of a dominant Bino admixture. 
The observed DM abundance in our Universe is $\Omega_{DM}^{obs}h^{2} = 0.120 \pm 0.001$ as measured by the PLANCK collaboration\,\cite{Planck:2018vyg}.
However, the annihilation cross-section for Bino-dominated neutralinos is too small and leads to overclosure of the Universe ($\Omega_{LSP} h^{2}>\Omega_{DM}^{obs}h^{2}$), except under special circumstances, such as annihilation through an $s$-channel resonance with a mass of $\sim 2 M_{\tilde{\chi}_{1}^{0}}$ and sfermion exchange.
Our primary focus in this work will be the first scenario, however, we will also study the implications of light sfermions on our results. When the LSP contributes to the invisible decay of the SM-like Higgs boson, the available resonances through which it can annihilate are the $Z$ and $h$ bosons. The regions of parameter space where the LSP mass lies within a window of 3-5\,GeV around half the $Z$ boson mass (45\,GeV) or half the $h$ boson mass (62.5\,GeV), are referred to as the $Z$ funnel region or the $h$ funnel region, respectively.

Several studies have explored the prospect of a light neutralino DM, in the constrained MSSM~(cMSSM) and the phenomenological MSSM~(pMSSM) considering the various experimental constraints at the time\,\cite{PhysRevD.37.719,Djouadi:1996mj,Belanger:2000tg,Belanger:2001am,Hooper:2002nq,Hisano:2002fk,Belanger:2003wb,Dreiner:2009ic,AlbornozVasquez:2010nkq,Calibbi:2011ug,Choudhury:2012tc,Arbey:2012na,Dreiner:2012ex,Boehm:2013qva,Choudhury:2013jpa,Ananthanarayan:2013fga,Calibbi:2013poa,Belanger:2013pna,Chakraborti:2014gea,Han:2014nba,Cahill-Rowley:2014boa,Bhattacherjee:2015sga,Belanger:2015vwa,Chakraborti:2015mra,Hamaguchi:2015rxa,Cao:2015efs,ATLAS:2015wrn,Bagnaschi:2015eha,Choudhury:2016lku,Barman:2016jov,Barman:2017swy,Abdughani:2017dqs,Roszkowski:2017nbc,Pozzo:2018anw,GAMBIT:2018gjo,Wang:2020dtb,KumarBarman:2020ylm,VanBeekveld:2021tgn,Barman:2022jdg}. 
The ATLAS and CMS Collaborations at the LHC have made available new results from searches of heavy Higgs bosons\,\cite{ATLAS:2020zms}, direct searches of charginos and neutralinos (collectively known as electroweakinos)\,\cite{CMS:2020bfa, ATLAS:2021moa, ATLAS:2021yqv, CMS:2022sfi}, as well as the invisible decay of the SM Higgs boson\,\cite{ATLAS:2022yvh} at Run2.
The XENON-1T, XENON-nT, PICO-60, PandaX-4T, and LUX-ZEPLIN (LZ) collaborations have also published limits on the DM direct detection~(DD) cross-sections $-$ both spin-dependent~(SD) and spin-independent~(SI)\,\cite{XENON:2018voc,XENON:2019rxp,PICO:2019vsc,PandaX-4T:2021bab,Aalbers:2022fxq,PandaX:2022xas,XENON:2023cxc}. 
Among these, the results from the LZ collaboration provide the most stringent bounds on the SI DD cross-sections\,\cite{Aalbers:2022fxq} for DM masses in the 10\,GeV - 1\,TeV range. With these new and improved results, revisiting the MSSM parameter space containing light neutralino DM, which can also contribute to the invisible decay of the Higgs boson, becomes crucial.

In Ref.\,\cite{Barman:2022jdg}, we investigated the current status of the pMSSM parameter space with ten free parameters, that accommodates a light neutralino DM 
satisfying the upper bound on the relic density, with Higgsino-like NLSP. We considered both positive and negative values for the Higgsino mass parameter $\mu$. 
The implications from the latest direct detection experiments in both spin-independent and spin-dependent interactions were studied in conjunction with the current bounds from Higgs invisible measurements, heavy Higgs searches as well as electroweakino searches at the LHC.
We found that the latest direct detection limit from the LZ collaboration puts the $\mu>0$ scenario under severe tension. For the $\mu<0$ scenario, the LZ bound along with the constraints from electroweakino searches at colliders exclude most of the parameter space, except for very light Higgsinos, having masses $125-145$\, GeV and $145-160$\,GeV in the $Z$ and $h$ funnel regions, respectively.



In this paper, we present comprehensive and exhaustive arguments for the results found in Ref.\,\cite{Barman:2022jdg}. To further concretise our findings, we study the interplay of direct detection and collider limits in a simplified scenario consisting of the SM extended by a spin-$1/2$ Majorana fermion DM in analogy with the neutralino $\tilde{\chi}_{1}^{0}$ DM in the pMSSM framework. 
We then investigate the light Higgsinos surviving in the negative $\mu$ scenario and the current analyses sensitive to them. We choose benchmarks from the different allowed regions of our parameter space and perform a dedicated analysis of the $3l+\met$ channel for light Higgsinos in the $\mu<0$ scenario using a machine learning algorithm, \texttt{XGBOOST}, to explore its potential sensitivity.
We further extend our previous work to study the impact of light staus. We present benchmarks where the Higgsino can decay to staus and perform an analysis with tau leptons in the final state to explore the sensitivity for Run-3.
Finally, we discuss how the status changes for a thermal neutralino in non-standard cosmological scenarios.

We organise this paper as follows: In Sec.~\ref{sec:parameter_space}, we summarize the Higgs and electroweakino sectors of the MSSM, most relevant to the present analysis. The parameter space of interest, scanning technique and the scan ranges, are also discussed in the same section. The impact of constraints from LEP measurements, flavor observables, and Higgs measurements at the LHC, is examined in Sec.~\ref{sec:constraints}. We devote Sec.~\ref{sec:DMconstraints} to exploring the implications from relic density bounds. The effect of constraints from direct detection measurements on the $Z$ and $h$ funnel regions are scrutinized in Sec.~\ref{ssec:Zfunnel} and \ref{ssec:hfunnel}. Sec.~\ref{ssec:simplified} examines a simplified scenario with the SM extended by a stable Majorana fermion. 
We discuss the effect of electroweakino constraints in Sec.\,\ref{sec:EWINOconstraints}.
Details and results from our collider analysis targeted on smaller Higgsino mass $\mu \lesssim 160~$GeV are presented in Sec.~\ref{sec:collider}. 
In Sec.\,\ref{sec:stau} and Sec.\,\ref{sec:non_std}, we respectively discuss how our results change when we have light staus and in scenarios of non-standard cosmology. We conclude in Sec.~\ref{sec:conclusion}.

\section{The ${\rm p}$MSSM framework and the parameter space}
\label{sec:parameter_space}

In the pMSSM framework, the lightest neutralino $\lspone$ is a well-motivated DM candidate, provided it is the LSP. The $\lspone$ eigenstate can be written in terms of the Bino~($\tilde{B}$), neutral Wino~($\tilde{W}_3$) and neutral Higgsinos~($\tilde{H}_1^0$ and $\tilde{H}_2^0$),
\begin{eqnarray}
\tilde{\chi}_1^0 = N_{11}\tilde{B} + N_{12}\tilde{W}^3 + N_{13}\tilde{H}_1^0 + N_{14}\tilde{H}_2^0,
\label{eq:chi10}
\end{eqnarray}
where, $N_{1i}$ represents the amount of Bino ($i=1$), Wino ($i=2$) and the Higgsino ($i=3,4$) admixtures. In the present work, we are interested in the region of the pMSSM parameter space where $\lspone$ is the LSP and \textit{`light'}, i.e., $M_{\lspone} \leq M_h/2$, such that it is kinematically feasible for the SM-like Higgs boson $h$ to decay into a pair of $\lspone$'s, thus contributing to its invisible decay. $\lspone$ interacts with other electroweakinos via the exchange of SM $Z/W^{\pm}$ bosons and the pMSSM Higgs bosons, with the coupling strengths determined by their electroweakino composition. These various interactions of the LSP become important factors in the calculation of its relic density and direct detection cross-sections. The $\lspone\lspone Z$ coupling can be expressed as follows\,\cite{Djouadi:2005gj},  
\begin{eqnarray}
g_{\tilde{\chi}_1^0\tilde{\chi}_1^0 Z}^L = -\frac{g}{2{\rm cos}\theta_W}\left(|N_{13}|^2-|N_{14}|^2\right), ~~~ g_{\tilde{\chi}_1^0\tilde{\chi}_1^0 Z}^R = -g_{\tilde{\chi}_1^0\tilde{\chi}_1^0 Z}^L,
\label{eq:chiZ_gen}
\end{eqnarray}
where $g$ is the SU(2) gauge coupling, $\theta_W$ is the Weinberg angle, and $L, R$ denotes the respective couplings with the left and right-handed $\tilde{\chi}_1^0$. Similarly, the coupling of $\lspone$ with the three neutral Higgs bosons of the MSSM Higgs sector can be expressed as\,\cite{Djouadi:2005gj}:
\begin{eqnarray}
g_{\tilde{\chi}_1^0\tilde{\chi}_1^0(h/H/A)}^{L} = g\left(N_{12}-{\rm tan}\theta_W N_{11}\right)\left(e_{h/H/A} N_{13}+d_{h/H/A} N_{14}\right), \nonumber \\
g_{\tilde{\chi}_1^0\tilde{\chi}_1^0(h/H)}^{R}=g_{\tilde{\chi}_1^0\tilde{\chi}_1^0(h/H)}^{L}, ~~~ g_{\tilde{\chi}_1^0\tilde{\chi}_1^0 A}^{R} = -g_{\tilde{\chi}_1^0\tilde{\chi}_1^0 A}^{L},
\label{eq:chih_gen}
\end{eqnarray}
where,
\begin{eqnarray}
e_h = -{\rm sin}\,\alpha, ~~ e_H = {\rm cos}\,\alpha, ~~ e_A = -{\rm sin}\,\beta, \nonumber\\
d_h = -{\rm cos}\,\alpha, ~~ d_H = -{\rm sin}\,\alpha, ~~ d_A = {\rm cos}\,\beta,
\label{eq:e_and_d}
\end{eqnarray}
where $\alpha$ is the mixing angle in the CP–even neutral Higgs sector and ${\rm tan}\beta$ is the ratio of the vacuum expectation values ({\it vevs}) of the two Higgs doublets.


We require the light CP-even neutral Higgs boson $h$ to be consistent with the properties of the observed SM-like Higgs boson. $h$ can decay via $h \to \lspone\lspone$, provided $M_{\lspone} \leq M_{h}/2$ and $|g_{\lspone\lspone h}| > 0$. As previously discussed, the former condition requires a Bino-dominated $\lspone$ in order to evade the lower bounds on $M_{\chonepm}$ from LEP measurements~\cite{OPAL:2003wxm}.
We also notice from Eqn.\,\ref{eq:chih_gen} that $g_{\tilde{\chi}_1^0\tilde{\chi}_1^0h}$ becomes zero when $\lspone$ is a pure Gaugino, referred to Bino and Wino collectively, or a pure Higgsino. Therefore, a Gaugino-dominated $\lspone$ must have some Higgsino admixture to couple with the SM-like Higgs boson, which typically entails a dominant Higgsino-like next-to-lightest supersymmetric particle~(NLSP). Furthermore, the lower bounds on the masses of Higgsinos are less stringent than that of the Winos. Accordingly, in the present study, we are led to the parameter space where $\lsptwo,~\lspthree$ and $\chonepm$ have a predominant Higgsino composition, while $\lspfour$ and $\chtwopm$ have a dominant Wino admixture.

Additionally, $\tilde{\chi}_1^0$ interacts with the SM fermions and sfermions.
Searches at the LHC have derived strong lower bounds on the first two generations of squarks. For example, a single non-degenerate squark is constrained above $\gtrsim 1200~$GeV for $M_{\lspone} \sim 60~$GeV~\cite{ATLAS:2020syg}.
Therefore, the effect of their interactions on the observables related to the LSP will be negligible and motivates fixing their mass parameters to high values, say 5\,TeV.

The sleptons are relatively weakly constrained from collider searches and their couplings with $\tilde{\chi}_1^0$ can be expressed as\,\cite{Djouadi:2005gj}, 
\begin{eqnarray}
g_{\tilde{\chi}_1^0l\tilde{l}} = \sqrt{2}g{\rm sin}\theta_W\left[Q_l\left(N_{11}{\rm cos}\theta_W+N_{12}{\rm sin}\theta_W\right) + \left(I_l^{3}-Q_l{\rm sin}^2\theta_W\right)\frac{N_{12}{\rm cos}\theta_W-N_{11}{\rm sin}\theta_W}{{\rm cos}\theta_W{\rm sin}\theta_W}\right],
\label{eq:chil_gen}
\end{eqnarray}
where $Q_l$ and $I_l^{3j}$ are the charge of the lepton and the third component of isospin of the lepton, respectively. 
Among the three generations of sleptons, the staus have the weakest limits from collider searches. Therefore, the presence of light staus can impact the parameter space of light neutralino thermal DM.
We perform our scan in two parts $-$ without and with light staus.

The relevant input parameters are: the Gaugino masses, $M_{1}$ (Bino) and $M_{2}$ (Wino), the Higgsino mass parameter $\mu$, the ratio of the Higgs vacuum expectation values $\tan\beta$, the pseudoscalar mass $M_A$, the mass of the third generation squarks $\{M_{\tilde{Q}_{3l}}$, $M_{\tilde{t}_{R}}$, $M_{\tilde{b}_{R}}\}$, the trilinear coupling of the stop, $A_{t}$, and the gluino mass parameter, $M_3$. We perform a random scan over these ten input parameters in the ranges specified below:
\begin{eqnarray}
& 30~{\rm GeV} < M_{1} < 100~ {\rm GeV}, ~1~{\rm TeV} < M_{2} < 3~ {\rm TeV}, \nonumber \\ 
& 100~{\rm GeV} < |\mu| <~2~ {\rm TeV}, \nonumber \\
& 2 <  \tan{\beta} < 50, ~100~{\rm GeV} < M_{A} < 6~{\rm TeV}, \nonumber \\
&3~{\rm TeV} < M_{\tilde{Q}_{3L}} < 20~{\rm TeV}, ~3~{\rm TeV} < M_{\tilde{t}_{R}} < 20~{\rm TeV}, \nonumber \\
& 3~{\rm TeV} < M_{\tilde{b}_{R}} < 20~{\rm TeV}, ~ -20~{\rm TeV} < A_{t} < 20~ {\rm TeV}, \nonumber \\
& 2~{\rm TeV} < M_{3} < 5~ {\rm TeV}.
\label{eq:scan}
\end{eqnarray}

For the first part of the scan, i.e., without the light staus, we decouple the first two generations of squarks and all the three generations of sleptons from the spectrum and set the following pMSSM input parameters to a fixed value,
\begin{eqnarray}
& M_{\tilde{Q}_{1,2L}} = M_{\tilde{u}_{1,2R}} = M_{\tilde{d}_{1,2R}} = 5\,{\rm TeV}, ~ A_{u/d/c/s/b} = 0, \nonumber \\
& M_{\tilde{L}_{1,2,3L}} = M_{\tilde{e}_{1,2,3R}} = 2\,{\rm TeV}, ~ A_{e/\mu/\tau} = 0.
\label{eq:fix}
\end{eqnarray}
Here, $\{M_{\tilde{Q}_{1,2L}}, M_{\tilde{u}_{1,2R}},M_{\tilde{d}_{1,2R}}\}$ are the first and second generation squark mass parameters, and $\{M_{\tilde{L}_{1,2,3L}}, M_{\tilde{e}_{1,2,3R}}\}$ are the left and right-handed slepton mass parameters. 

We perform separate scans for the positive and negative values of $\mu$ to examine the role of $sgn(\mu)$ on the results. It is also worth noting that within the parameter space of our interest, DM relic density and direct detection constraints restrict $M_{\lspone}$ to the $Z$ and $h$ funnel regions only\,\cite{KumarBarman:2020ylm,VanBeekveld:2021tgn,Carena:2018nlf}. 
We begin by performing a random scan over the specified parameter space.
Subsequently, in order to sufficiently populate the funnel regions, we perform a dedicated scan where we dynamically tune $M_1$ such that $M_{\tilde{\chi}_1^0}$ is within $M_{Z}/2 \pm 5~$GeV or $M_{h}/2 \pm 3~$GeV, where $M_h$ is the mass of the Higgs boson estimated by \texttt{FeynHiggs} for each point in the parameter space. Additionally, we extract the pole mass of the top quark, $M_{t}$, randomly from a gaussian distribution with a central value of 173.21\,GeV and a standard deviation of 0.55\,GeV\,\cite{Olive_2014}\,\footnote{Note that using the more recent measured value of $M_t=172.69$\,GeV\,\cite{Workman:2022ynf} does not affect our final results, since it is always possible to adjust the stop mass parameters, which we scan over, to obtain the measured value of the Higgs boson mass.}

To study the effect of light staus, we perform a second scan where we vary the stau mass parameters, in a way that the staus become the NLSPs. 
We can accommodate single left-handed (LH) or right-handed (RH) staus with masses around 100-150\,GeV according to Ref.\,\cite{ATLAS:2019gti}. 
According to Eqn.\,\ref{eq:chil_gen}, the coupling of sleptons with the LSP neutralino depends on the third component of isospin of the slepton, which is higher for RH staus as compared to the LH ones. Therefore, RH staus will have greater impact on the relic density.
To avoid constraints from additional light LH staus and light sneutrino, we consider only light RH staus. This is a minimal extension to our earlier scan\,\cite{Barman:2022jdg}, which is
achieved by lowering the parameter $M_{\tilde{e}_{3R}}$, and varying it between 85\,GeV and 500\,GeV, while keeping $M_{\tilde{L}_{3L}}$ and $A_\tau$ fixed at 2\,TeV and 0, respectively. 

We use \texttt{FeynHiggs 2.18.1}~\cite{Heinemeyer:1998yj,Heinemeyer:1998np,Degrassi:2002fi,Frank:2006yh,Hahn:2013ria,Bahl:2016brp,Bahl:2017aev,Bahl:2018qog} to generate the particle spectrum of the SUSY particles and of the Higgs bosons for each set of input parameters\,\footnote{The input parameters are read and written in the SLHA file as on-shell parameters by the \texttt{FeynHiggs} code. We find no significant changes in our result when we use a different spectrum generator, such as \texttt{SoftSUSY-4.1.17}, which provides the output SLHA with $\overline{DR}$ parameters.} and branching fractions for the decay of the Higgs bosons. \texttt{MicrOMEGAS 5.2.13}~\cite{Belanger:2004yn,Belanger:2006is,Belanger:2008sj,Belanger:2010gh,Belanger:2013oya,Belanger:2020gnr} is used to compute the LEP, flavor physics, and dark matter observables, as further discussed in Sec.~\ref{sec:constraints}, where we also describe the various relevant constraints and their impact on the scanned parameter space. 

\section{Constraints from LEP, flavor observables and the Higgs sector}
\label{sec:constraints}

As previously discussed, we associate the lightest CP-even Higgs boson $h$ with the discovered Higgs boson at the LHC, and require that the masses and branching of the two match within experimental and theoretical uncertainties. Measurements at the LHC have put the mass of the Higgs boson at $125.38 \pm 0.14~$GeV\,\cite{CMS:2020xrn}. We require that the mass of $h$, as computed by \texttt{FeynHiggs 2.18.1}, be in the range $122~\rm{GeV} - 128~$GeV. 
Considering theoretical uncertainties stemming from the dependence on the renormalisation scheme and scale, from the assumption of zero external momentum in two-loop corrections, and also absence of higher order corrections, we allow for a conservative 3\,GeV window around the experimentally measured value\,\cite{Allanach:2004rh,Heinemeyer:2007aq,Borowka:2015ura}\,\footnote{Additionally, we can compute the error in estimating the mass of the Higgs boson using \texttt{FeynHiggs 2.18.1}, denoted as $\Delta_{M_h}^{FH}$. We discuss in a later section the impact of a precise Higgs mass condition, defined as requiring $M_h \pm \Delta_{M_h}^{FH} \in \left[125.38-2\times  0.14\,\rm{GeV}, 125.38+2\times0.14\,\rm{GeV}\right]$, i.e., within 2$\sigma$ of the experimentally measured mass of the Higgs boson.}.
At low values of tan\,$\beta$, consistency with the Higgs boson mass constraint requires large $A_t$ and stop masses. However, large $A_t$ can give rise to color and charge-breaking minima (CCB)\,\cite{Camargo-Molina:2013sta,Chowdhury:2013dka,Blinov:2013fta}, where the scalar partners of top quarks having color and electric charges develop a non-zero {\it vev} and the corresponding minima is lower than the minima of the Higgs field. This can be evaded, given, $|X_t|<\sqrt{6 M_{t_1}M_{t_2}}$\,\cite{Chowdhury:2013dka}, where $X_t=A_t-\mu/\rm{tan}\,\beta$ and $M_{t_{1,2}}$ represents the stop masses. In the present study, we see that the CCB condition has no significant effect on the allowed parameter space. 

We also apply limits on the invisible decay width of $Z$-boson $\Gamma_{Z} \leq 2~$MeV\,\cite{ALEPH:2005ab}, chargino mass $M_{\lspone} \geq 103.5~$GeV\,\cite{OPAL:2003wxm}, and cross-section of neutralino pair production $\sigma(\lsptwo\lspone) \leq 0.1~$pb in final states with jets $+\met$\,\cite{OPAL:2003wxm}, as obtained from LEP. We impose constraints on various flavor physics observables, such as, $Br(b\rightarrow s\gamma) = (3.32\pm 0.16) \times 10^{-4}$\,\cite{HFLAV:2016hnz}, $Br(B_s\rightarrow \mu^+\mu^-) = 3.0^{+0.67}_{-0.63} \times 10^{-9}$\,\cite{CMS:2014xfa}, and $Br(B\rightarrow \tau\nu)= 1.28\pm 0.25$\,\cite{Belle:2010xzn}, allowing $2\sigma$ uncertainty around the best-fit values. We have used \texttt{MicrOMEGAS 5.2.13}\,\cite{Belanger:2004yn,Belanger:2006is,Belanger:2008sj,Belanger:2010gh,Belanger:2013oya,Belanger:2020gnr} to calculate both the LEP and flavor physics observables. 

We also impose the Higgs signal strength constraints on the 
parameter space 
using the \texttt{HiggsSignal 2.6.2}\,\cite{Bechtle:2013xfa, Stal:2013hwa, Bechtle:2014ewa} package, while limits from the heavy Higgs boson searches at the LHC are imposed using the \texttt{HiggsBounds 2.10.0}\,\cite{Bechtle:2008jh,Bechtle:2011sb,Bechtle:2012lvg,Bechtle:2013wla,Bechtle:2015pma} package. The parameter space is also required to satisfy the most stringent upper bound on the Higgs invisible branching ratio $Br(h \to \rm{inv}) \leq 11\%$, as measured by the ATLAS collaboration~\cite{ATLAS:2024lyh}, which is stronger than the current CMS bound (15\%\,\cite{CMS:2023sdw}). Hereafter, constraints on the mass, signal strength~(imposed through \texttt{HiggsSignal 2.6.2}), and the invisible branching ratio of the Higgs boson are collectively referred to as the constraints on ``Higgs properties". We summarize the constraints in Table~\ref{tab:constraints1}.  

\begin{table*}[!htb]
\centering
\begin{tabular}{|m{0.05\textwidth}|m{0.3\textwidth}|m{0.2\textwidth}|m{0.35\textwidth}|} \hline 
\textbf{Sr. No.} & \textbf{Observable} & \textbf{Calculated by} & \textbf{Constraint}\\
\hline \hline 
(1) & Light Higgs boson mass & \texttt{FeynHiggs 2.18.1} & 122\,GeV$<M_h<$128\,GeV \\
\hline
(2) & Higgs signal strength & \texttt{HiggsSignal 2.6.2} & 111 channels, $p$-value$>0.05$ \\
\hline
\multirow{2}{*}{(3)} & \multirow{2}{*}{Heavy Higgs bosons} & \multirow{2}{*}{\texttt{HiggsBounds 2.10.0}} & Constraints from collider searches of heavy \\
 &                             &                                              & Higgs bosons implemented in \texttt{HiggsBounds} \\
\hline
(4) & Invisible decay of Higgs boson & \texttt{FeynHiggs 2.18.1} & Br$(h\rightarrow\tilde{\chi}_1^0\tilde{\chi}_1^0)<0.11$ \\
\hline
(5) & Invisible decay of $Z$ boson from LEP & \texttt{MicrOMEGAS 5.2.13} & $\Gamma(Z\rightarrow \text{invisible})<2$\,MeV\\
\hline
(6) & Chargino mass limit from LEP & \texttt{FeynHiggs 2.18.1} & $M_{\chi_1^{\pm}}>103$\,GeV\\
\hline
(7) & LEP limits on neutralino & \multirow{2}{*}{\texttt{MicrOMEGAS 5.2.13}} & $\sigma(e^+e^-\rightarrow\tilde{\chi}_1^0\tilde{\chi}_2^0)\times{\rm Br}(\tilde{\chi}_2^0\rightarrow\tilde{\chi}_1^0+{\rm jets}) + $ \\ 
& in dijet + MET final states & & $\sigma(e^+e^-\rightarrow\tilde{\chi}_1^0\tilde{\chi}_3^0)\times{\rm Br}(\tilde{\chi}_3^0\rightarrow\tilde{\chi}_1^0+{\rm jets}) <0.1$\,pb\\
\hline
\multirow{3}{*}{(8)} & \multirow{3}{*}{Flavour observables} & \multirow{3}{*}{\texttt{MicrOMEGAS 5.2.13}} & $3.00\times10^{-4}<{\rm Br}(b\rightarrow s\gamma)<3.64\times10^{-4}$ \\
&    &    & $1.66\times10^{-9}<{\rm Br}(B_s\rightarrow \mu^+\mu^-)<4.34\times10^{-9}$ \\
&    &    & $0.78<\frac{({\rm Br}(B\rightarrow \tau\nu))_{obs}}{({\rm Br}(B\rightarrow \tau\nu))_{SM}}<1.78$ \\
\hline
\hline         
\end{tabular}
\caption{Summary of constraints from Higgs sector, LEP, and flavor observables.}
\label{tab:constraints1}
\end{table*}

\begin{figure*}[hbt!]
    \centering
    \includegraphics[width=0.48\textwidth]{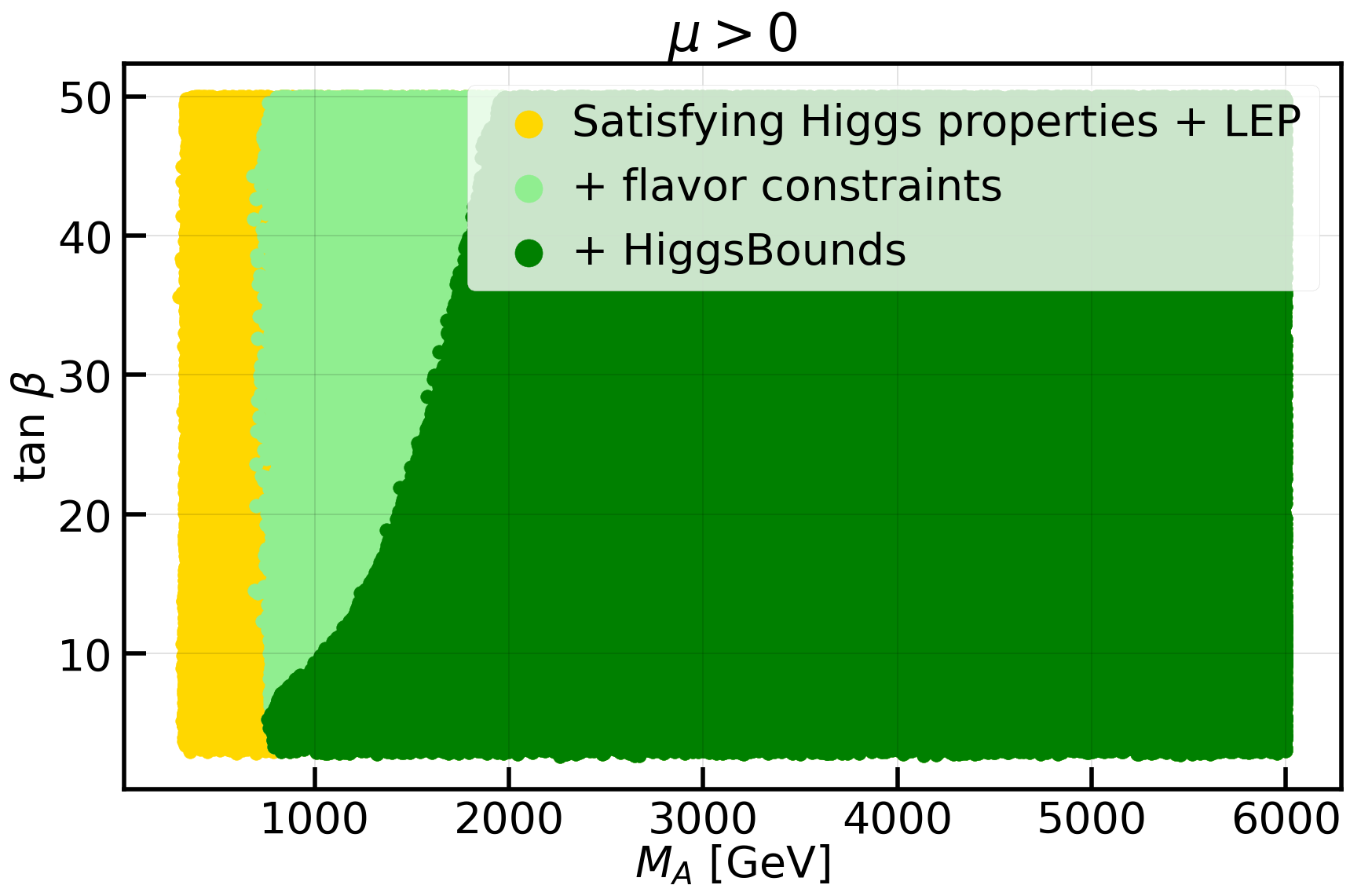}~\includegraphics[width=0.48\textwidth]{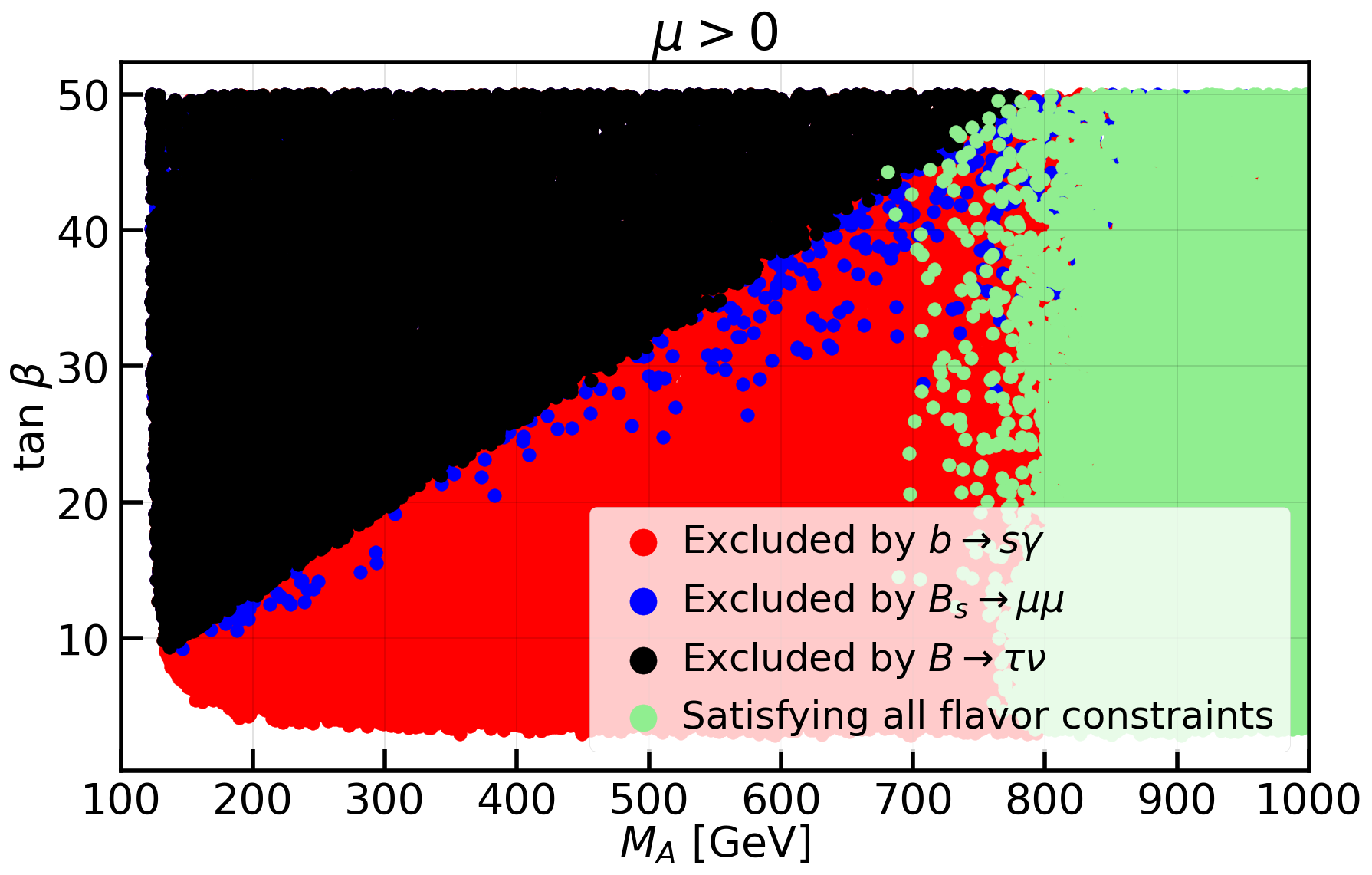}\\
    \includegraphics[width=0.48\textwidth]{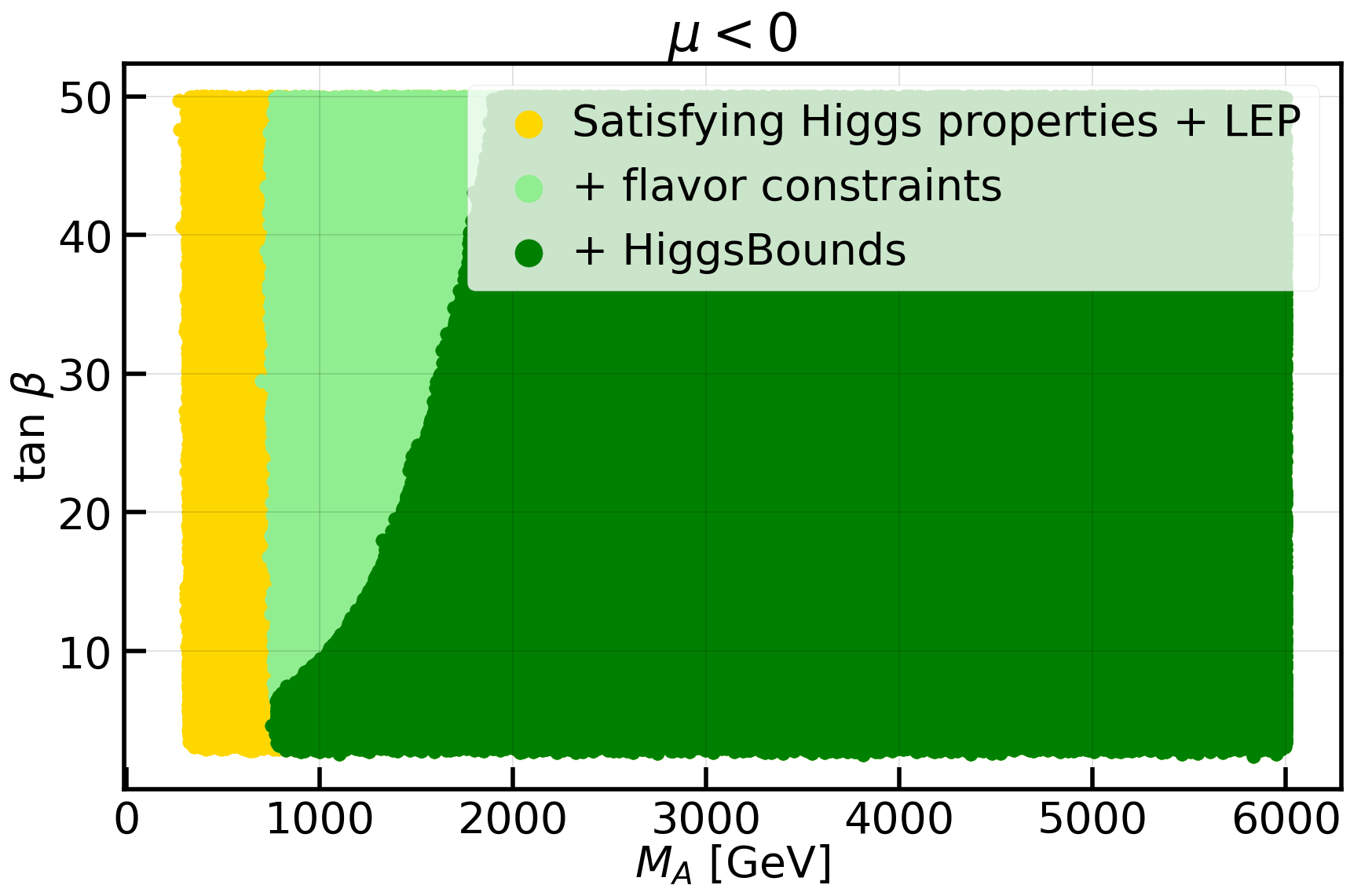}~\includegraphics[width=0.48\textwidth]{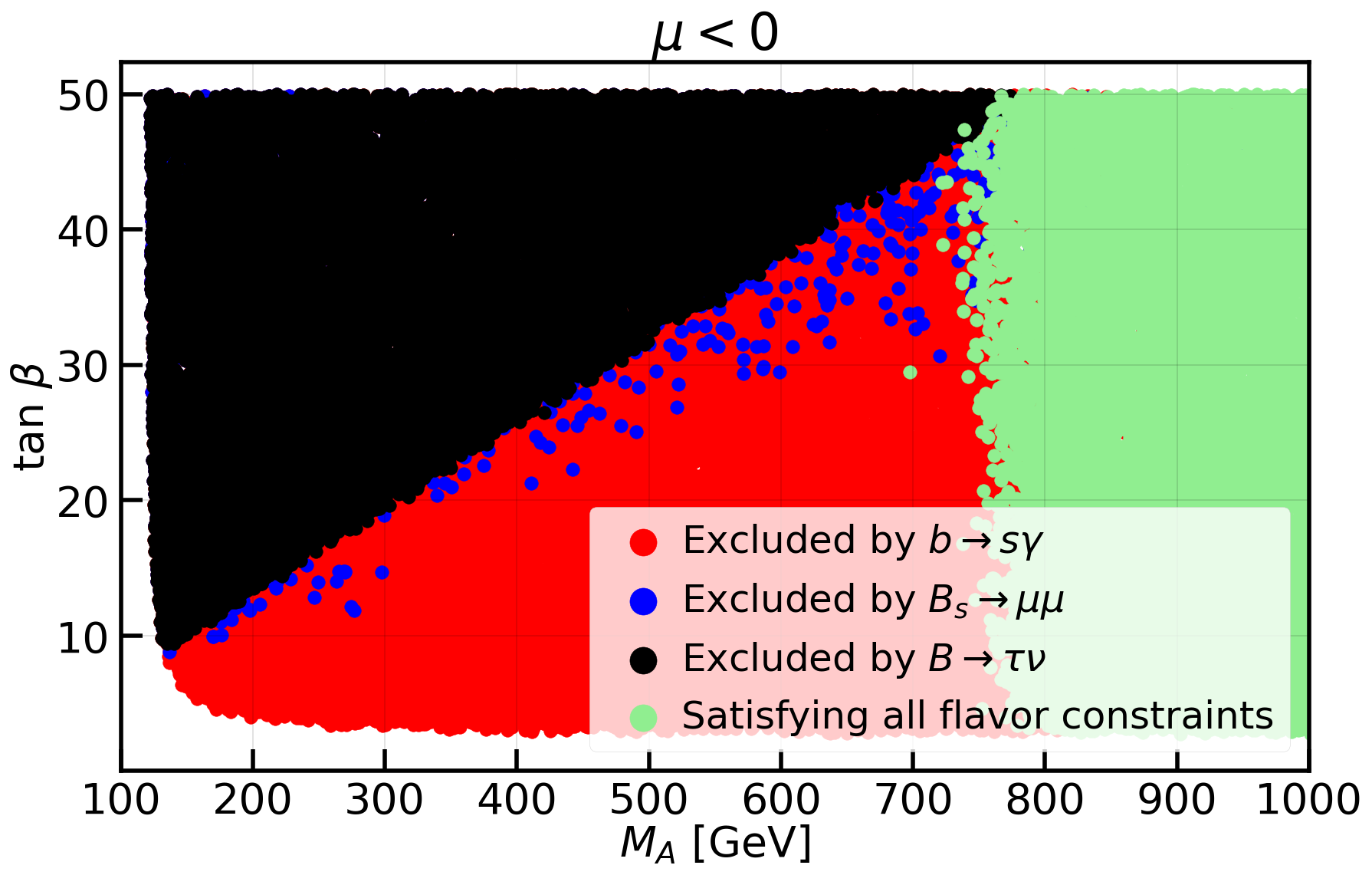}
    \caption{{\it Left:} Parameter space in the $M_A$-${\rm tan}\beta$ plane satisfying the Higgs properties, LEP constraints, flavor constraints, and results of heavy Higgs boson searches implemented in \texttt{HiggsBounds}; {\it Right:} Parameter space in the $M_A$-${\rm tan}\beta$ plane zoomed-in to show the effect of the different flavor observables for $M_A<1$\,TeV.
    The {\it top} and {\it bottom} panels are for $\mu>0$ and $\mu<0$ respectively.}
    \label{fig:constraint1}
\end{figure*}

We show the scanned points in the $M_A - \tan\beta$ plane for $\mu > 0$~(top) and $\mu < 0$~(bottom) and the effect of various constraints in Fig.~\ref{fig:constraint1}. In the left panels, we show the parameter points allowed by the Higgs boson measurements and LEP constraints~(yellow), flavor physics bounds~(light-green), and the constraints from heavy Higgs searches included via \texttt{HiggsBounds}~(dark-green), applied consecutively. In the right panels, we zoom in on the $M_A \leq 1~$TeV region, showing the parameter space excluded by different flavor-changing processes. The flavor physics observables together exclude points with low $M_A~$($\lesssim 700~$GeV). The low $\tan\beta$ region is excluded by $Br(b \to s\gamma)$ while the other two flavor observables are more impacting in the high $\tan\beta$ region. Points in the high $\tan\beta$ region for $700~\textrm{GeV} \lesssim M_A \lesssim 2~$TeV are excluded by constraints from heavy Higgs boson searches at the LHC applied via the \texttt{HiggsBounds} package. Among the various heavy Higgs searches considered, results from the search for heavy Higgs bosons decaying into a pair of tau leptons at $\sqrt{s}=13~$TeV with $\mathcal{L}=139~\textrm{fb}^{-1}$ data~\cite{ATLAS:2020zms} by the ATLAS collaboration resulted in the most stringent constraints on the parameter space of our interest. We further observe that the LEP constraints~(Table\,\ref{tab:constraints1} (2), (3), and (4) applied together), \texttt{HiggsSignal} and the upper limit on the invisible branching of $h$ to invisible final states do not show any specific trend in excluding the parameter space in the $M_A$-${\rm tan}\beta$ plane. All these observations are irrespective of the sign of the $\mu$ parameter.


\section{Dark Matter constraints}
\label{sec:DMconstraints}

As discussed previously, the lightest supersymmetric particle in the pMSSM, here $\tilde{\chi}_1^0$, is a viable DM candidate. It can have a thermal production in the early Universe, which freezes-out. 
In the standard cosmology, we require the relic density of the LSP $\Omega_{\lspone}$ to be equal to the observed DM relic density as measured by the PLANCK collaboration $\Omega^{obs}_{DM}h^2 = 0.120\pm0.001$\,\cite{Aghanim:2018eyx}, which assuming a $2\sigma$ interval can vary from 0.118\,$\lesssim \Omega^{obs}_{DM}h^2 \lesssim$\,0.122. Lifting up the requisite that the neutralino LSP forms 100\% of the observed DM relic owing to the possibility of multicomponent DM, we can modify the relic density constraint to $\Omega_{\lspone} \lesssim 0.122$. We use the \texttt{MicrOMEGAS 5.2.13}\,\cite{Belanger:2004yn,Belanger:2006is,Belanger:2008sj,Belanger:2010gh,Belanger:2013oya,Belanger:2018ccd,Belanger:2020gnr} package to compute the relic density of $\tilde{\chi}_1^0$. 

In addition to the relic density constraint, we need to take into consideration the limits from dark matter direct detection~(DD) experiments which constrain the spin-dependent DM-neutron~(SDn), DM-proton~(SDp) and spin-independent~(SI) DM-nucleon interaction cross-sections as a function of mass of the DM.  We use \texttt{MicrOMEGAS 5.2.13} to compute these cross-sections and then compare them with the 90\% confidence level~(CL) upper limits quoted by the 
PICO-60 (SDp\,\cite{PICO:2019vsc}), PandaX-4T 
(SDn\,\cite{PandaX:2022xas}), and LZ (SI\,\cite{Aalbers:2022fxq}) experiments, since these are the strongest available bounds for each category in the DM mass range of 10\,GeV to 1\,TeV at present. 
The LZ collaboration sets an upper limit on the SI cross-section of a DM particle with mass in the $Z$-funnel to be $1.06\times 10^{-47}$\,cm$^2$, and in the $h$-funnel to be  $1.54 -1.64 \times 10^{-47}$\,cm$^2$ for $M_h\in122-128$\,GeV.
We tabulate the DM related constraints applied on the parameter space of our interest in Table\,\ref{tab:constraints2}.

\begin{table*}[!htb]
\centering
\begin{tabular}
{|m{0.05\textwidth}|m{0.3\textwidth}|m{0.18\textwidth}|m{0.37\textwidth}|}
\hline
\textbf{Sr. No.} & \textbf{Observable} & \textbf{Calculated by} & \textbf{Constraint}\\
\hline\hline
(9) & Relic density & \texttt{MicrOMEGAS 5.2.13} & $\Omega < 0.122$ (PLANCK)\\
\hline
\multirow{2}{*}{(10)} & \multirow{2}{*}{Limits on direct detection} & \multirow{3}{*}{\texttt{MicrOMEGAS 5.2.13}} & Spin-dependent proton: PICO-60 \\
& \multirow{2}{*}{cross-sections scaled with $\xi$ (Eqn.\,\ref{eq:xi})} &  & Spin-dependent neutron: PandaX-4T \\
&  &  & Spin-independent: LUX-ZEPLIN (LZ) \\
\hline
\hline         
\end{tabular}
\caption{Summary of DM relic density and direct detection constraints on the LSP neutralino.}
\label{tab:constraints2}
\end{table*}

The DD limits from the experimental collaborations are placed assuming that a single DM candidate constitutes the entire relic. Therefore, if the neutralino DM is underabundant, i.e., $\Omega_{\lspone} < 0.118$, then the DD limits are applied on scaled cross-sections. The scaling factor $\xi$ is taken unity when the LSP relic is within the experimental uncertainty, i.e., $0.118<\Omega_{LSP}<0.122$.
For $\Omega_{LSP}<0.118$, it is scaled by the ratio of the central value of the observed relic density to the computed relic density for $\lspone$, as follows:
\begin{equation}
    \xi = \frac{\Omega_{\lspone}}{0.120}
    \label{eq:xi}
\end{equation}

The recent upper limit on the SI cross-section $\sigma_\textrm{SI}$ derived by the LZ collaboration is roughly $\sim 3-4$ times stronger than the previous most stringent limits from PandaX-4T in the region of DM masses considered in this work\,\footnote{The LZ collaboration provides slightly stronger limits ($\sim 1.5$ times better) than the recent XENON-nT experiment\,\cite{XENON:2023cxc}.}. To demonstrate the role of the LZ result, we divide the constraints on our scanned parameter space into \textbf{``Before LZ''} which includes constraints from LEP, flavor, Higgs properties, heavy Higgs searches using \texttt{HiggsBounds}, relic density, and the DD experiments XENON-1T, PICO-60, and PandaX-4T, and \textbf{``After LZ''} with the constraint from the LZ experiment.

Although our scan is over a ten-dimensional parameter space, not all parameters contribute to the individual observables. The relic density is determined by the annihilation channels of the DM, which in the present scenario will dominantly proceed through the $s$-channel diagrams involving the $Z$ and the Higgs bosons as propagators. Scattering between the neutralino and the SM quarks and gluons, which forms the basis of the DD experiments, will involve the same propagators in the $t$-channel. It is worth noting that diagrams involving squark exchange do not play an important role due to strong lower limits on squark masses from searches at the LHC.
Therefore, the most important parameters for the DM constraints from relic density and the DD experiments are $-$ $M_1$, $M_2$, $|\mu|$, ${\rm tan}\beta$, and $M_A$, since these affect the couplings of the DM with the $Z$ and Higgs bosons as shown in Eqns.\,\ref{eq:chiZ_gen} and \ref{eq:chih_gen}. Large couplings are excluded by the DD experiments, whereas small values of couplings are excluded by the observed relic density constraint unless the mass of DM lies within a narrow window around half the mediator mass resulting in resonant enhancement of the DM annihilation cross-section. 

\begin{figure*}[hbt!]
    \centering
    \includegraphics[width=0.48\textwidth]{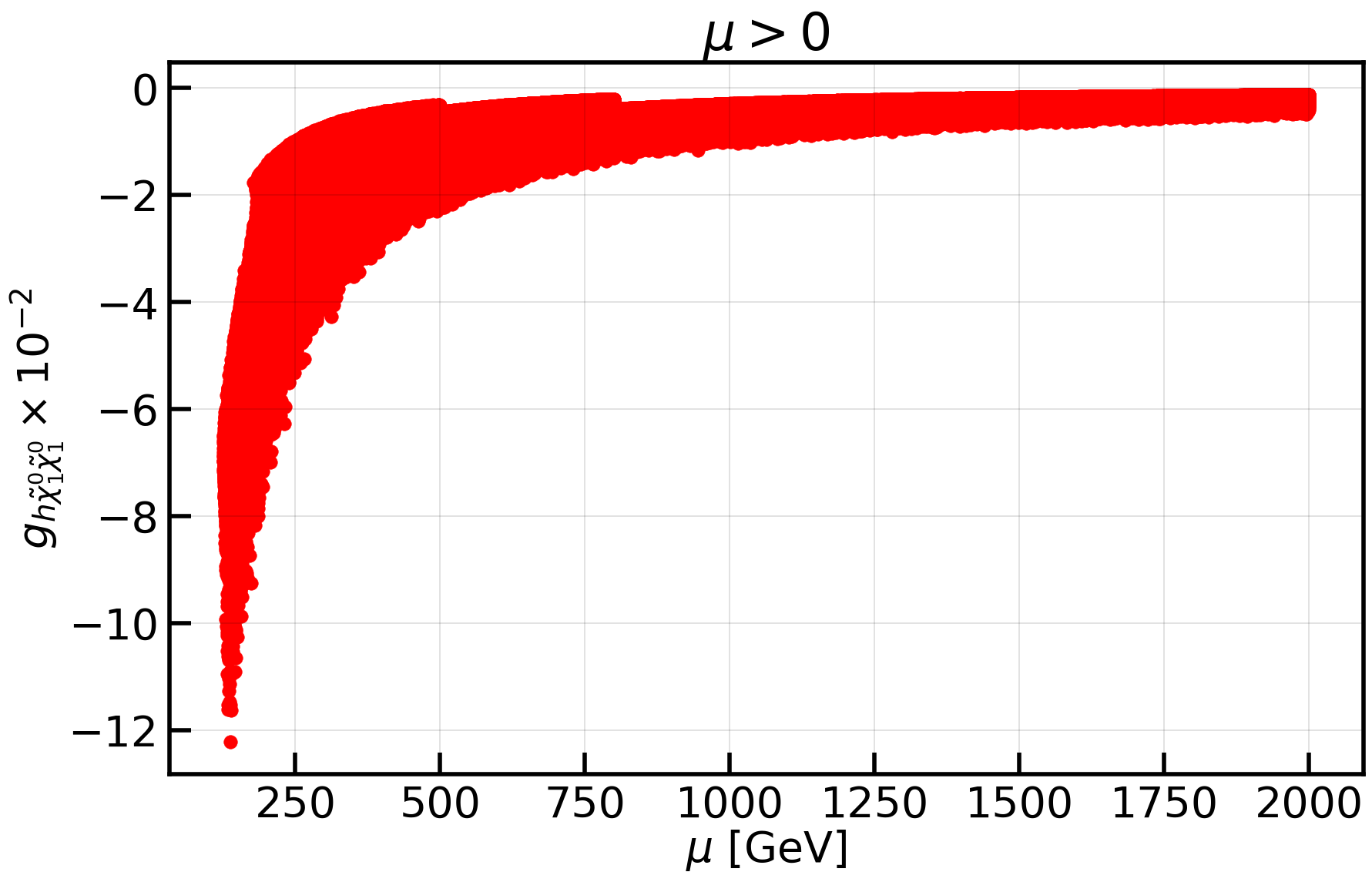}~
    \includegraphics[width=0.48\textwidth]{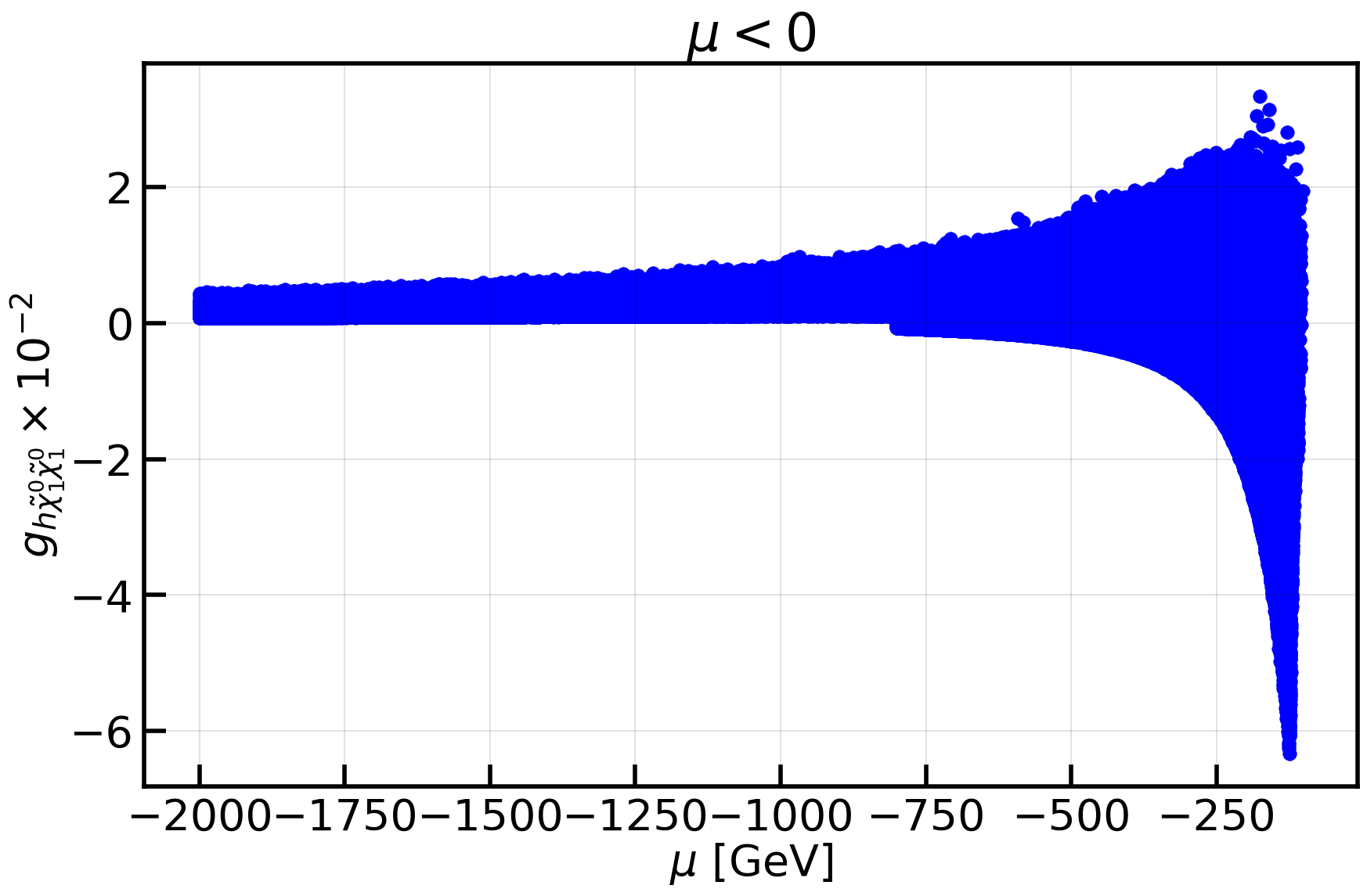}\\
    \includegraphics[width=0.48\textwidth]{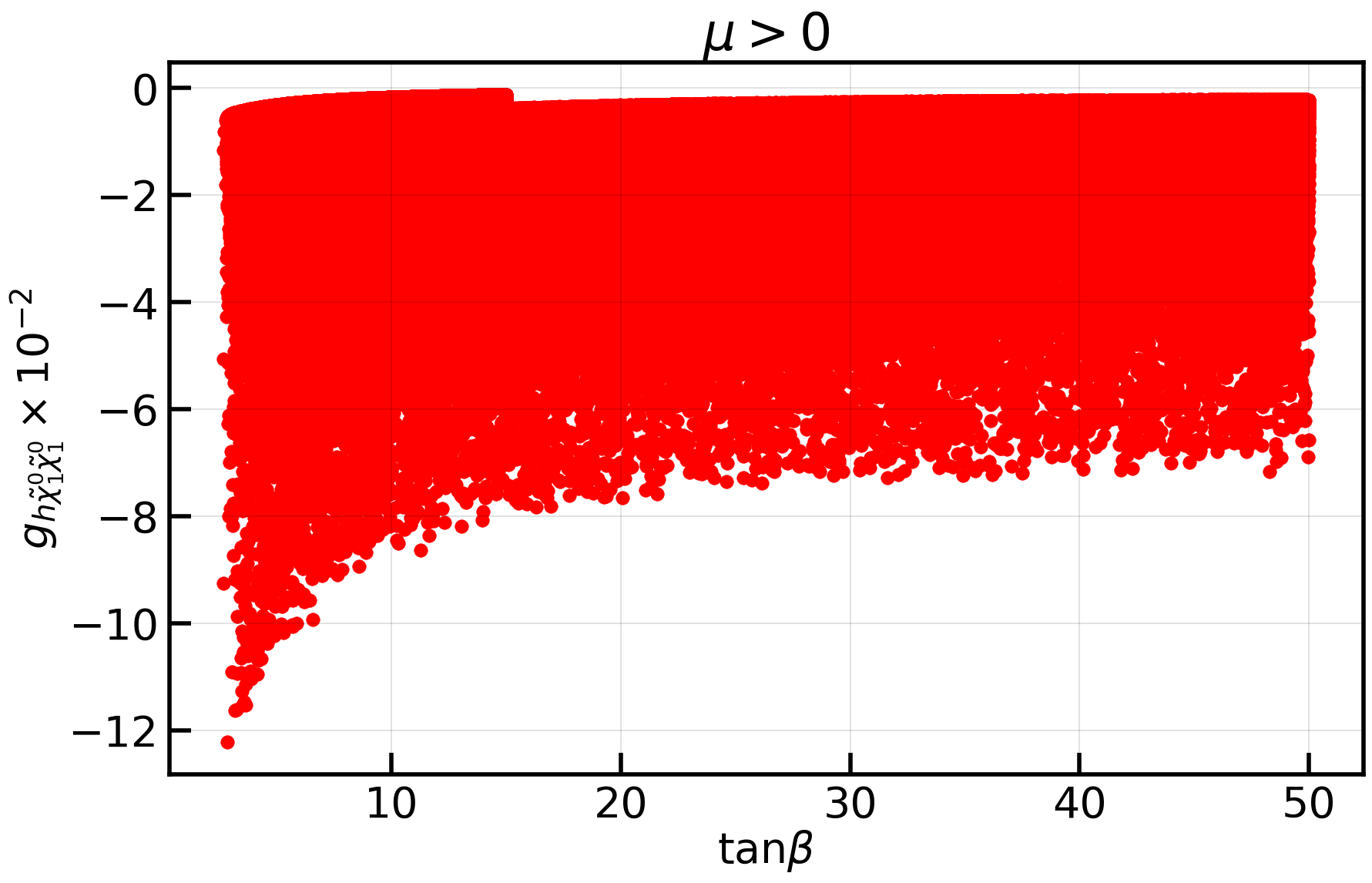}~
    \includegraphics[width=0.48\textwidth]{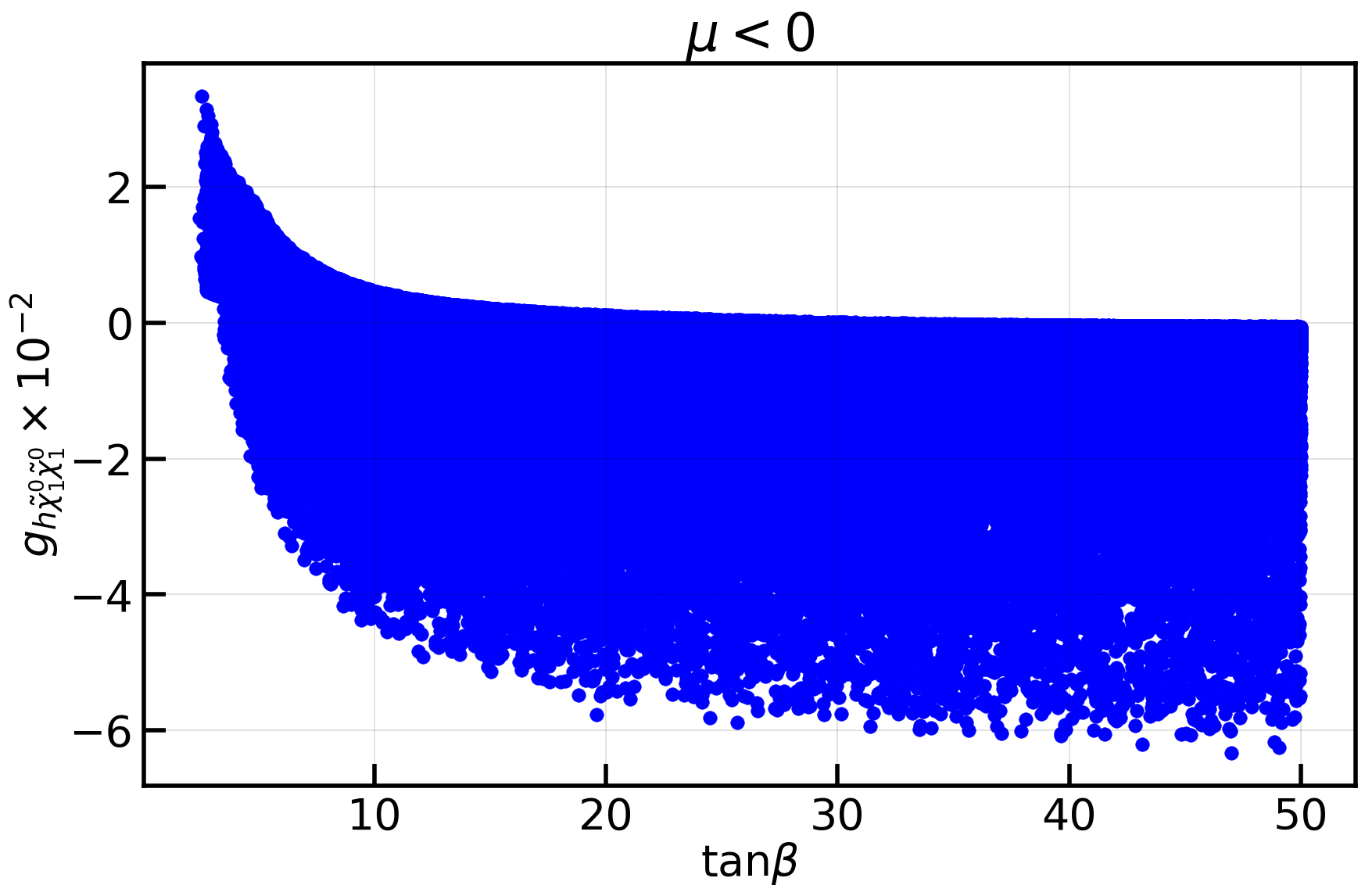} 
    \caption{Variation of the $g_{h\tilde{\chi}_1^0\tilde{\chi}_1^0}$ coupling with $\mu$ ({\it top}) and tan$\beta$ ({\it bottom}) for $\mu>0$ ({\it left}) and $\mu<0$ ({\it right}) for points satisfying Higgs constraints, LEP and flavor constraints.}
    \label{fig:couplings}
\end{figure*}

In the $Z$ funnel, the coupling depends only on the Higgsino components in the LSP, which decreases as we move to higher values of $\mu$. 
Rewriting and simplifying Eqn.\,\ref{eq:chih_gen} for the lighter CP-even Higgs boson in the limit $M_A\gg M_Z$ (where cos\,$\alpha\,\sim\,$sin\,$\beta$), we have
\begin{eqnarray}
g_{h\tilde{\chi}_1^0\tilde{\chi}_1^0} 
\approx -g\left(N_{12}-{\rm tan}\theta_W N_{11}\right)\left({\rm sin}\beta N_{14}-{\rm cos}\beta N_{13}\right)
\end{eqnarray}
Fig.\,\ref{fig:couplings} shows the coupling of the SM Higgs boson with the lightest neutralino as a function of the $\mu$ parameter ({\it top} panel) and tan$\beta$ ({\it bottom} panel) for both positive ({\it left} panel) and negative ({\it right} panel) values of $\mu$. For positive $\mu$, the coupling is always negative and can increase in magnitude with decreasing value of $\mu$ and tan$\beta$. For negative $\mu$, the maximal value  of the coupling increases for small $|\mu|$. Moreover, it can have either sign depending on the value of tan\,$\beta$ $-$ the coupling is negative at large tan\,$\beta$ and it increases with decreasing tan$\beta$, eventually becoming positive at low tan$\beta$, around $\sim 10$.
Therefore, we can have large magnitude of coupling for both high and low values of tan$\beta$.

The SI DD cross-section will further receive contribution from the heavy CP-even Higgs boson in the pMSSM ($H$) present in the $t$-channel, which has the following coupling to $\tilde{\chi}_1^0$ (from Eqn.\,\ref{eq:chih_gen}):
\begin{eqnarray}
    g_{H\tilde{\chi}_1^0\tilde{\chi}_1^0} 
    \approx g\left(N_{12}-{\rm tan}\theta_W N_{11}\right)\left({\rm sin}\beta N_{13}+{\rm cos}\beta N_{14}\right)&,~~M_A\gg M_Z,
    \label{eq:chi10_H}
\end{eqnarray}
which has a sign opposite to the sign of $\mu$. It couples to the SM up-type and down-type quarks as follows:
\begin{eqnarray}
    g_{Huu} = i\frac{m_u}{v}\frac{{\rm sin}\alpha}{{\rm sin}\beta} \propto -{\rm cot}\beta& , ~~M_A\gg M_Z\nonumber\\
    g_{Hdd} = i\frac{m_d}{v}\frac{{\rm cos}\alpha}{{\rm cos}\beta} \propto {\rm tan}\beta& , ~~M_A\gg M_Z
    \label{eq:quark_H}
\end{eqnarray}
The DM-quark scattering cross-section involves the product of the coupling $g_{H\tilde{\chi}_1^0\tilde{\chi}_1^0}$ with $g_{Huu}$ or $g_{Hdd}$.
For $\mu>0$, the $h$ and $H$ contributions have opposite signs for up-type quarks, whereas for down-type quarks they add up, moreover, $g_{Hdd}$ is tan$\beta$ enhanced. 
The constructive interference between the contributions for down-type quarks is more effective than the destructive interference between the up-type quark contributions from $h$ and $H$.
For $\mu<0$, when tan$\beta$ is large and $g_{h\tilde{\chi}_1^0\tilde{\chi}_1^0}$ is negative, contributions from the two Higgs bosons destructively interfere for down-type quarks, and add up for up-type quarks. Since the cancellation is for coupling with down-type quarks, it is more effective for larger values of tan$\beta$. For small tan$\beta$, $g_{h\tilde{\chi}_1^0\tilde{\chi}_1^0}$ turns positive, and follows the same trend as for $\mu>0$. 

Having discussed the trends of the various couplings of the LSP DM, let us have a closer look at the $Z$ and $h$ funnels of both positive and negative $\mu$ and study how the recent LZ limit has affected these scenarios.

\subsection{$Z$ funnel}
\label{ssec:Zfunnel}

\begin{figure}[hbt!]
    \centering
    \includegraphics[width=0.5\textwidth]{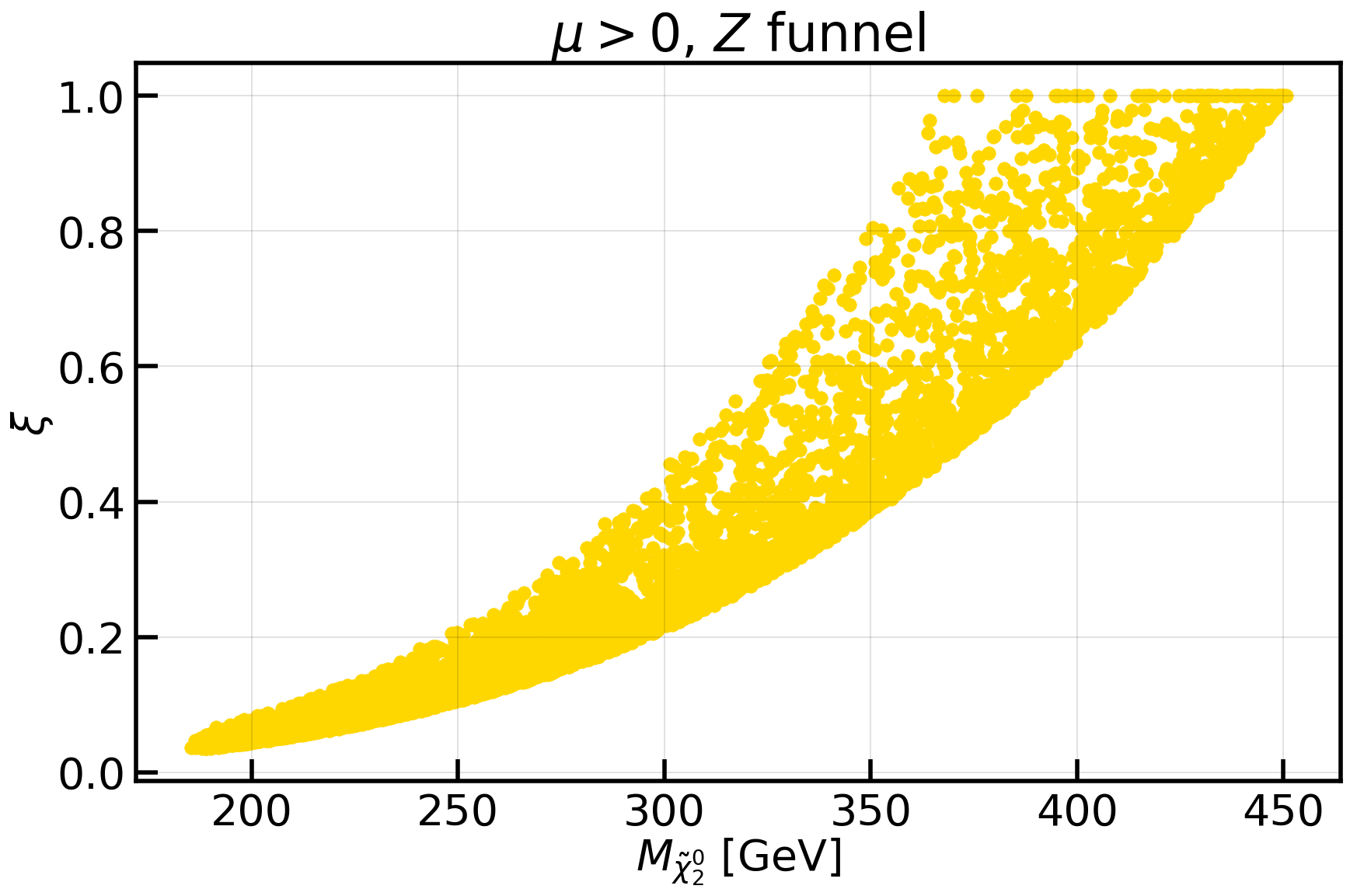}~\includegraphics[width=0.5\textwidth]{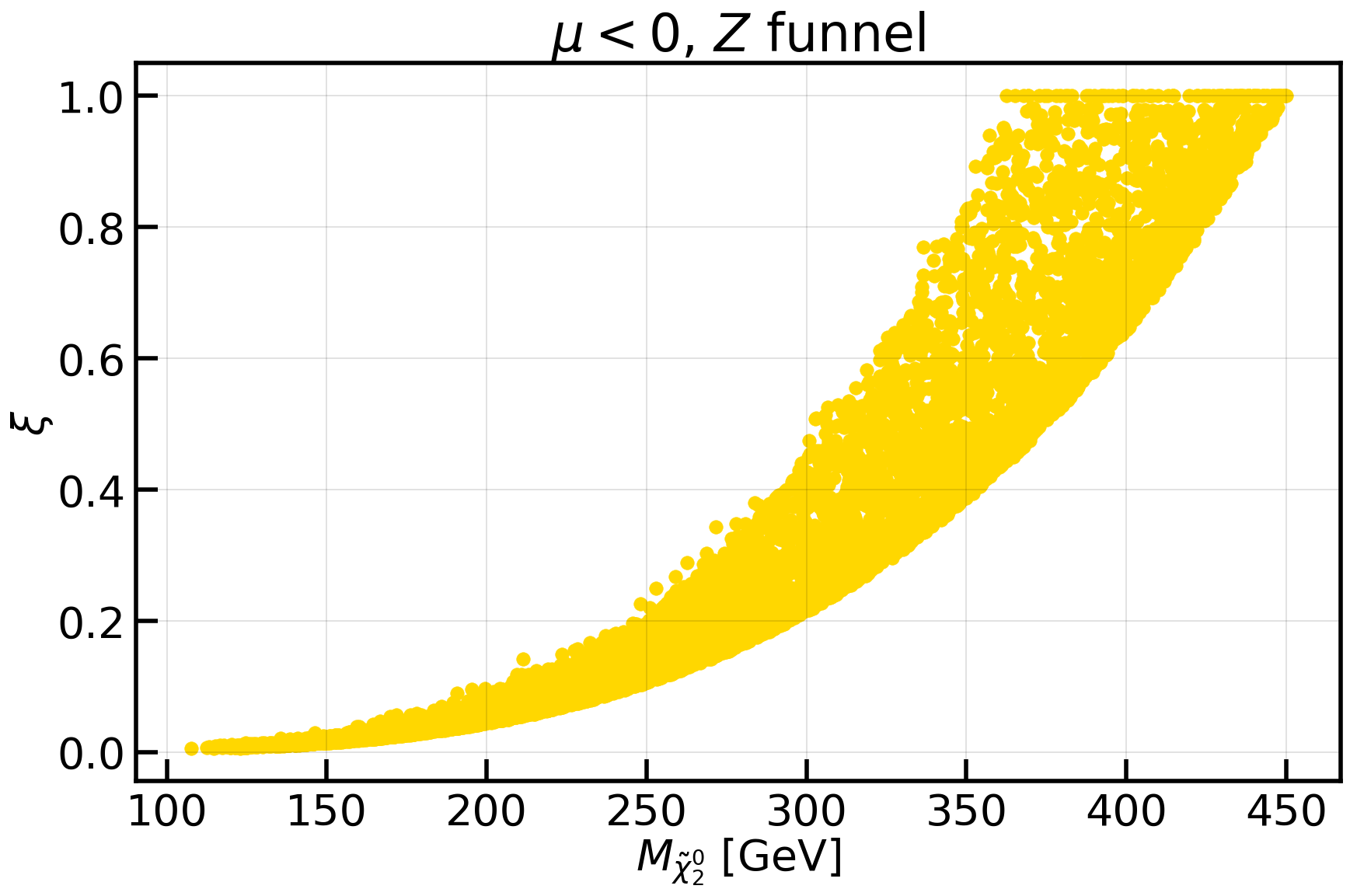}
    \caption{Fraction of DM satisfied by $\tilde{\chi}_1^0$ ($\xi=\frac{\Omega_{\rm LSP}}{0.120}$) as a function of $M_{\tilde{\chi}_2^0}$ for both $\mu>0$ ({\it left}) and $\mu<0$ ({\it right}) with the \textbf{``Before LZ''} set of cuts.}
    \label{fig:Zfunnel-xi}
\end{figure}

In the $Z$ funnel, the observed relic density bound restricts $|\mu|$ to small values since the $g_{Z\tilde{\chi}_1^0\tilde{\chi}_1^0}$ coupling depends only on the Higgsino components of $\tilde{\chi}_1^0$. 
Fig.\,\ref{fig:Zfunnel-xi} shows the fraction of DM satisfied by the LSP, $\xi$, as a function of the NLSP neutralino mass ($M_{\tilde{\chi}_2^0}$) for both $\mu>0$ and $\mu<0$ with the \textbf{``Before LZ''} set of cuts. In both cases, beyond $M_{\tilde{\chi}_2^0} \sim 450$\,GeV, DM becomes overabundant assuming the standard cosmological model.

\begin{figure}[hbt!]
    \centering
    \includegraphics[width=0.48\textwidth]{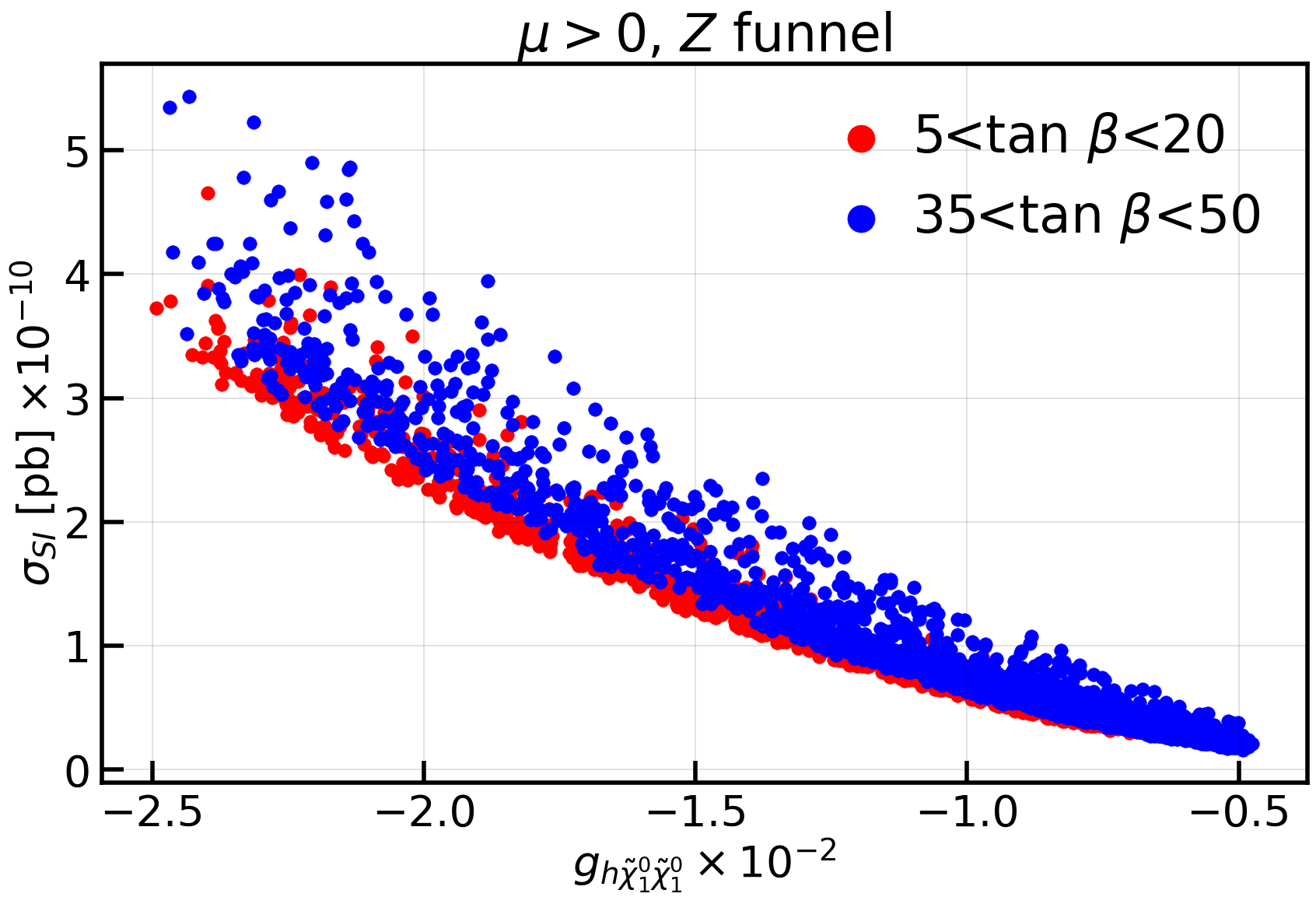}~
    \includegraphics[width=0.48\textwidth]{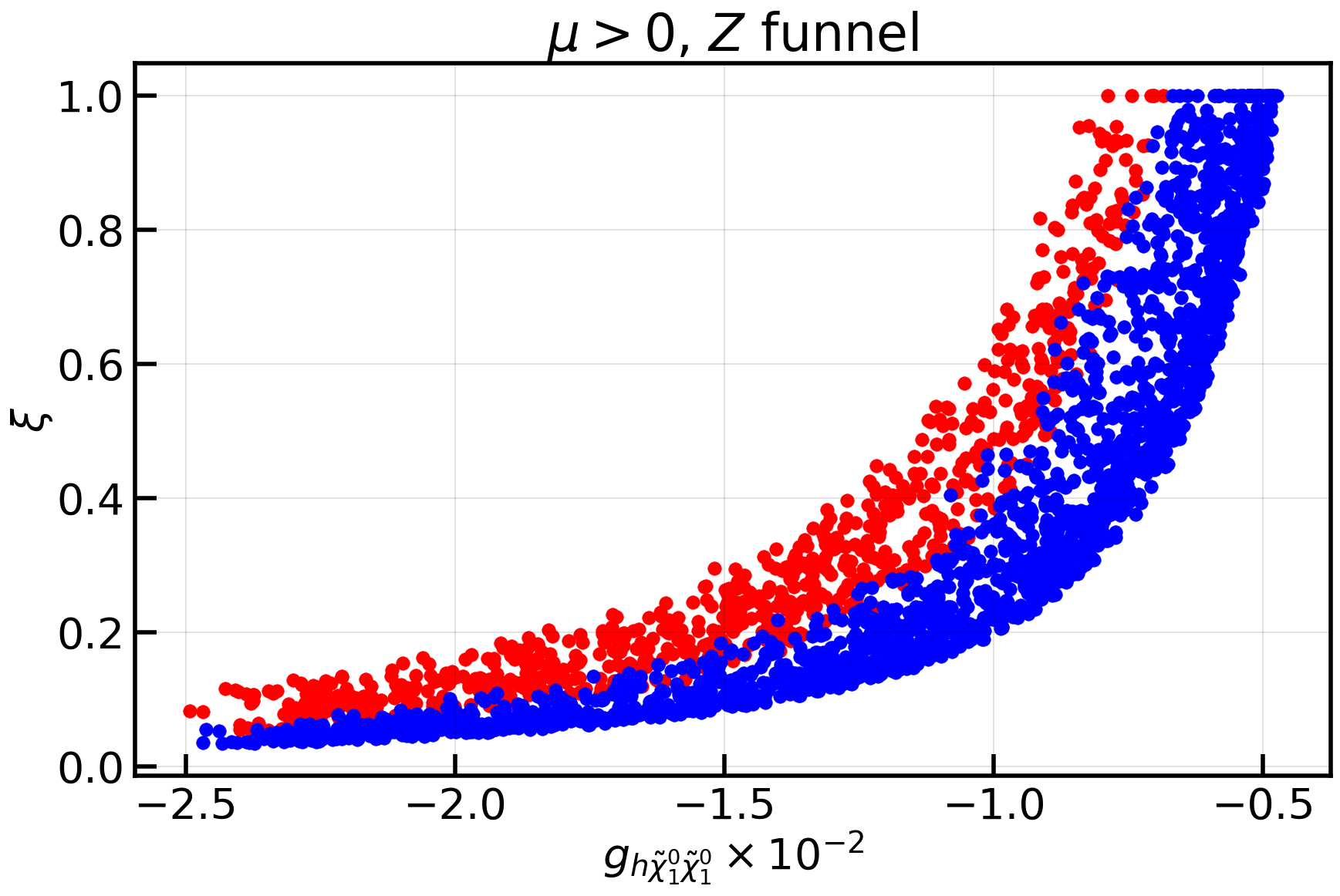}
    \caption{Variation of $\sigma_{SI}$ ({\it top left}) and $\xi$ ({\it top right}) with the DM-light Higgs boson coupling for two different ranges of tan\,$\beta$ for the $\mu>0$ $Z$ funnel.}
    \label{fig:fig4}
\end{figure}

For interpreting the constraints from the LZ experiment, we need to consider the $g_{h\tilde{\chi}_1^0\tilde{\chi}_1^0}$ coupling and the heavy Higgs contribution as well, which we discussed earlier. For $\mu>0$, in the low tan$\beta$ region, $g_{h\tilde{\chi}_1^0\tilde{\chi}_1^0}$ increases and in the high tan$\beta$ region, the $g_{Hdd}$ coupling gets enhanced, both increasing the SI DD cross-section. 
The scaled SI DD cross-section has two components $-$ the SI DD cross section and the fraction of DM relic density produced by $\tilde{\chi}_1^0$.
We plot the variation of these individual quantities as a function of the DM-$h$ coupling in Fig.\,\ref{fig:fig4} {\it top panel} for the $\mu>0$ scenario in the $Z$ funnel. We divide the points in two regions of tan\,$\beta$ $-$ one from 5-20 ({\it red}) and another from 35-50 ({\it blue}).
With decreasing magnitude of the coupling $g_{h\tilde{\chi}_1^0\tilde{\chi}_1^0}$, $\sigma_{SI}$ decreases as expected. However, for the same value of the coupling, higher tan\,$\beta$ values have slightly larger values of $\sigma_{SI}$, which is due to the tan\,$\beta$ enhanced heavy Higgs contribution. For $\xi$, we observe the opposite trend with the coupling and tan\,$\beta$.

\begin{figure}[hbt!]
    \centering    
    \includegraphics[width=0.5\textwidth]{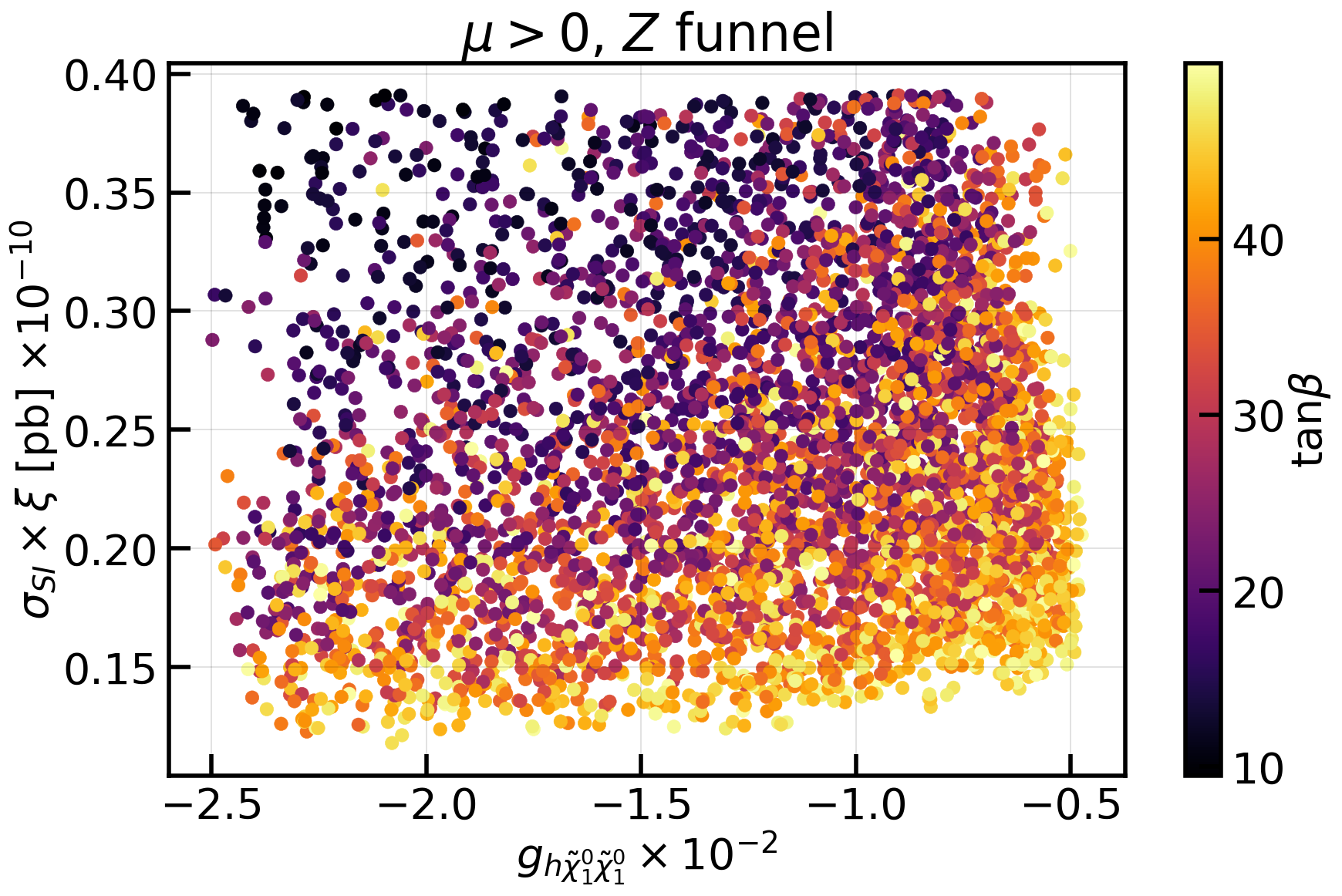}~
    \includegraphics[width=0.5\textwidth]{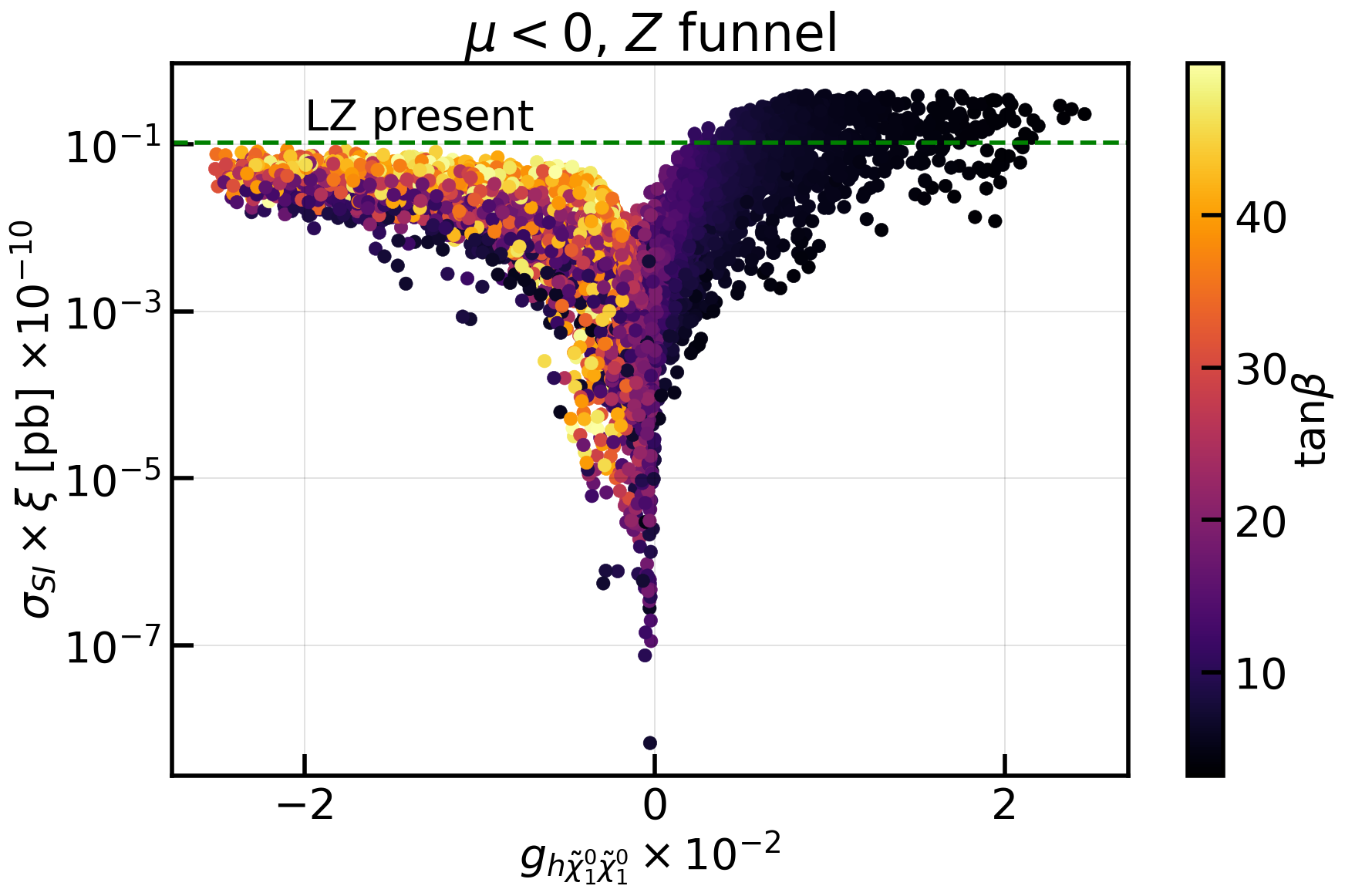}
    \caption{Variation of the scaled SI DM-nucleon cross-section ($\sigma_{SI}\times\xi$) with $g_{h\tilde{\chi}_1^0\tilde{\chi}_1^0}$ for both $\mu>0$ ({\it left}) and $\mu<0$ ({\it right}) with the \textbf{``Before LZ''} set of cuts.}
    \label{fig:Zfunnel-SI}
\end{figure}

Therefore, their product does not vary much, as shown in the {\it left panel} of Fig.\,\ref{fig:Zfunnel-SI} which displays the variation of $\sigma_{SI}\times\xi$ with $g_{h\tilde{\chi}_1^0\tilde{\chi}_1^0}$ for $\mu>0$, with tan$\beta$ in the {\it colorbar}.
All of these points lie above the LZ bound and hence, all points are excluded.
Since we require the product to be smaller than the LZ limit, we are interested in lower values of $\sigma_{SI}$ and $\xi$. 
The lower edges in both the plots of Fig.\,\ref{fig:Zfunnel-SI} are not scattered and follow a trend. 
We also observe that we cannot achieve any lower value of $\xi$, and therefore, further populating the parameter space will not change our conclusion regarding the exclusion of the $Z$ funnel for the $\mu>0$ scenario.
This puts the $Z$ funnel of $\mu>0$ region under severe tension in the pMSSM, where we do not find any region of parameter space satisfying the relic density constraint and the LZ DD limit of $\sigma_{SI}\times\xi\lesssim 0.106\times10^{-10}$\,pb, simultaneously.

The {\it right} panel of Fig.\,\ref{fig:Zfunnel-SI} shows a plot similar to the    one in the {\it left} panel  but  for $\mu<0$.
For $\mu<0$, we have seen that $g_{h\tilde{\chi}_1^0\tilde{\chi}_1^0}$ can attain very small values as it crosses zero coupling. Since we are in the $Z$ funnel, the relic density is determined by $g_{Z\tilde{\chi}_1^0\tilde{\chi}_1^0}$ and smaller values of $|g_{h\tilde{\chi}_1^0\tilde{\chi}_1^0}|$ are  allowed. These lead to very small $\sigma_{SI}\times\xi$, well below the present LZ limit. 
Moreover, for large tan$\beta$, we have negative values of $g_{h\tilde{\chi}_1^0\tilde{\chi}_1^0}$ which leads to tan$\beta$ enhanced cancellations from the heavy Higgs contribution. 
Therefore, the negative couplings have relatively smaller values of $\sigma_{SI}\times\xi$, as compared to the positive couplings, even when the magnitude of the coupling is the same. 
The entire region of parameter space with negative couplings, where there is an interference between the $H$ and $h$ contributions, satisfies the LZ limit.
The future SDn DD experiments, which constrain the $g_{Z\tilde{\chi}_1^0\tilde{\chi}_1^0}$ coupling, play a much more crucial role in probing the parameter space in the $Z$ funnel of $\mu<0$, as we will later see in Fig.\,\ref{fig:result_negmu}.\\


\subsection{$h$ funnel}
\label{ssec:hfunnel}

As we move to the Higgs funnel, the LZ limit becomes slightly weaker than in the $Z$ funnel. 
Moreover the relic density bound can be satisfied for Higgsinos heavier than $\sim 450$\,GeV, unlike the $Z$ funnel. This is because at high $\mu$, the coupling $g_{h\tilde{\chi}_1^0\tilde{\chi}_1^0}$ can be large provided tan\,$\beta$ is small (see Fig.\,\ref{fig:couplings}).

\begin{figure}[hbt!]
    \centering
    \includegraphics[width=0.5\textwidth]{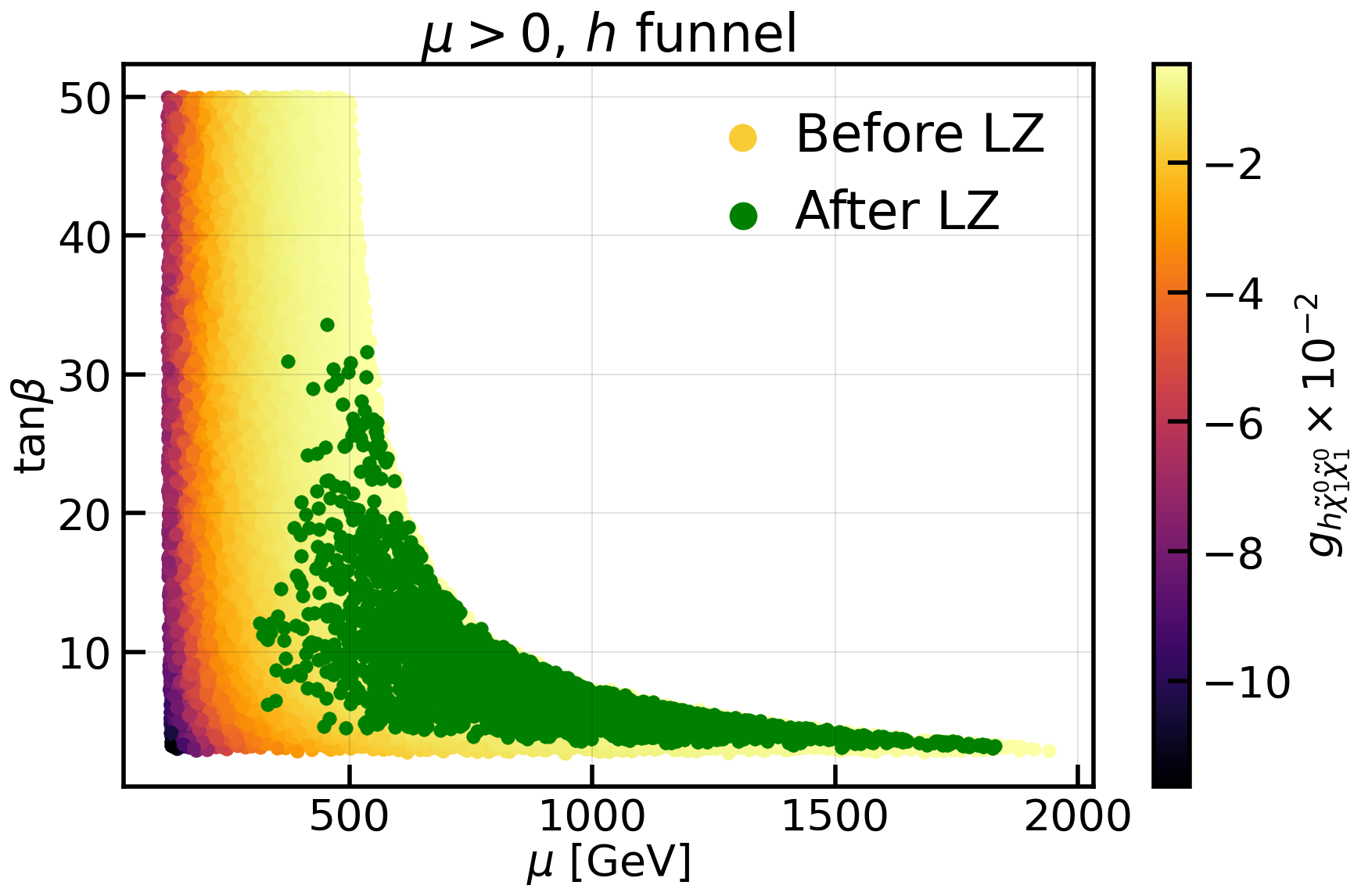}~
    \includegraphics[width=0.5\textwidth]{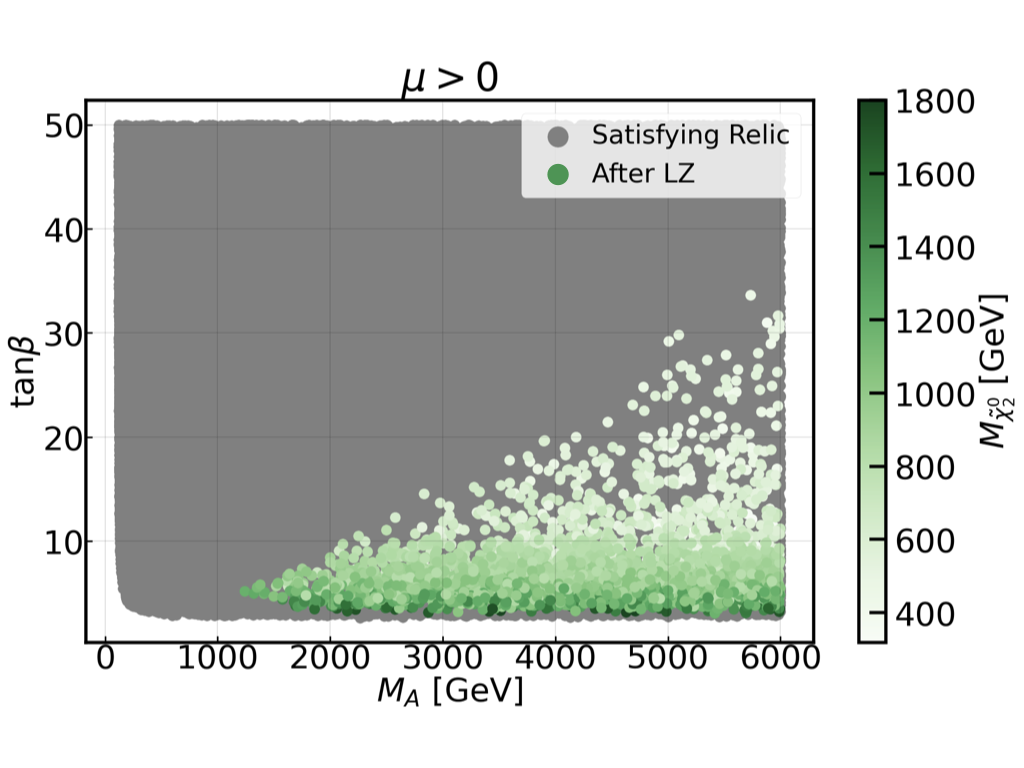}
    \caption{{\it Left:} The $\mu>0$ $h$ funnel parameter space in the $\mu$-tan$\beta$ plane before ({\it yellow}) and after ({\it green}) applying the LZ limit~\cite{Aalbers:2022fxq} with $g_{h\tilde{\chi}_1^0\tilde{\chi}_1^0}$ in the {\it colorbar} for ``Before LZ''. {\it Right:} The full $\mu>0$ parameter space in the $M_A$-tan$\beta$ plane that satisfy the relic density constraint ({\it grey}) and that satisfy all the constraints including LZ ({\it green}) with $M_{\tilde{\chi}_2^0}$ in the {\it colorbar}.
    }
    \label{fig:hfunnel_posmu}
\end{figure}

The {\it left} panel of Fig.\,\ref{fig:hfunnel_posmu} shows the parameter space points in the $h$ funnel of $\mu>0$ in the $\mu$-tan$\beta$ plane with the colorbar showing the value of $g_{h\tilde{\chi}_1^0\tilde{\chi}_1^0}$. 
The \textbf{``Before LZ''} set of constraints, especially the bound on relic density restricts large $\mu$ to have only small tan$\beta$ values. The points shown in {\it green} in the left panel of Fig.\,\ref{fig:hfunnel_posmu} are allowed by the LZ limit. These points predominantly occupy regions characterised by smaller $g_{h\lspone\lspone}$, where $\mu$ is greater than around 400~GeV and $\tan\beta$ is smaller than 35\,\footnote{These upper and lower bounds are provided just to give a qualitative idea, and might change with further scanning of points at the edge. However, since these points lie very close to the LZ bound, the values of these bounds on $\mu$ and $\tan\beta$ won't be significantly affected.}. Note that even though the points at higher values of tan$\beta$ have very small couplings, they do not survive the LZ constraint due to the added tan$\beta$ enhanced heavy Higgs contribution.

To further demonstrate the significant impact of the LZ result on the positive $\mu$ scenario, the {\it right} panel of Fig.\,\ref{fig:hfunnel_posmu} shows the $\mu>0$ parameter space in the $M_A$-tan$\beta$ plane satisfying the relic density constraint in {\it grey}.
The parameter space obtained after imposing both the relic density and LZ bounds survives all the other set of cuts applied till this point, i.e., the constraints from Higgs properties, LEP, flavor and DD bounds on SDn and SDp cross-sections of the DM. These points are shown in shades of {\it green}. 
The {\it colorbar} in the {\it right} panel of Fig.\,\ref{fig:hfunnel_posmu} shows that the minimum allowed $M_{\tilde{\chi}_2^0}$ value is around 350\,GeV.
The LZ bound on SI DD constrains large $g_{h\tilde{\chi}_1^0\tilde{\chi}_1^0}$ values, ruling out lighter Higgsinos (see Fig.\,\ref{fig:couplings}).
Moreover, Higgsinos below 500\,GeV are found at large values of tan\,$\beta$ where the coupling $g_{h\tilde{\chi}_1^0\tilde{\chi}_1^0}$ is reduced and large $M_A$ as can be seen in Fig.\,\ref{fig:hfunnel_posmu}.
Higher values of tan$\beta$ are allowed as we go to large $M_A$ where the heavy Higgs contribution decreases.

\begin{figure}[hbt!]
    \centering
    \includegraphics[width=0.49\textwidth]{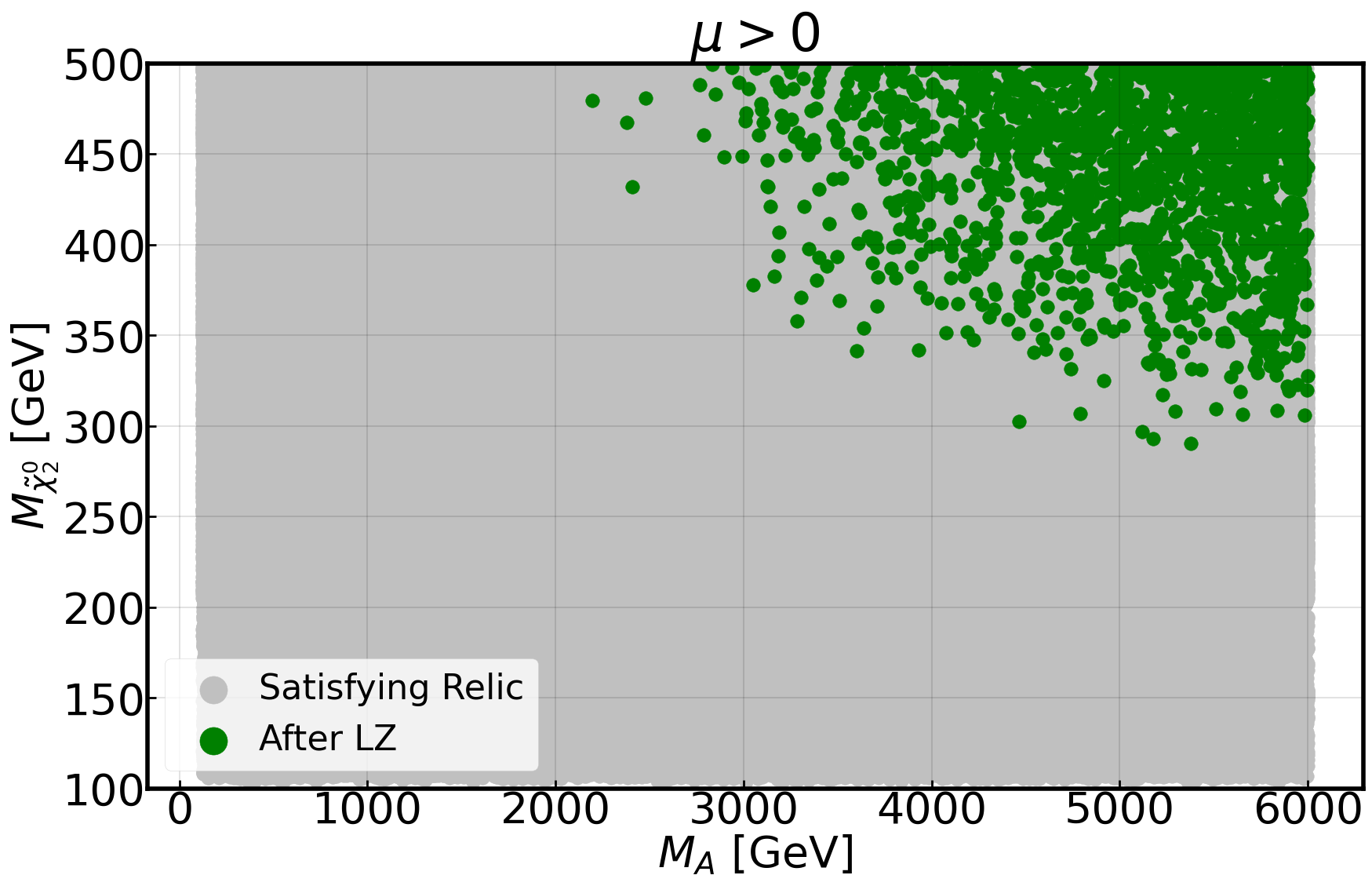}    \includegraphics[width=0.49\textwidth]{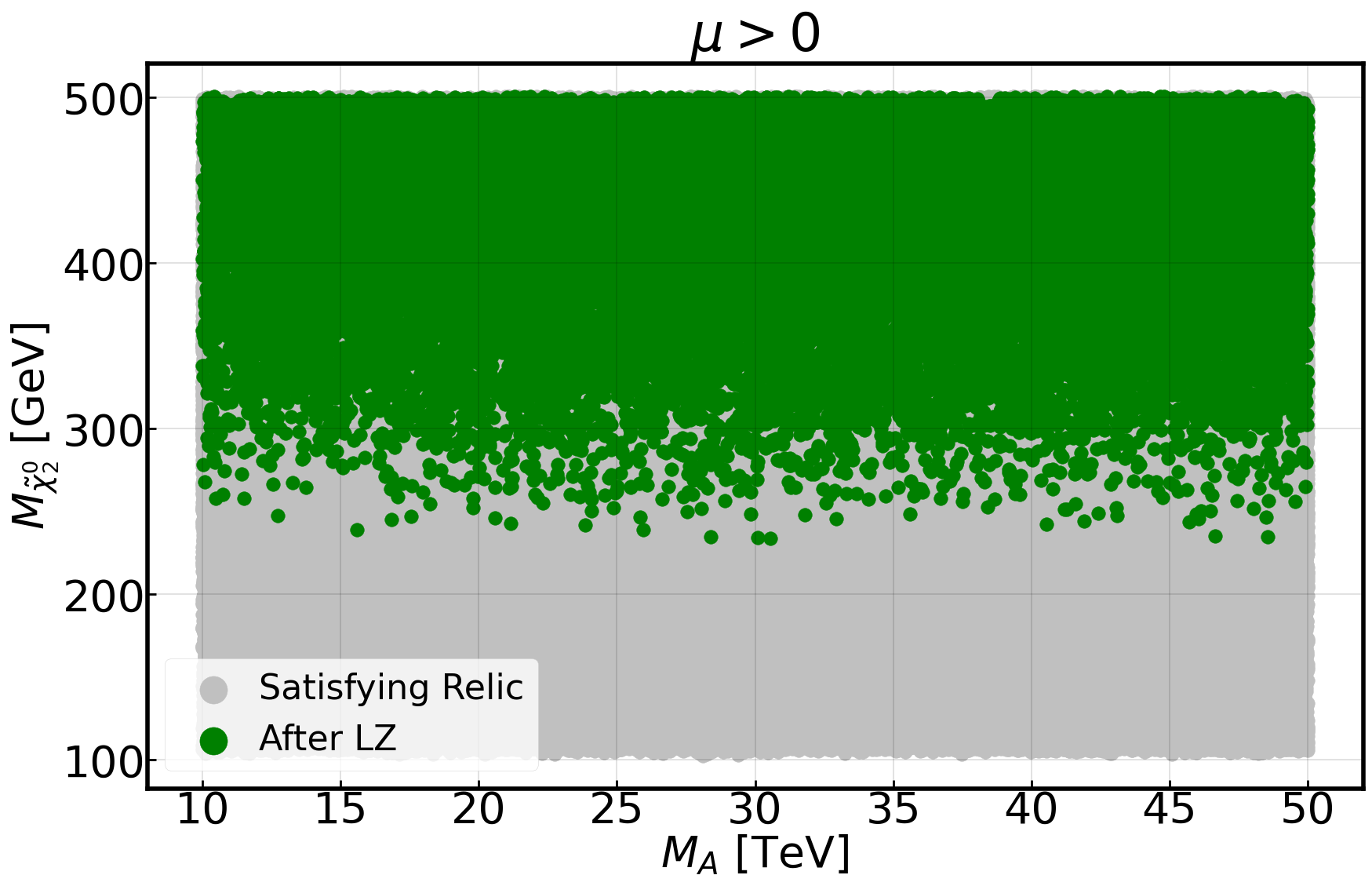}
    \caption{
    Minimum allowed Higgsino mass after applying the DM constraints for a range of $M_A$ between 100\,GeV to 6\,TeV ({\it left}) and 10 to 50\,TeV ({\it right}). The points satisfying the relic density constraints are shown in {\it grey}, while the {\it green colored} points are allowed by the other current constraints including the LZ DD bound.
    }
    \label{fig:largeMA}
\end{figure}

We have scanned $M_A$ up to 6\,TeV, and a natural question which arises is whether even larger $M_A$ can decouple the effect of $H$ and extend the allowed parameter space to include larger values of tan\,$\beta$.
This motivates us to perform a dedicated scan in the 10-50\,TeV $M_A$ region. Fig.\,\ref{fig:largeMA} shows the parameter space in the $M_A - M_{\tilde{\chi}_2^0}$ plane for low $M_A$ ({\it left}) and high $M_A$ ({\it right}) regions. For low $M_A$ (100\,GeV - 6\,TeV), the allowed points populate the upper right corner of the parameter space and the minimum value of $M_{\tilde{\chi}_2^0}$ decreases with increasing $M_A$. This trend is due to the heavy Higgs contribution which becomes smaller as we go to higher $M_A$ values, and hence points with smaller values of $M_{\tilde{\chi}_2^0}$ survive the DM constraints.
For our high $M_A$ scan (10-50\,TeV), the minimum value of $M_{\tilde{\chi}_2^0}$ is almost constant over the $M_A$ range. However, higgsino masses below 200\,GeV are still not allowed when we go to very high $M_A$ values.

\begin{figure}[hbt!]
    \centering
    \includegraphics[width=0.47\textwidth]{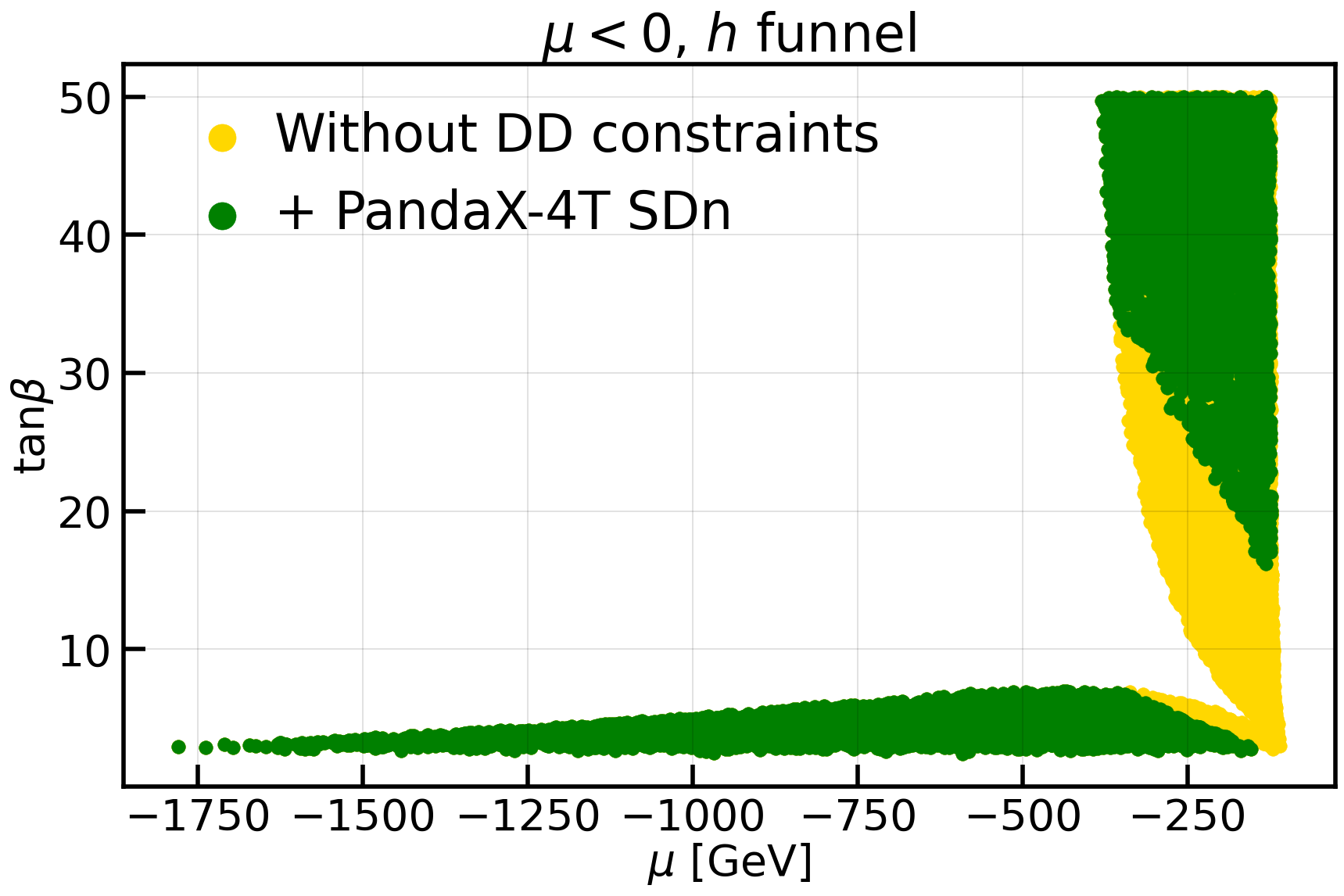}~
    \includegraphics[width=0.47\textwidth]{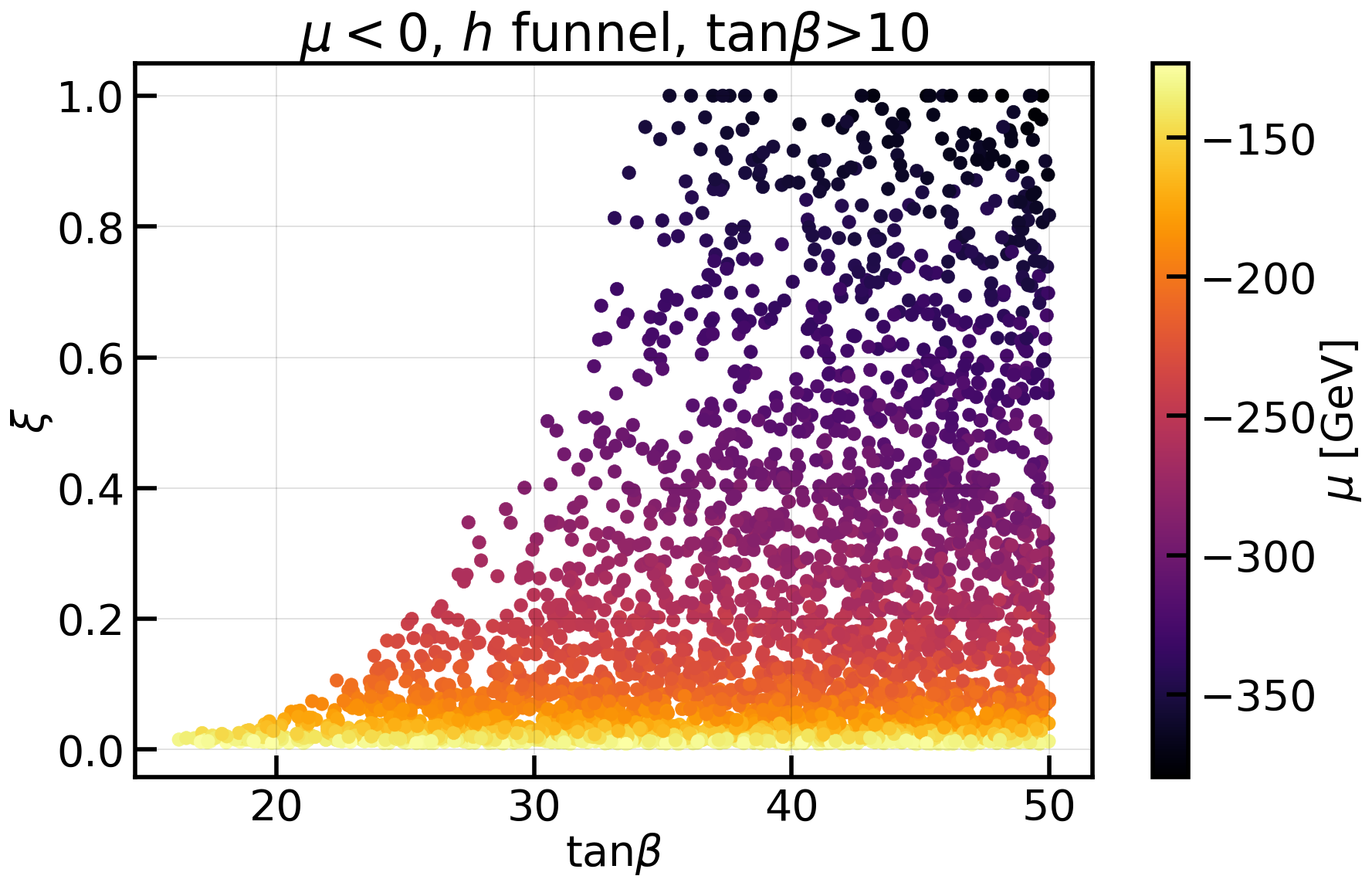}
    \caption{{\it Left}: Parameter space points in the $h$ funnel of $\mu<0$ in the $\mu$-tan$\beta$ plane when all the constraints except the DD constraints are applied (in {\it yellow}), and in addition when the strongest available SDn DD constraint from the PandaX-4T experiment is applied (in {\it green}). {\it Right}: $\xi$ values of the points passing the PandaX-4T SDn bound with tan$\beta>10$ as a function of tan$\beta$ with the colorbar showing $\mu$. 
    }
    \label{fig:hfunnel_negmu}
\end{figure} 

In the {\it left} panel of Fig.\,\ref{fig:hfunnel_negmu}, we show the parameter points in the $h$ funnel region of the $\mu < 0$ scenario in the $\mu-\tan\beta$ plane. The points are depicted under two conditions: firstly, all the constraints except the DD constraints are applied ({\it yellow}), and secondly, the most stringent SDn DD constraint from PandaX-4T is additionally applied ({\it green}). Note that the LZ limit is not applied on this parameter space yet. We observe that $\tan\beta$ is restricted to smaller values for large $|\mu|$, similar to the $\mu > 0$ scenario.
The SDn DD cross-section depends on the coupling of the LSP with the $Z$ boson, which grows with decreasing magnitude of $\mu$, and therefore, the PandaX-4T SDn bound excludes low values of $|\mu|$. Regions of large tan$\beta$ evade this bound, although $g_{Z\tilde{\chi}_1^0\tilde{\chi}_1^0}$ does not have any significant dependence on tan$\beta$. The tan$\beta$ dependence comes from the scaling factor $\xi$,  since the tan$\beta$ dependent $g_{h\tilde{\chi}_1^0\tilde{\chi}_1^0}$ coupling determines the relic density in the Higgs funnel.

We show values of $\xi$ for the points passing the PandaX-4T SDn bound with tan$\beta>10$ as a function of tan$\beta$ in the {\it right} panel of Fig.\,\ref{fig:hfunnel_negmu} with the colorbar showing $\mu$. The purpose of this is to show how points with low $\mu$ and high tan\,$\beta$ survive the PandaX-4T bound on SDn cross-sections. It was shown in Fig.\,\ref{fig:couplings} that for $\mu<0$, large magnitudes of $g_{h\tilde{\chi}_1^0\tilde{\chi}_1^0}$ are possible for large tan$\beta$ and small $\mu$. This makes $\xi$ small and reduces the scaled SDn direct detection cross-section, even for low values of $|\mu|$, where otherwise the SDn cross-sections are high due to large $g_{Z\tilde{\chi}_1^0\tilde{\chi}_1^0}$ coupling. 
These high coupling values at high tan$\beta$ also survive the SI DD bounds due to destructive contribution from the heavy Higgs boson. 
This region is particularly interesting since it involves light Higgsinos, which provide important benchmarks for Run-3 of LHC.
For larger $|\mu|$, the $g_{h\tilde{\chi}_1^0\tilde{\chi}_1^0}$ coupling is smaller and larger tan$\beta$ is required to increase $\xi$ in order to satisfy the bound on SDn DD cross-section. 

\begin{figure}[hbt!]
    \centering
    \includegraphics[width=0.5\textwidth]{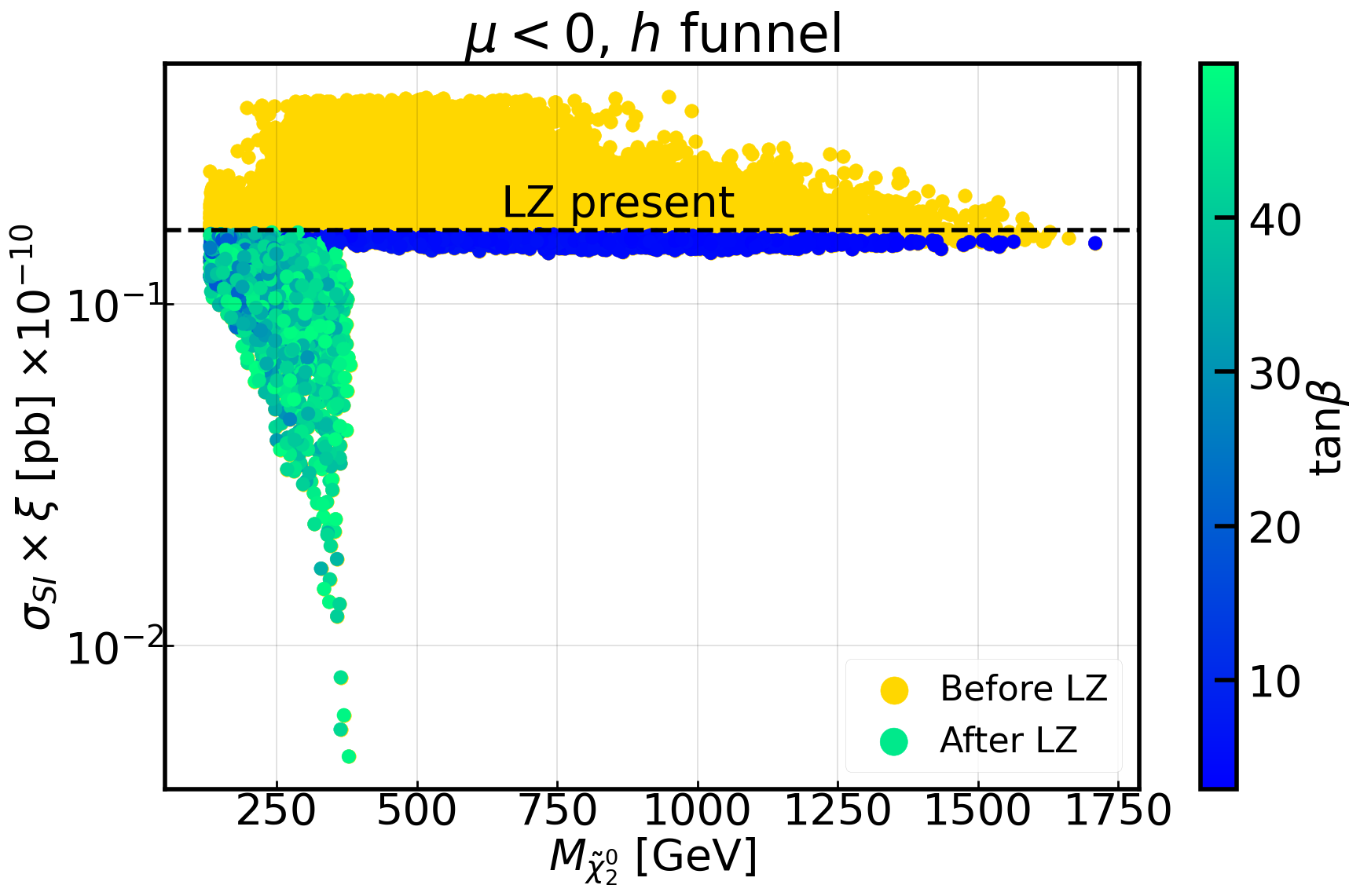}~
    \includegraphics[width=0.5\textwidth]{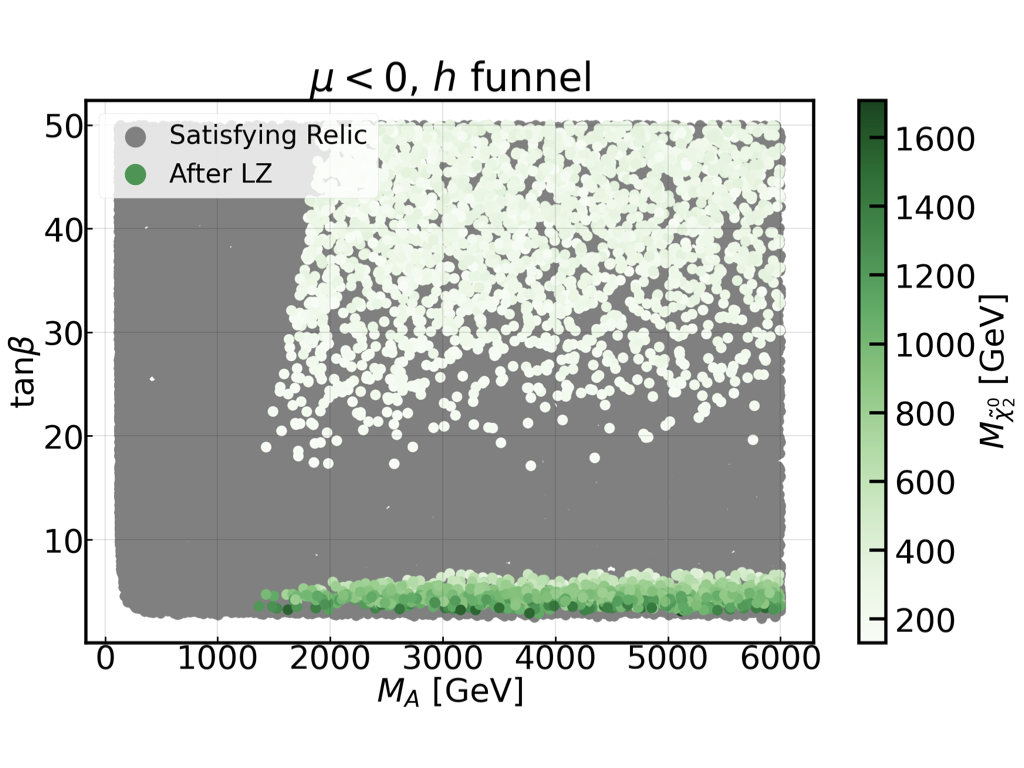}
    \caption{{\it Left}: The $h$ funnel of the $\mu<0$ in the $M_{\tilde{\chi}_1^0}$-$\sigma_{SI}\times\xi$ plane before and after the LZ limit. {\it Right}: The $h$ funnel of the $\mu<0$ in the $M_A$-tan$\beta$ plane satisfying various cuts, and allowed points with $M_{\tilde{\chi}_2^0}$ in the {\it colorbar}.
    }
    \label{fig:hfunnel_negmu_MA}
\end{figure} 

To understand the effect of the LZ result, the {\it left} panel of Fig.\,\ref{fig:hfunnel_negmu_MA} shows the parameter space in the $M_{\tilde{\chi}_2^0}$-$\sigma_{SI}\times\xi$ plane for the $h$ funnel for $\mu < 0$  before and after the LZ result. This includes the SDn constraint from PandaX-4T, which we discussed in Fig\,\ref{fig:hfunnel_negmu}.
The {\it colorbar} in the plot indicates the corresponding $\tan\beta$ values.
LZ excludes a significant part of the parameter space, especially for high $\mu$ values corresponding to large $M_{\tilde{\chi}_2^0}$, where only a narrow strip of allowed region with very small tan\,$\beta$ remains. With only a 20\% improvement in the LZ limit in the future, this region can be fully probed.
The very low SI DD cross-sections found for Higgsinos in the mass range 200-400\,GeV is due to the destructive interference between the $h$ and $H$ contributions which is more important at large tan\,$\beta$.
The {\it right} panel of the same figure shows the allowed parameter space in the $M_A$-tan$\beta$ plane, with $M_{\tilde{\chi}_2^0}$ in the {\it colorbar}.
The gap in the allowed region around tan\,$\beta$ values of 6-18 is due to the recent bound on SDn cross-section by the PandaX-4T collaboration, as we observed previously in Fig.\,\ref{fig:hfunnel_negmu}. 


The upshot for $\mu>0$ scenario is that after the DM constraints, the $Z$-funnel is excluded by the LZ result, however, a region of the $h$-funnel survives these constraints. In this region, the lightest allowed Higgsino is around 350 GeV. For the $\mu<0$ scenario, we found allowed points both in the $Z$ and $h$ funnel, and Higgsinos as light as 125\,GeV survive DM constraints. The major factor that creates a difference between the results of positive and negative $\mu$ is the effect of the heavy Higgs bosons in the DD cross-sections. 

\subsection{Comparison with a simplified model}
\label{ssec:simplified}

The previous section has shown the importance of the heavy Higgs boson contribution to the SI DD cross-section and its dependence on tan\,$\beta$ in pMSSM. 
As we move towards heavier Higgsinos, the relic density upper bound can only be satisfied at very low values of tan\,$\beta$ for both $\mu>0$ and $\mu<0$. In this region, the effect of $H$ is decoupled, and the lighter Higgs boson plays the dominant role. 
It is worth studying whether our results for high $\mu$ in pMSSM with the DM constraints, especially the interplay of relic density and the recent LZ upper limit on the SI DD cross-section, generalise to any BSM theory consisting of a Majorana fermion coupling with only the light Higgs boson, which resembles the discovered Higgs boson at LHC. 

We consider a simplified model where the SM is extended by a single Majorana fermion, $\chi$, which is the DM candidate. It has coupling with the SM Higgs boson ($g_{\chi h}$) and has the following Lagrangian:
\begin{equation}
    \mathcal{L} = \mathcal{L}_{SM} + g_{\chi h} \bar{\chi^C}\chi h.
    \label{eq:lagrangian_simplified}
\end{equation}
We scan over the mass of the DM and its coupling with $h$, i.e., in the $M_{\chi}-g_{\chi h}$ plane. The Higgs boson mass is fixed at 125\,GeV, and the total width of $h$ is calculated from the model. Since this minimal model has only two input parameters, $M_{\chi}$ and $g_{\chi h}$, the total width of the Higgs boson has negligible variation and attains a value around $4.02$\,MeV when the model is implemented in \texttt{MicrOMEGAS 5.2.13}.

\begin{figure}[hbt!]
    \centering
    \includegraphics[width=0.5\textwidth]{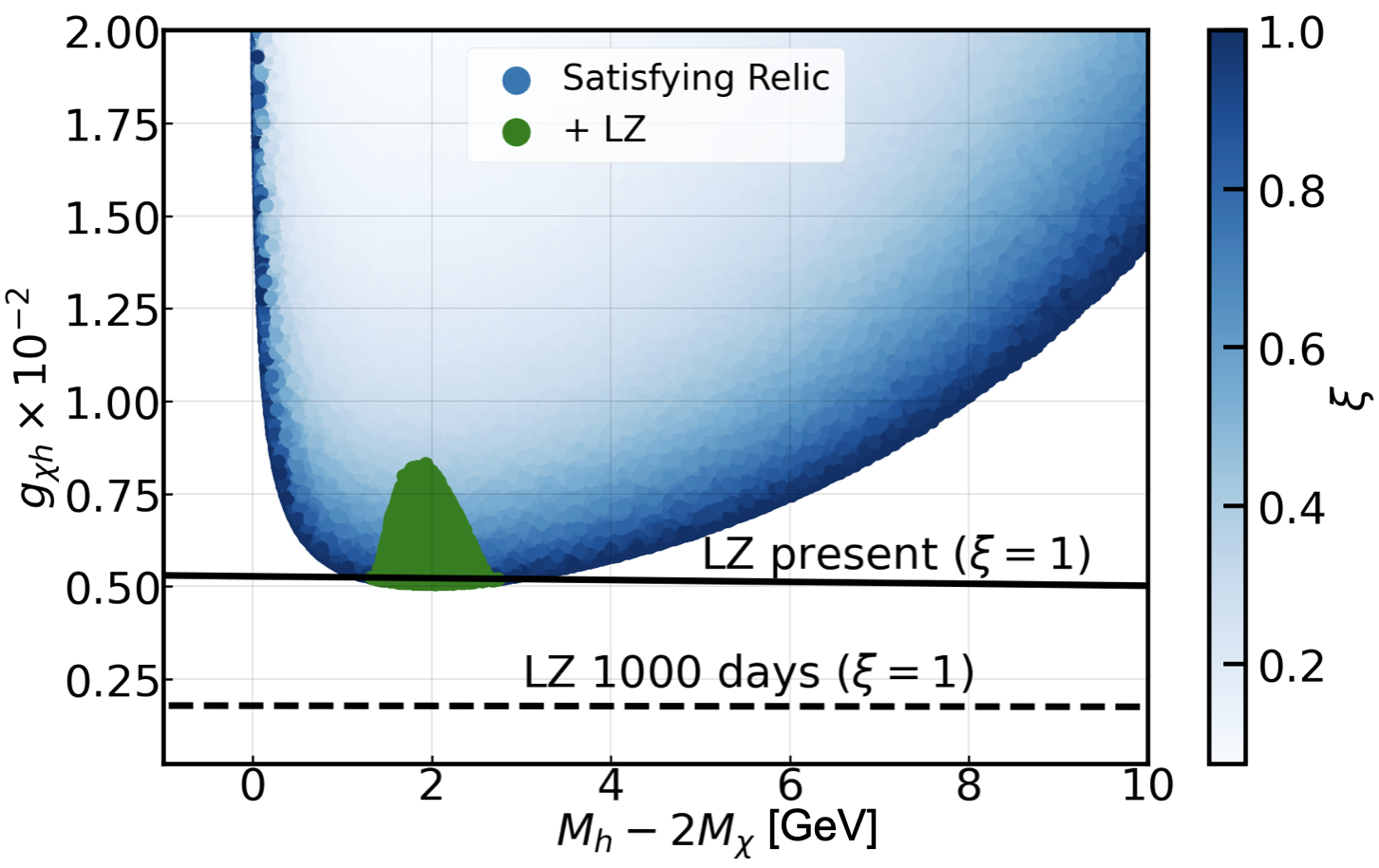}~
    \includegraphics[width=0.5\textwidth]{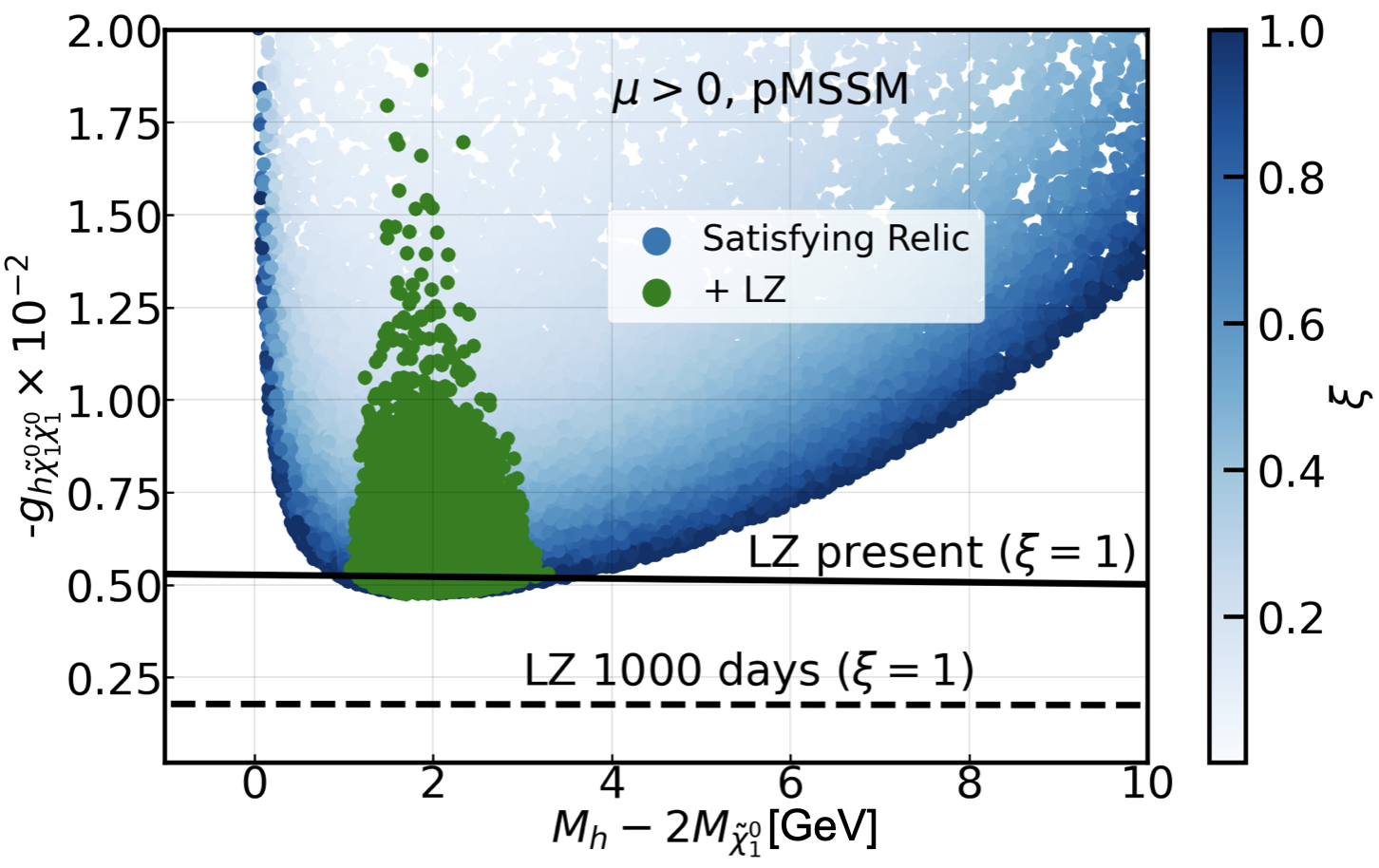}    
    \caption{Result from the scan of a simplified model where we extend the SM by a Majorana fermion coupled to the $h$ boson ({\it left}) and the $\mu>0$ scenario of pMSSM ({\it right}) in the $g_{\chi h}-(M_h-2\times M_{\chi})$ plane. The {\it blue} points satisfy the relic density upper bound with $\xi$ shown in the colorbar,
    and the {\it green} points also satisfy the LZ limit.}
    \label{fig:simplified}
\end{figure}

The DM relic density depends on the couplings involved in the annihilation process, the difference between twice the DM mass and the resonance in the funnel region, and the width of the mediator particle. The coupling of the Higgs boson to the SM particles and the width of the Higgs boson do not vary much in this set-up, the only variables are therefore $M_h-2M_{\chi}$ and $g_{\chi h}$. 
The contribution of DM to the total width of $h$ is negligible.
Moreover, the scattering cross-section of $\tilde{\chi}$ on nucleons relevant for DD depends on the same coupling. We present our results of the scan
in the plane of $M_h-2M_{\chi}$ versus $g_{\chi h}$.
The {\it left} panel of Fig.\,\ref{fig:simplified} shows the region of parameter space surviving the relic density constraint in {\it blue} with the {\it colorbar} showing the fraction of DM constituted by this additional fermion, and the allowed region of parameter space after the LZ DD bound in {\it green}. 
We find that the coupling allowed by both relic density and DD is minimum when $M_{h} - 2M_{\chi} \sim 2$\,GeV. Furthermore, $g_{h\chi}$ cannot be much smaller than $\sim~0.005$ in order to satisfy these two constraints.
The region of parameter space that satisfies current constraint can be probed by improving the DD limit by $\sim$11\%, which should be achieved in a few days of running of LZ.

The result for the simplified scenario makes it clear that the relic density and the recent LZ limit strongly constrain the coupling and mass plane of the DM when it connects to the SM particles through a Higgs portal.
Therefore, even in a BSM theory like pMSSM, where we have a wide range of input parameters, the parameter space is equally constrained.
The {\it right} panel of Fig.\,\ref{fig:simplified} shows an analogous plot for the $\mu>0$ scenario of pMSSM for heavier Higgsinos ($\mu\gtrsim 850$\,GeV).
For $\mu<0$ scenario of pMSSM where $|\mu|$ is kept higher than $\sim 850$\,GeV, we get a similar result.
The key difference that we observe in the pMSSM for large $|\mu|$ is that few points with larger coupling ($g_{\chi h}\sim 0.01-0.025$) are still allowed by the LZ limit, unlike the simplified scenario.
This is partly due to the larger variation of the Higgs boson mass in the pMSSM scan, while in the simplified model, it is fixed to 125\,GeV. Indeed a tighter constraint on the Higgs boson mass in the pMSSM affects slightly the range of allowed values for both the coupling and the mass difference  $M_h-2M_{\tilde{\chi}_1^0}$  as will be discussed in Fig.\,\ref{fig:sigmaSI_higgsmass}. Moreover, even for the same Higgs boson mass, coupling and DM mass, the total width of the Higgs boson in the pMSSM is different than in the simplified model  due to  the higher order corrections in $\Gamma(h\rightarrow b\bar{b})$ coming from $\Delta m_b$ corrections\,\cite{Carena:1999py,Hall:1993gn,Guasch:2003cv,Dawson:2011pe}. Therefore, some of the points with larger couplings marginally pass the LZ limit in the pMSSM because the smaller width implies a lower value for the relic density, thereby reducing $\xi$.
However, the constraints from the Higgs signal strength measurements do not allow $\Gamma_h$ to deviate significantly from the SM prediction, which limits its effect on our results.  


\section{Electroweakino constraints and the allowed parameter space}
\label{sec:EWINOconstraints}

The electroweakinos can be directly produced at the colliders, where the NLSP Higgsinos ($\tilde{\chi}_1^\pm$/$\tilde{\chi}_2^0$/$\tilde{\chi}_3^0$) can decay to final states involving the LSP neutralino ($\tilde{\chi}_1^0$) along with $W$, $Z$ or $h$ bosons, which can have both leptonic and hadronic decays. 
We use the \texttt{SModelS\,2.2.1}\,\cite{Kraml:2013mwa,Ambrogi:2017neo,Dutta:2018ioj,Heisig:2018kfq,Ambrogi:2018ujg,Khosa:2020zar,Alguero:2020grj,Alguero:2021dig} package to implement the electroweakino search constraints on our scanned parameter space. 
Recently many analyses have updated their results with the full Run-2 data and this version of \texttt{SModelS} includes results from the recent search for electroweakinos in the leptonic final states at CMS\,\cite{CMS:2020bfa} and ATLAS\,\cite{ATLAS:2021moa} and in the hadronic final states at ATLAS\,\cite{ATLAS:2021yqv}. 
The constraints from the recent searches play a significant role in excluding a large range of $M_{\tilde{\chi}_1^{\pm}}$, $M_{\tilde{\chi}_2^0}$ and $M_{\tilde{\chi}_3^0}$, extending the sensitivity to higher masses, especially with the ATLAS analysis of the hadronic final states.

\begin{figure}[hbt!]
    \centering
    \includegraphics[height=5.4cm]{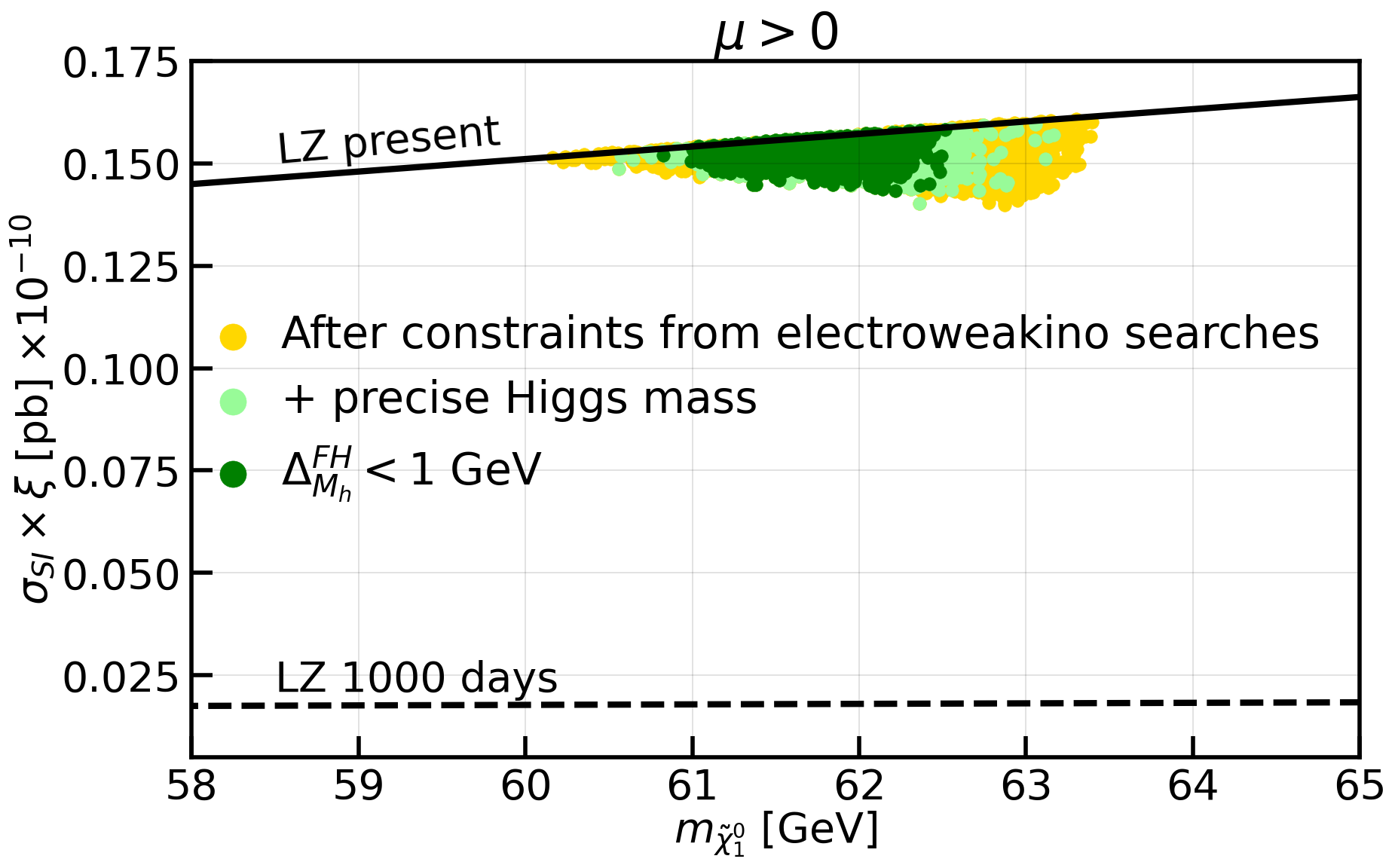}
    \includegraphics[height=5.4cm]{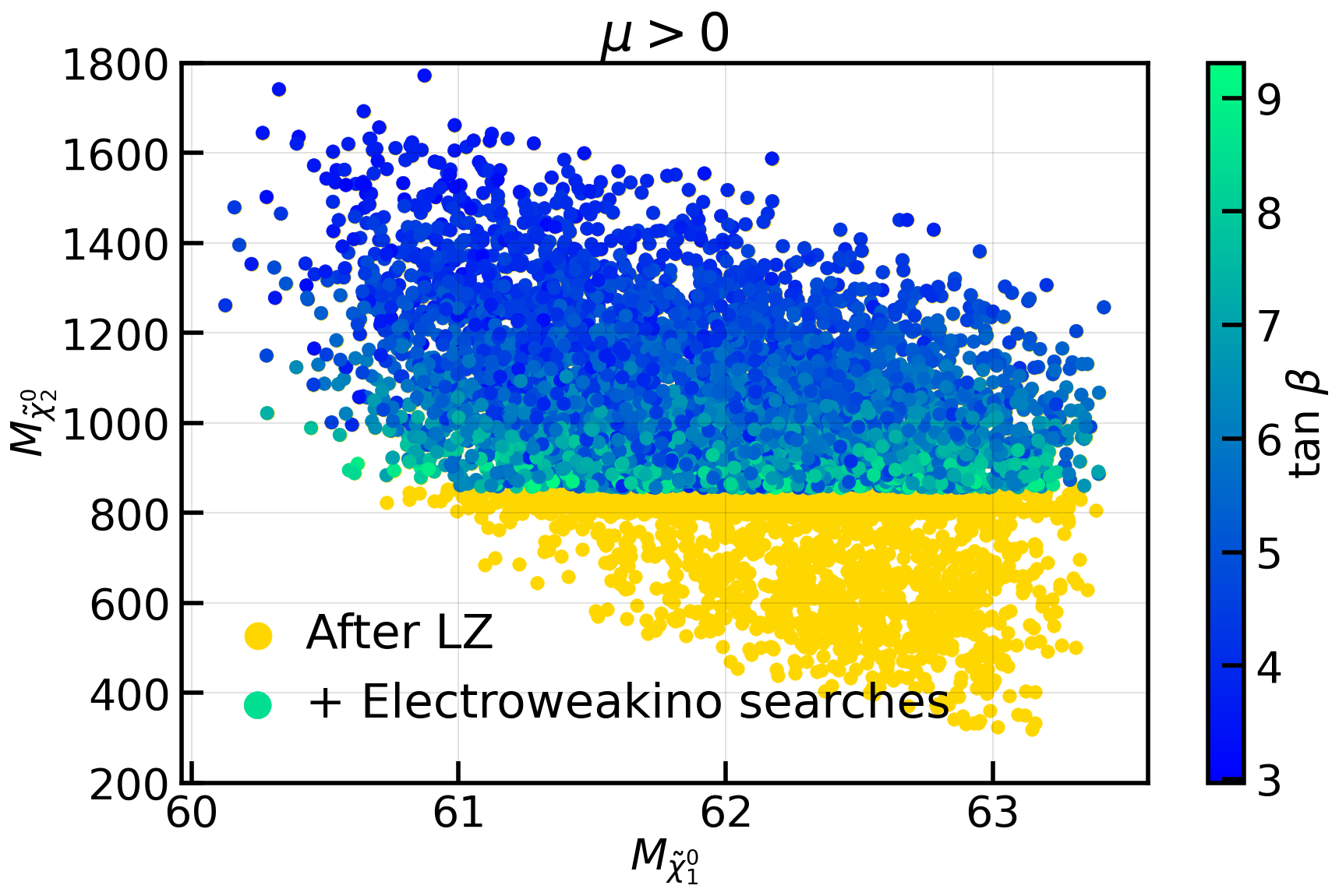}
    \caption{{\it Left:} Scaled SI DM-nucleon cross-section ($\sigma_{SI}\times\xi$) for $\mu>0$ as a function of the mass of the LSP neutralino DM in the region of parameter space satisfying LEP, flavor, Higgs constraints, relic density, DM DD constraints from the XENON-1T, PICO-60, PandaX-4T, and LZ experiments,
    the electroweakino limits implemented in \texttt{SModelS}, and the precise Higgs mass condition with $\Delta_{M_h}^{FH}<1$\,GeV; {\it Right:} Parameter space satisfying all the constraints defined as \textbf{``After LZ''} and the region surviving the electroweakino searches in the $M_{\tilde{\chi}_1^0}$-$M_{\tilde{\chi}_2^0}$ plane for $\mu>0$, where the {\it colorbar} shows tan\,$\beta$.} 
    \label{fig:result_posmu}
\end{figure}

We apply the limits from electroweakino searches on the parameter space surviving all the constraints discussed previously in Sections 3 and 4, which are defined as \textbf{``After LZ''}. 
We identify regions of the parameter space surviving all the constraints: $A)$ high values of $|\mu|$ ($\gtrsim 800$\,GeV) with low tan\,$\beta$ ($\lesssim 10$) in the $h$ funnel of both positive and negative $\mu$, and B) low values of $|\mu|$ ($\lesssim 200$\,GeV) in both the $Z$ and $h$ funnels of negative $\mu$. We perform dedicated scans over these regions again with an additional sample of size $\sim10^8$, which makes the total number of points scanned for our analysis to be $\sim 3 \times 10^8$. From these $3\times 10^8$ points, around 41\% pass the Higgs boson mass condition. This is due to the fact that we perform our scans with very high values of $M_{\tilde{Q}_{3L}}$, $M_{\tilde{t}_{R}}$, and $M_{\tilde{b}_{R}}$, starting from 3\,TeV and vary $A_t$ from -20\,TeV to 20\,TeV, which helps in satisfying the Higgs mass condition.

Let us first discuss the results in the positive $\mu$ scenario.
Fig.\,\ref{fig:result_posmu} ({\it left} panel) shows the scaled SI DM-nucleon cross-section ($\sigma_{SI}\times\xi$) with the mass of the LSP neutralino DM for the allowed parameter space for $\mu > 0$. In the {\it right} panel of Fig.\,\ref{fig:result_posmu}, we show these points in the $M_{\tilde{\chi}_1^0}$-$M_{\tilde{\chi}_2^0}$ plane with $\tan\beta$ in the {\it colorbar}.
In Section\,\ref{sec:DMconstraints}, we have discussed that the LZ limit excludes the $Z$ funnel region for $\mu>0$. As a result, both the panels of Fig.\,\ref{fig:result_posmu} have allowed points only in the $h$ funnel region.

The {\it right} panel of Fig.\,\ref{fig:result_posmu} reveals that on applying the electroweakino constraints implemented in \texttt{SModelS}, $M_{\tilde{\chi}_2^0}\lesssim850$\,GeV are excluded by various collider searches. 
The allowed region of higgsinos heavier than 850\,GeV populates very small values of tan\,$\beta$, which results from the DM constraints as we have discussed earlier in Section\,\ref{ssec:hfunnel}.
The region of parameter space surviving the constraints from electroweakino searches can be probed by improving the LZ limit by 20\% which can be achieved with just a few more days of running, as can be seen from the {\it left} panel of Fig.\,\ref{fig:result_posmu}. 
For the simplified scenario, note that only 11\% improvement in the LZ bound is required to probe the allowed parameter region, which is smaller than that required for the pMSSM parameter space. The difference in the two scenarios is again due to the variation in the Higgs boson total decay width.
The coupling $g_{h\tilde{\chi}_1^0\tilde{\chi}_1^0}$ for heavier Higgsinos becomes constant, and since the surviving regions are restricted to low tan$\beta$ values, there is no significant contribution from the heavy Higgs. This parameter space cannot have any smaller values of $\sigma_{SI}\times\xi$, as the relic density constraint does not allow for lower values of the coupling.

\begin{figure}[hbt!]
    \centering
    \includegraphics[width=0.48\textwidth]{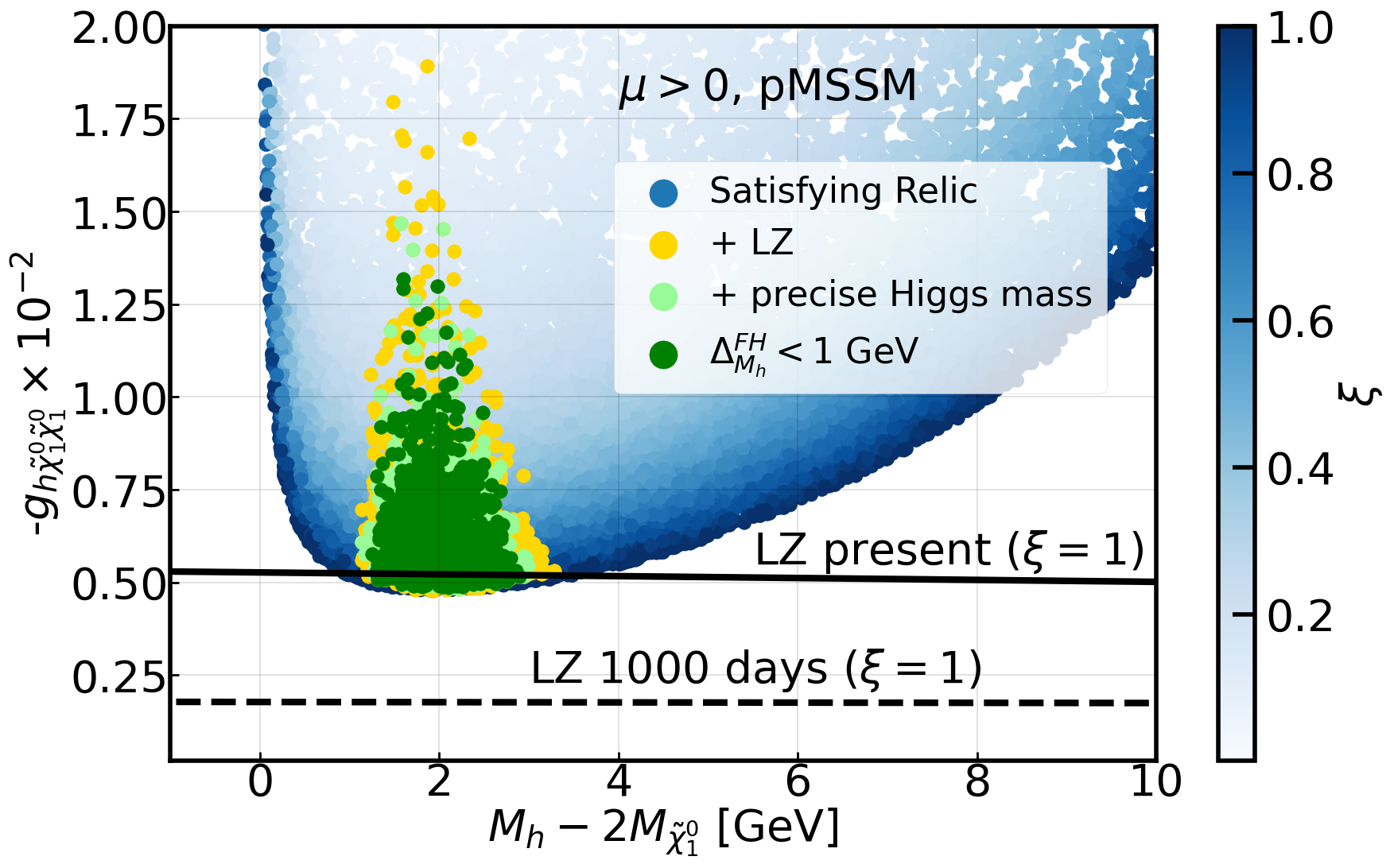}
    \caption{
    The allowed regions for the $\mu>0$ scenario of pMSSM in the $g_{\chi h}-(M_h-2\times M_{\chi})$ plane to show the impact of the precise Higgs mass condition.}
    \label{fig:sigmaSI_higgsmass}
\end{figure}

\begin{figure}[hbt!]
    \centering
    \includegraphics[width=0.47\textwidth]{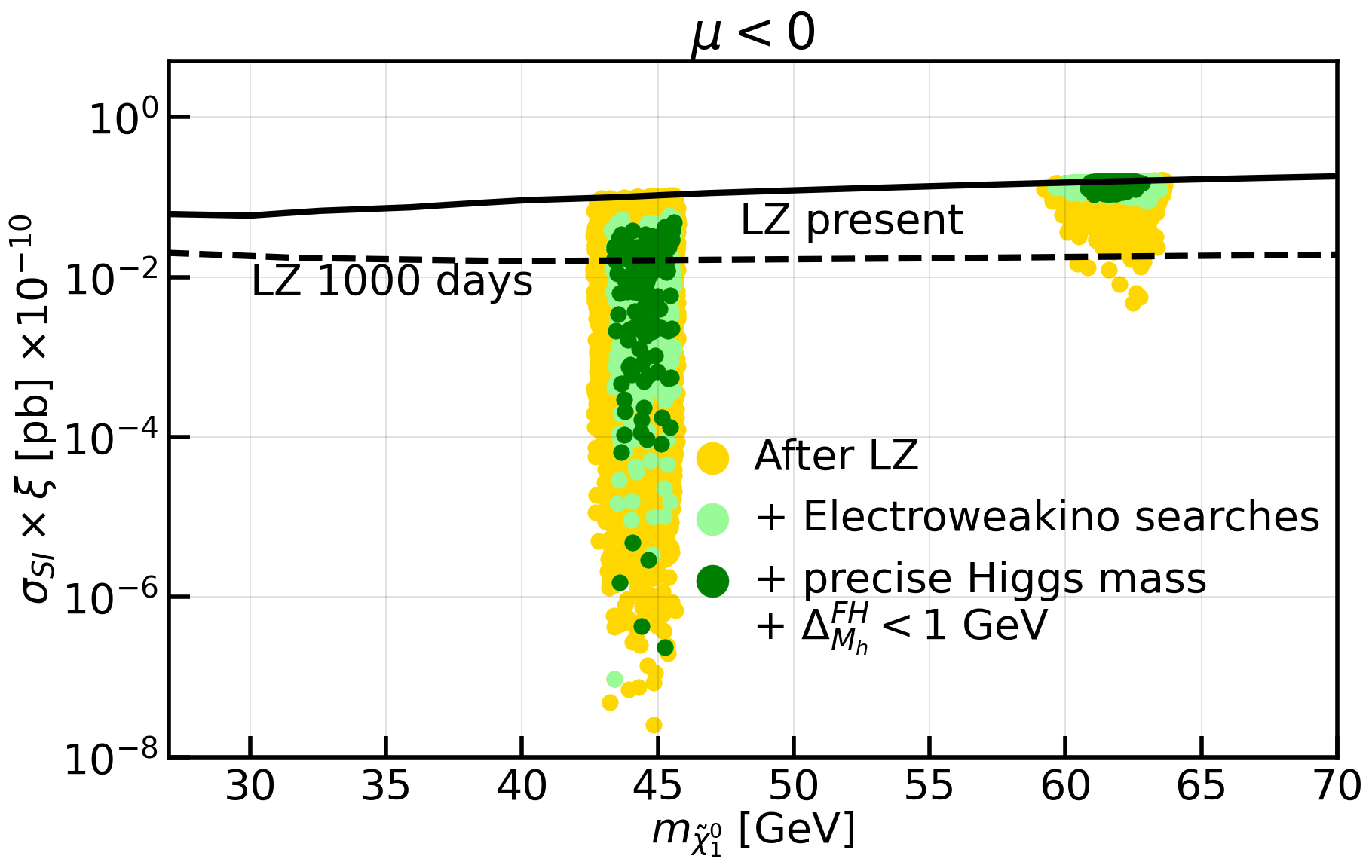}
    \includegraphics[width=0.47\textwidth]{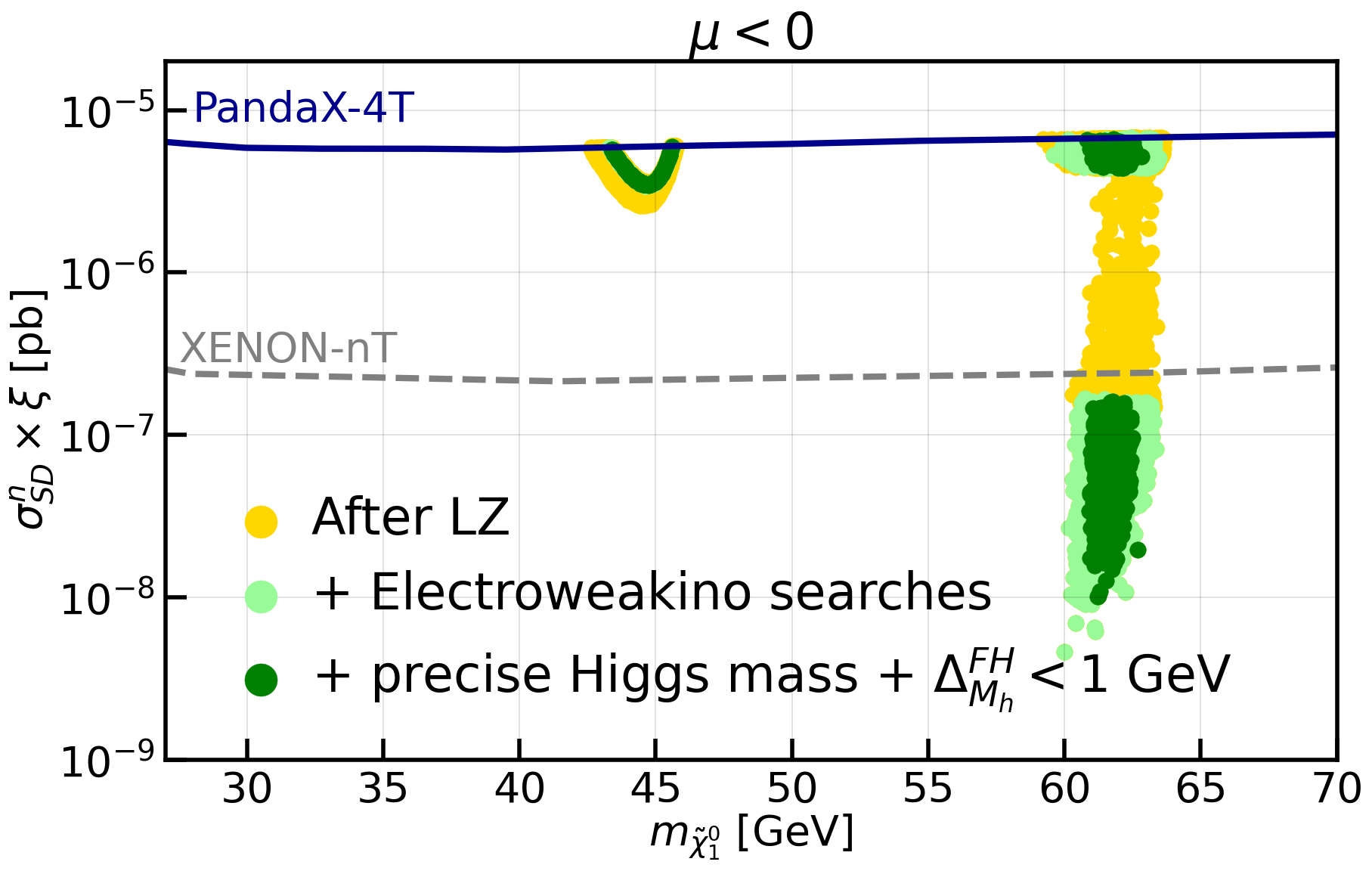}\\
    \includegraphics[width=0.47\textwidth]{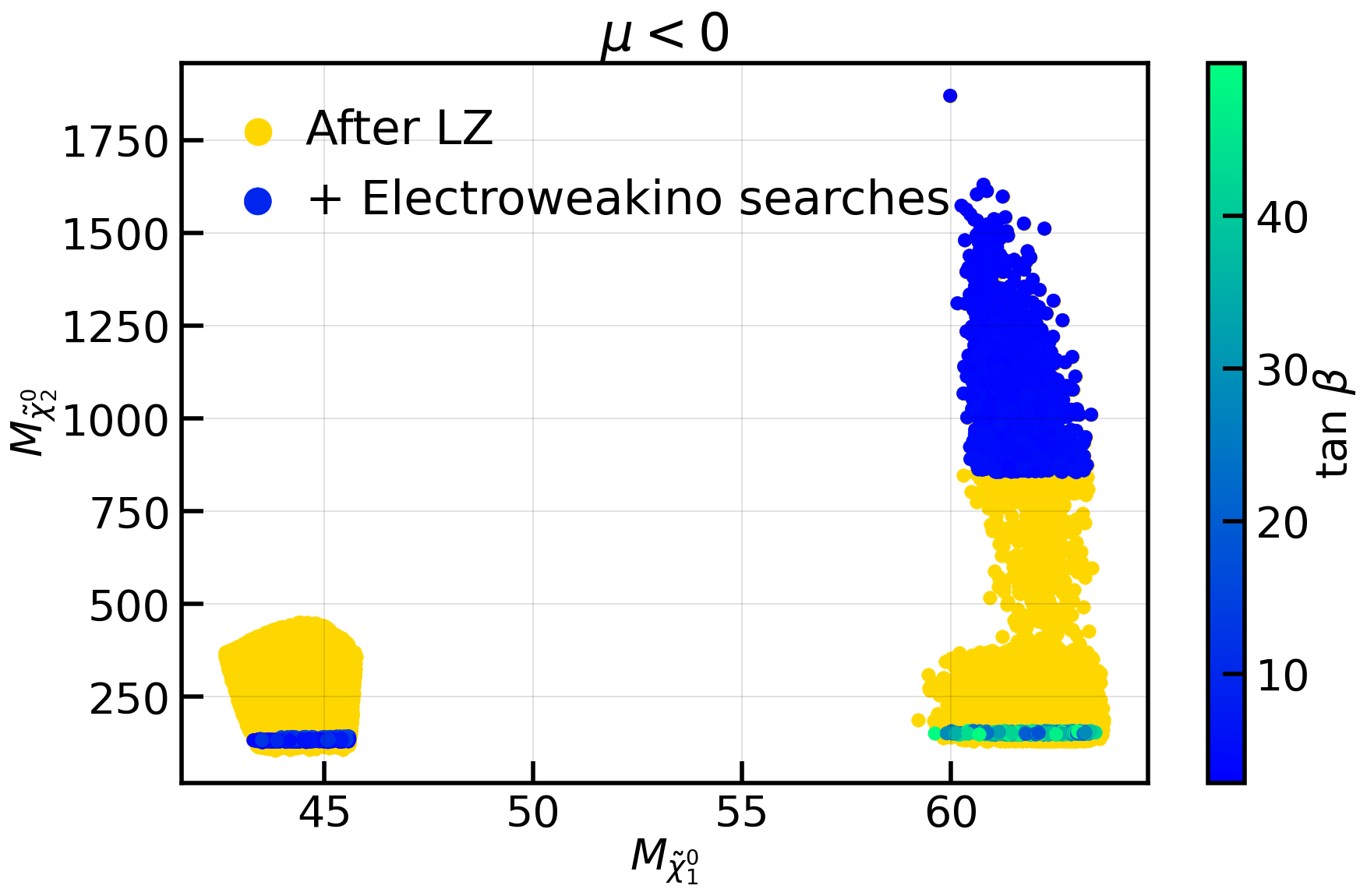}
    \caption{{\it Top:} Scaled SI DM-nucleon cross-section ($\sigma_{SI}\times\xi$, {\it left}) and scaled SD DM-neutron cross-section ($\sigma_{SDn}\times\xi$, {\it right}) for $\mu<0$ as a function of the mass of the LSP neutralino DM in the region of parameter space satisfying LEP, flavor, Higgs constraints, relic density, and DM DD constraints from the XENON-1T, PICO-60, PandaX-4T, and LZ experiments (shown in {\it yellow} and marked \textbf{``After LZ''}), after applying the electroweakino limits implemented in \texttt{SModelS}, and after the precise Higgs mass condition with $\Delta_{M_h}^{FH}<1$\,GeV;
    {\it Bottom:} Parameter space satisfying all the constraints defined as \textbf{``After LZ''} and the regions surviving the electroweakino searches in the $M_{\tilde{\chi}_1^0}$-$M_{\tilde{\chi}_2^0}$ plane for $\mu<0$, where the {\it colorbar} shows tan\,$\beta$. 
    }
    \label{fig:result_negmu}
\end{figure}

Although the Higgs mass can have large theoretical uncertainties, the experimentally measured value of the Higgs mass has a very small error bar. This impacts the allowed mass range of the neutralino, since the relic density is very sensitive to the mass difference between the neutralino from the Higgs funnel. In  Fig\,\ref{fig:result_posmu} the dark green points illustrate the impact of imposing  the precise Higgs mass condition with $\Delta_{M_h}^{FH}<1$\,GeV on the allowed range of the neutralino DM mass. Similar results on the range of allowed neutralino masses are obtained when performing  a dedicated scan where we impose that the uncertainty on the Higgs mass predicted from \texttt{FeynHiggs} is below 1\,GeV and $M_h \pm \Delta_{M_h}^{FH} \in \left[125.38-2\times  0.14\,\rm{GeV}, 125.38+2\times0.14\,\rm{GeV}\right]$, i.e., within 2$\sigma$ of the experimentally measured mass of the Higgs boson. 
In Fig.\,\ref{fig:sigmaSI_higgsmass}, we show the impact of restricting the Higgs mass close to the experimentally measured value on the coupling-$M_h-2M_{\tilde{\chi}_1^0}$ plane. Moreover, we find that the allowed points mostly concentrate near $M_h-2M_{\tilde{\chi}_1^0}\sim 1-3$\,GeV, irrespective of the 3\,GeV Higgs mass window or the more precise Higgs mass condition from \texttt{FeynHiggs}. 
In the next section, we choose our benchmark points from this region where the theoretically estimated Higgs mass is not far from the experimental value.

For the negative $\mu$ scenario, the {\it top} panels of Fig.\,\ref{fig:result_negmu} show the scaled SI DM-nucleon cross-section ($\sigma_{SI}\times\xi$, {\it left}) and scaled SD DM-neutron cross-section ($\sigma_{SDn}\times\xi$, {\it right}) as a function of the mass of the LSP neutralino DM. We observe that we have regions of parameter space which survive all the constraints in both the $Z$ and the $h$ funnels. The {\it top left} panel of Fig.\,\ref{fig:result_negmu} shows that the allowed region in the $h$ funnel is well within the reach of the next few days of LZ data which can improve the limit on SI DD cross-section by 80\%, and from the {\it top right} panel, we infer that the allowed parameter space in the $Z$ funnel can be probed by the SDn result projected by the XENON-nT collaboration.

The {\it bottom} panel of Fig.\,\ref{fig:result_negmu} shows the parameter space in the $M_{\tilde{\chi}_1^0}$-$M_{\tilde{\chi}_2^0}$ plane for $\mu<0$, with the {\it colorbar} representing tan\,$\beta$, for the points allowed by all the constraints. The electroweakino searches restrict the allowed parameter space to either heavy Higgsinos, having masses $\gtrsim 850$\,GeV, or to a narrow region of parameter space with light Higgsinos with masses in the range of 125-145\,GeV in the $Z$-funnel and 145-160\,GeV in the $h$-funnel, many of which have very small $R$-values\,\footnote{$R$-value is the ratio of the signal cross-section and the experimentally allowed upper bound on the cross-section of a BSM process in a particular final state. A smaller $R$-value indicates that the parameter space point is allowed and lies way outside the current limit, whereas a $R$-value greater than 1 indicates that the signal is excluded.}. We further investigate the allowed region of such light Higgsinos in the following section.
Fig.\,\ref{fig:result_posmu} and Fig.\,\ref{fig:result_negmu} are similar to Figs.\,1 and 2 of Ref.\,\cite{Barman:2022jdg}, they are shown here for completeness. We also show the impact of using a tighter constraint on the Higgs mass in Figs.\,\ref{fig:result_posmu} and \ref{fig:result_negmu}. The {\it right} panel of Fig.\,\ref{fig:result_posmu} and the {\it bottom} panel of Fig.\,\ref{fig:result_negmu} also show the tan\,$\beta$ range of the allowed regions in the {\it colorbar}.

The future lepton colliders like ILC and CEPC will be crucial for precision measurements of Higgs boson. The projected upper limit on the invisible branching of the Higgs boson is 0.4\% at ILC\,\cite{Asner:2013psa} and 0.3\% at CEPC\,\cite{An:2018dwb}. Although these can probe a significant part of the allowed parameter space in the $\mu<0$ case, as shown in Fig.\,\ref{fig:h_inv} in Appendix\,\ref{app:h_inv}, we still have regions with Br($h\rightarrow$invisible)$<0.003$ in both the $Z$ and $h$-funnels. 
In the $\mu<0$ case, the partial decay width of the $Z$ boson to $\tilde{\chi}_1^0$ ($\Gamma_{\rm inv}^{\rm new}$) is always less than 0.1\,MeV for the allowed parameter region that we obtain. Therefore, we do not expect the Giga-$Z$ option of ILC, which is expected to have a modest improvement over LEP\,\cite{Carena:2003aj}, to be sensitive to this region.

\begin{figure}[hbt!]
    \centering
    \includegraphics[width=0.5\textwidth]{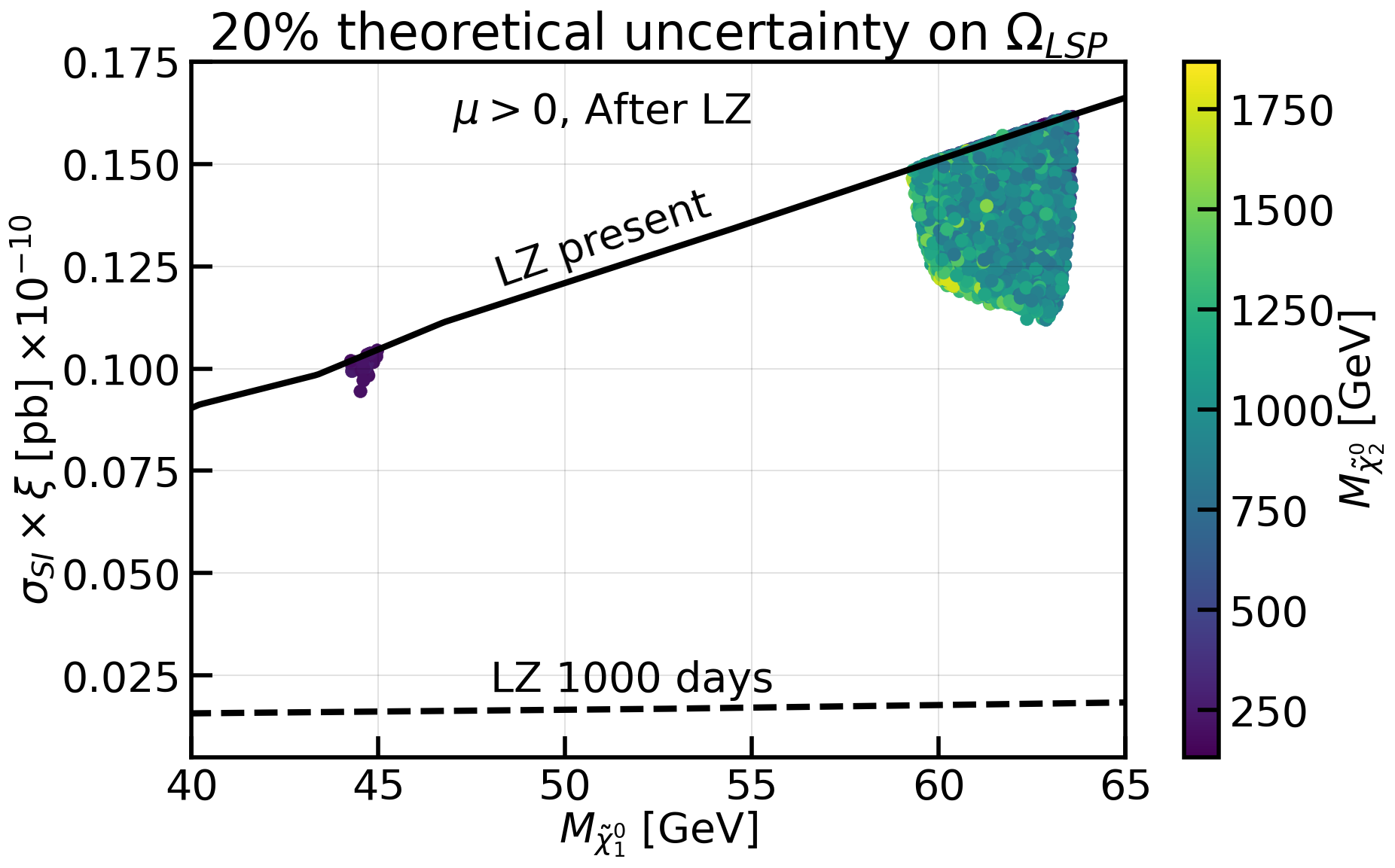}
    \caption{Scaled SI DM-nucleon cross-section ($\sigma_{SI}\times\xi$) for $\mu>0$, with 20\% overestimation of relic density due to theoretical uncertainty, as a function of the mass of the LSP neutralino DM in the region of parameter space satisfying LEP, flavor, Higgs constraints, relic density, and DM DD constraints from the XENON-1T, PICO-60, PandaX-4T, and LZ experiments, defined as \textbf{``After LZ''}.
    }
    \label{fig:result_posmu_20pc}
\end{figure}

The relic density can have theoretical uncertainties in its calculation\,\cite{Baro:2007em,Baro:2009na,Banerjee:2021hal}. 
If we overestimate the relic density by few percent as compared to its actual value due to the theoretical uncertainties, the scaling factor, $\xi$, reduces, resulting in a lower SI DD cross-section.
Fig.\,\ref{fig:result_posmu_20pc} shows the scaled SI DD cross-section ($\sigma_{SI}\times\xi$) with the DM mass of the parameter space for $\mu>0$ surviving the \textbf{``After LZ''} set of constraints assuming that the relic density is overestimated by 20\%, with the {\it colorbar} showing the mass of $\tilde{\chi}_2^0$. We observe that a small allowed region has opened up in the $Z$ funnel which survives the present LZ result. However, the Higgsinos have masses above $\sim200$\,GeV, and get excluded by the collider bounds from electroweakino searches, as we have seen from the {\it bottom} panel of Fig.\,\ref{fig:result_negmu}. In the $h$ funnel, the allowed region extends further down to $\sigma_{SI}\times\xi\sim 0.112\times 10^{-10}$\,pb as compared to our result with no theoretical uncertainty on the relic density, where the allowed region in the $h$ funnel had the lowest scaled SI DD cross-section of around 0.137$\times10^{-10}$\,pb. Still, this region is well within the projected limit from the full 1000 days of the LZ experiment.

Let us now focus on the regions of light Higgsinos allowed by the present electroweakino searches.

\section{Collider analyses for probing the light Higgsinos}
\label{sec:collider}

\begin{figure}[hbt!]
    \centering
    \includegraphics[width=0.5\textwidth]{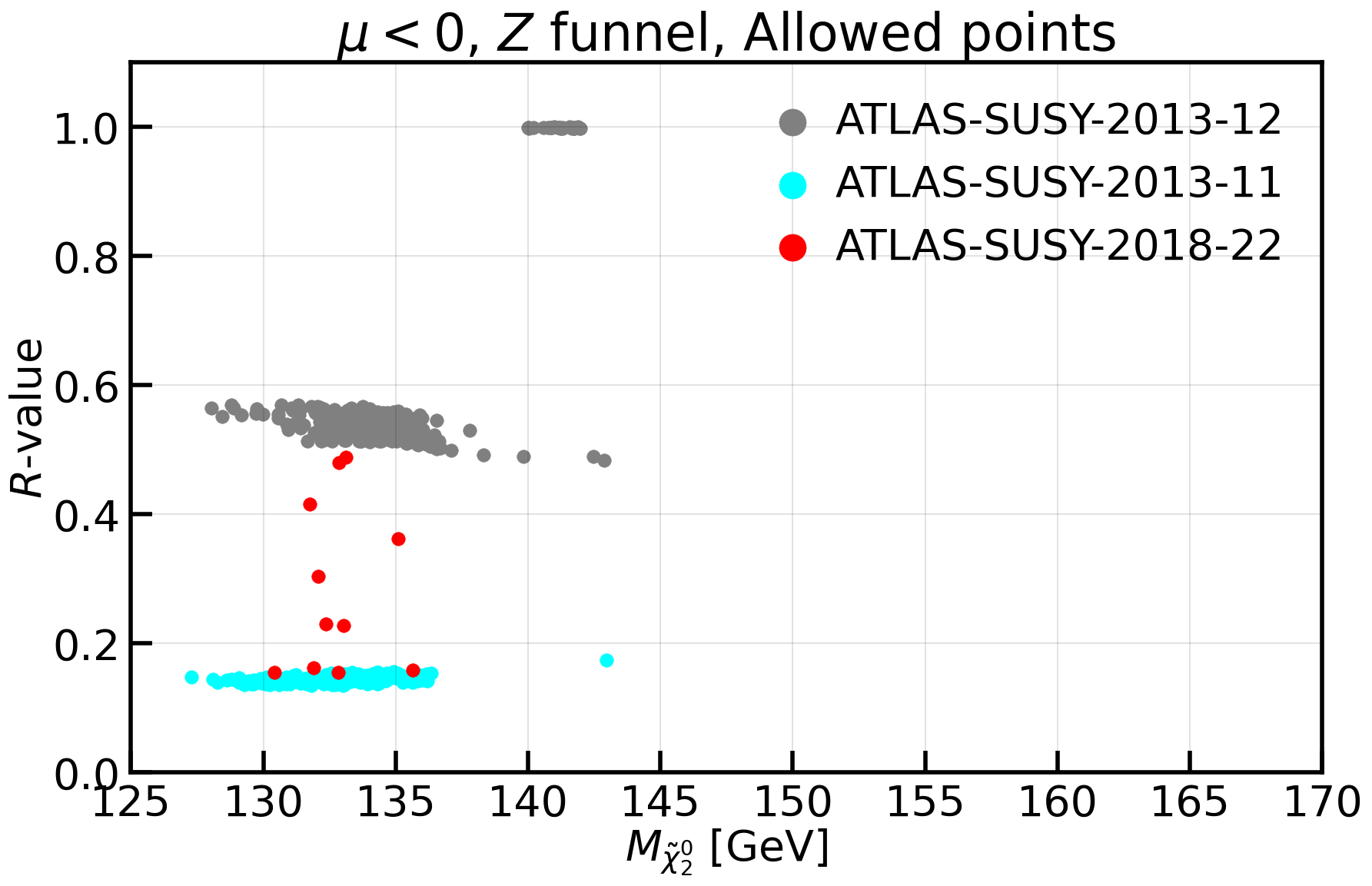}~
    \includegraphics[width=0.5\textwidth]{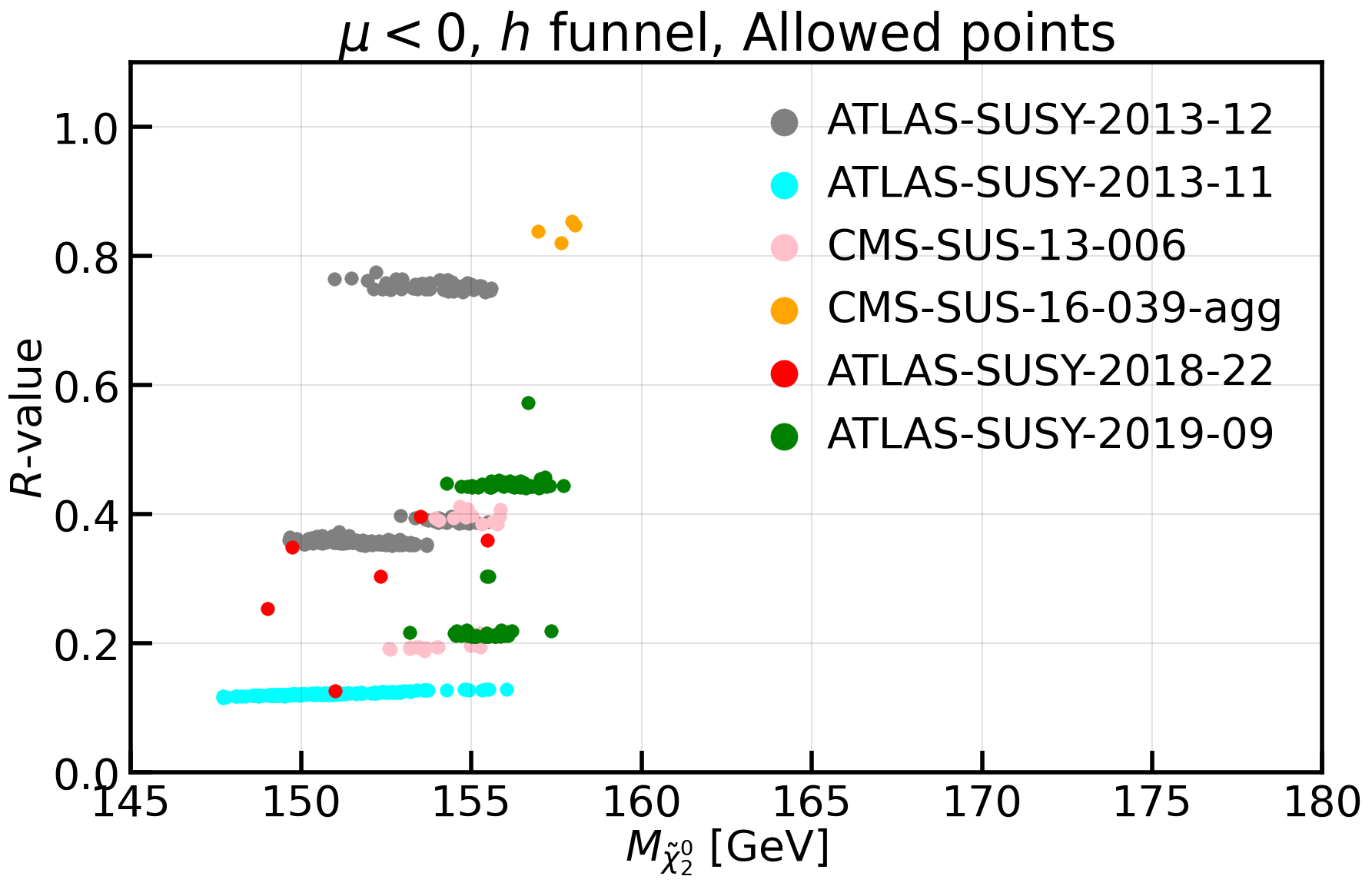}
    \caption{
    $R$-values from \texttt{SModelS-2.2.1} of the most sensitive analyses versus the $M_{\tilde{\chi}_2^0}$ for the light Higgsino scenarios allowed for $\mu<0$ in the $Z$ funnel ({\it left}) and $h$ funnel ({\it right}). The different colors depict the most sensitive analysis for each  point.
    }
    \label{fig:analysis}
\end{figure}

In this section, we study the surviving regions of parameter space with lighter Higgsinos in the standard cosmological scenario.
In Fig.\,\ref{fig:analysis}, taken from Ref.\,\cite{Barman:2022jdg}, we present the $R$-value as given by the \texttt{SModelS} package for the most sensitive analyses on the allowed parameter space in the $Z$~(left) and $h$ funnel (right) of the negative $\mu$ scenario.
This includes ATLAS analyses for the final states 2 leptons ($e$,$\mu$) + MET and 3 leptons ($e$,$\mu$,$\tau$) + MET with 20.3\,fb$^{-1}$ of data, along with final states having jets + MET and 3 leptons + MET with an integrated luminosity of 139\,fb$^{-1}$. The relevant CMS analyses include searches for electroweakinos with decays to leptons, $W$, $Z$, and Higgs bosons with a luminosity of 19.5\,fb$^{-1}$ and multilepton final states with 35.9\,fb$^{-1}$.
We observe that many of these points have very low $R$-values which indicate low sensitivity of the collider searches.
The sensitivity drops around the region where the mass difference between the $\tilde{\chi}_1^\pm$/$\tilde{\chi}_2^0$ and the $\tilde{\chi}_1^0$ is close to the mass of the $Z$ boson. 
Since in this region the decay products of the NLSP electroweakinos are produced at rest, the $\tilde{\chi}_1^0$ does not carry significant momentum, thereby, decreasing the effectiveness of the $p_T^{miss}$ variable to differentiate the signal from SM backgrounds. Therefore, the experimental results suffer from low sensitivity near the $Z$ boson mass threshold.

Representative benchmarks from each of the allowed regions of the parameter space are presented in Table\,\ref{tab:benchmarks}. These benchmarks have very small uncertainty in the Higgs boson mass as estimated by \texttt{FeynHiggs} ($\Delta_{M_h}^{FH}\lesssim \mathcal{O}(1)~$GeV), and have \texttt{SModelS} $R$-values below 0.5\,\footnote{These benchmarks survive the electroweakino searches implemented in the latest version of \texttt{SModelS-2.3.0}\,\cite{MahdiAltakach:2023bdn}. We discuss the impact of combination of analyses implemented in \texttt{SModelS-2.3} on our benchmark points in Appendix\,\ref{app:smodels}.}. They are also allowed when tested with \texttt{CheckMATE\,2}\,\cite{Dercks:2016npn}, another package that implements the constraints from electroweakino searches. We find that the Tevatron searches for light charginos\,\cite{Mario_P_Giordani_2006} are also not sensitive to these benchmarks.

\begin{table}[hbt!]
    \resizebox{0.97\textwidth}{!}{
    \centering
    \begin{tabular}{c|c|c|c|c|c}
    \hline
    \multicolumn{4}{c|}{Benchmarks ({\it mass parameters in} GeV)} & $M_{h} [\Delta_{M_h}^{FH}]~$[GeV] & $\sigma_{SI}\times\xi\times10^{-10}$ [pb] \\\hline
    \multirow{2}{*}{$\mu>0$} & \multirow{2}{*}{$h$-funnel} & \multirow{2}{*}{BP1} & $M_t=173.21$, $M_1=62.5$, $M_2=2000$, $\mu=1000$, tan$\beta=5$, $M_A=3000$, & \multirow{2}{*}{125.38~[$\pm$0.97]} & \multirow{2}{*}{0.151} \\
    &  & & $M_{\tilde{Q}_{3L}}=10000$, $M_{\tilde{t}_{R}}=10000$, $M_{\tilde{b}_{R}}=10000$, $A_t=10000$, $M_3=3000$ & &  \\\hline
    \multirow{6}{*}{$\mu<0$} & \multirow{2}{*}{$Z$-funnel} & \multirow{2}{*}{BP2}
                & $M_t=173.21$, $M_1=44$, $M_2=2000$, $\mu=-124$, tan$\beta=5$, $M_A=3000$, & \multirow{2}{*}{125.88~[$\pm$0.96]} & \multirow{2}{*}{$7.46\times10^{-4}$} \\
        &       &  & $M_{\tilde{Q}_{3L}}=10000$, $M_{\tilde{t}_{R}}=10000$, $M_{\tilde{b}_{R}}=10000$, $A_t=10000$, $M_3=3000$ &  & \\\cline{2-6}
        & \multirow{4}{*}{$h$-funnel} & \multirow{2}{*}{BP3} & $M_t=173.21$, $M_1=68$, $M_2=2000$, $\mu=-150$, tan$\beta=50$, $M_A=3000$,& \multirow{2}{*}{125.67~[$\pm$0.63]} & \multirow{2}{*}{0.143} \\
        &    &  & $M_{\tilde{Q}_{3L}}=5000$, $M_{\tilde{t}_{R}}=5000$, $M_{\tilde{b}_{R}}=5000$, $A_t=-5000$, $M_3=3000$ & & \\\cline{3-6}
        &   & \multirow{2}{*}{BP4} & $M_t=173.21$, $M_1=61$, $M_2=2000$, $\mu=-1000$, tan$\beta=4.5$, $M_A=3000$, & \multirow{2}{*}{125.15~[$\pm$0.99]} & \multirow{2}{*}{0.150} \\
        &   &  & $M_{\tilde{Q}_{3L}}=10000$, $M_{\tilde{t}_{R}}=10000$, $M_{\tilde{b}_{R}}=10000$, $A_t=10000$, $M_3=3000$ &  & \\\hline
    \end{tabular}}
    \caption{Parameters corresponding to four benchmark points satisfying all the present constraints from the $\mu>0$ and $\mu<0$ scenarios along with their scaled SI DD cross-sections. The mass of the Higgs boson $M_{h}$ and the uncertainty in $M_h$ computed by FeynHiggs~($\Delta_{M_h}^{FH}$) are also shown.}
    \label{tab:benchmarks}
\end{table}

For probing the benchmarks with heavy Higgsinos, BP1 and BP4, the hadronic decay channels of the $W$ and $Z$ bosons are more sensitive than the leptonic ones. The current ATLAS result for the hadronic final state excludes Higgsinos below 850\,GeV ({\it right} panel of Fig.\,\ref{fig:result_posmu}). Assuming that the upper limit on the cross-section improves by a factor of $\sqrt{\mathcal{L}}$ with increasing luminosity, $\mathcal{L}$, Run-3 will be able to probe Higgsinos up to a mass of 900-925\,GeV and HL-LHC will further increase the sensitivity to $\sim 1100$\,GeV. Therefore, BP1 and BP4 can be interesting benchmarks to be probed in the HL-LHC runs.

Our benchmarks for light Higgsinos survive the electroweakino constraints as implemented in recasting frameworks like \texttt{SModelS} and \texttt{CheckMATE}. However, it is not guaranteed that all the recent relevant public results from the experimental collaborations have been added in the database of these packages. This motivates a detailed analysis for such light Higgsinos, which we present in the subsequent section.




To estimate the prospects for probing the region with light charginos and neutralinos at the LHC, 
we perform an analysis of the low mass Higgsino-like electroweakinos in the leptonic $3l+\met$ final state at $\sqrt{s}=14$\,TeV using the \texttt{XGBOOST}\,\cite{xgboost}
framework. 
We study the process $pp\rightarrow\tilde{\chi}_1^{\pm}\tilde{\chi}_2^0/\tilde{\chi}_1^{\pm}\tilde{\chi}_3^0,~\tilde{\chi}_1^{\pm}\rightarrow ff'\tilde{\chi}_1^0,~\tilde{\chi}_2^0/\tilde{\chi}_3^0\rightarrow f\bar{f}\tilde{\chi}_1^0$ with $M_{\tilde{\chi}_1^{\pm}}=125.1$\,GeV, $M_{\tilde{\chi}_2^0}=129.9$\,GeV, $M_{\tilde{\chi}_3^0}=133.5$\,GeV, and $M_{\tilde{\chi}_1^0}=44.6$\,GeV (BP2 from Table\,\ref{tab:benchmarks})
where $f, f'$ are SM fermions.
We restrict to the leptonic final state which is cleaner for a lighter benchmark, such as ours. The SM background processes studied in the analyses are summarised in Table\,\ref{tab:backgrounds} with their respective cross-sections and simulation details.

\begin{table*}[!htb]
\centering
\begin{tabular}{c|c|c|c}
Background & Cross section [pb] & Generated using & Total generated \\
\hline
$lll\nu$ & 0.4684$\times 1.2$ & \texttt{MadGraph\,2.7.3} & $9.98\times10^6$ \\
$WZ$, leptonic, $2j$ matched & 1.253$\times 1.2$ & \texttt{MadGraph\,2.7.3} & $4.97\times10^6$ \\
$ZZ$, leptonic, $2j$ matched & 0.1186$\times 1.2$ & \texttt{MadGraph\,2.7.3} & $1.25\times10^6$ \\
$t\bar{t}$, leptonic & 55.36$\times 1.74$ & \texttt{MadGraph\,2.7.3} & $6\times10^7$ \\
$VVV$, inclusive & 0.2678$\times 1.2$ & \texttt{MadGraph\,2.7.3} & $2.5\times10^6$ \\
$Wh$, inclusive & 1.504\,\cite{higgsxs} & \texttt{Pythia\,8.306} & $5\times10^6$ \\
$Zh$, inclusive & 0.883\,\cite{higgsxs} & \texttt{Pythia\,8.306} & $5\times10^6$ \\
ggF $h\rightarrow ZZ$, leptonic & 0.0137 & \texttt{Pythia\,8.306} & $5\times10^6$ \\
VBF $h\rightarrow ZZ$, leptonic & 0.00115 & \texttt{Pythia\,8.306} & $5\times10^6$ \\
$t\bar{t}h$, inclusive & 0.6113\,\cite{higgsxs} & \texttt{Pythia\,8.306} & $5\times10^6$ \\
$t\bar{t}W$, leptonic & 0.01387$\times$1.22 & \texttt{MadGraph\,2.7.3} & $2.5\times10^6$ \\
$t\bar{t}Z$, leptonic & 0.00644$\times$1.23 & \texttt{MadGraph\,2.7.3} & $2.5\times10^6$ \\
\hline        
\end{tabular}
\caption{Details of the background simulation and cross-sections}
\label{tab:backgrounds}
\end{table*}

We perform an analysis of the $3l+\met$ final state where we require exactly three leptons satisfying $p_T>25,25,20$\,GeV and $|\eta|<2.4$, and we have put a veto on $b$-jets with $p_T>30$\,GeV and $|\eta|<2.5$. In our signal benchmark, BP2, since we do not have any on-shell $Z$-boson, we also veto events where the invariant mass of a pair of same flavor opposite sign (SFOS) leptons lie within 10\,GeV of the $Z$ mass. After these preselections, we train our signal and background samples using \texttt{XGBOOST} with a set of the following variables:

\begin{itemize}
    \item Transverse momenta ($p_T$) of the three leptons
    \item Transverse mass ($M_T$) and contransverse mass ($M_{CT}$) of each of the three leptons with the $\met$
    \item Minimum and maximum values of $\Delta R$ between opposite sign lepton pairs along with their $\Delta\eta$ values
    \item Invariant mass of the opposite sign lepton pairs with minimum and maximum $\Delta R$
    \item Missing transverse momentum
    \item Number of jets in the event with the $p_T$ of the two leading jets
    \item Scalar sum of $p_T$ of all the jets in the event ($H_T$)
    \item Invariant mass of the three leptons
\end{itemize}

In benchmark BP3, we study the process $pp\rightarrow\tilde{\chi}_1^{\pm}\tilde{\chi}_2^0/\tilde{\chi}_1^{\pm}\tilde{\chi}_3^0,~\tilde{\chi}_1^{\pm}\rightarrow W^\pm\tilde{\chi}_1^0,~\tilde{\chi}_2^0/\tilde{\chi}_3^0\rightarrow Z\tilde{\chi}_1^0$ with $M_{\tilde{\chi}_1^{\pm}}=125.9$\,GeV, $M_{\tilde{\chi}_2^0}=155.3$\,GeV, $M_{\tilde{\chi}_3^0}=157.4$\,GeV, and $M_{\tilde{\chi}_1^0}=62.2$\,GeV (benchmark 3 from Table\,\ref{tab:benchmarks}).
For this benchmark, we apply the preselections described for BP2 above. 
In this benchmark, we have an on-shell $Z$ boson in the final state. We, therefore, select those events where the invariant mass of a pair of same flavor opposite sign (SFOS) leptons lie within 10\,GeV window of the $Z$ mass, and we define these two leptons as the SFOS pair of leptons.
We use the following variables for training the \texttt{XGBOOST} framework:

\begin{itemize}
    \item Transverse momenta ($p_T$) of the three leptons
    \item Transverse mass ($M_T$) and contransverse mass ($M_{CT}$) of the lepton, which is not part of the SFOS pair of leptons, with the $\met$
    \item $\Delta R$ and $\Delta\eta$ between the SFOS lepton pair
    \item $\Delta \phi$ and $\Delta \eta$ between the SFOS lepton pair system and the unpaired lepton
    \item $\Delta \phi$ between the SFOS lepton pair system and $\met$
    \item $\Delta \phi$ between the unpaired lepton and $\met$
    \item Missing transverse momentum
    \item Number of jets in the event with the $p_T$ of the two leading jets
    \item Scalar sum of $p_T$ of all the jets in the event ($H_T$)
    \item Invariant mass of the three leptons
\end{itemize}

We train our \texttt{XGBOOST} model using the following hyperparameters:
\begin{center}
\texttt{`objective':`multi:softprob', `colsample\_bytree':0.3, `learning\_rate':0.1, \\ `num\_class':12, `max\_depth':7, `alpha':5, `eval\_metric':`mlogloss', \\ `num\_round':1000, `early\_stopping\_rounds':3}
\end{center}
We divide our total sample in two parts $-$ one for training and one for validation. The background events are merged with a weight factor calculated using the fraction of the number of events expected at the LHC for a particular luminosity and the number of events generated for each background process. The weights are then normalised such that the sum of the weights of all the background processes becomes unity. For each epoch, we train on the training data and test the training on the validation sample.
The model minimises its loss function unless the loss on the validation sample does not decrease in three consecutive iterations. The \texttt{XGBOOST} models are separately trained with 21 kinematic variables for BP2 and 18 kinematic variables for BP3. These trained models are then used to discriminate the signal benchmarks from each background class by computing the significance of observing the signal over the background events. 
At the $\sqrt{s}=14$\,TeV LHC with 137\,fb$^{-1}$ of integrated luminosity~($\mathcal{L}$), Table\,\ref{tab:significance} shows the expected number of our two signal benchmark points and background events for a threshold of 0.9 on our \texttt{XGBOOST} output.

\begin{table*}[!htb]
\centering
\begin{tabular}{c|c|c|c}
\multicolumn{2}{c|}{Number of events for $\mathcal{L}=137$\,fb$^{-1}$} & BP2 & BP3 \\
\hline
\multirow{11}{*}{Backgrounds} & $lll\nu$ & 205.6 & $-$\\
& $WZ$, leptonic, $2j$ matched & $-$ & 46.7\\
& $ZZ$, leptonic, $2j$ matched & 14.7 & 5.8\\
& $t\bar{t}$, leptonic & 677.6 & 21.8\\
& $VVV$, inclusive & 13.0 & 2.3\\
& $Wh$, inclusive & 46.5 & 1.4\\
& $Zh$, inclusive & 7.4 & 1.4\\
& ggF $h\rightarrow ZZ$, leptonic & 2.2 & 0.002\\
& VBF $h\rightarrow ZZ$, leptonic & 0.2 & 6.0$\times10^{-4}$\\
& $t\bar{t}h$, inclusive & 8.2 & 0.3\\
& $t\bar{t}W$, leptonic & 9.2 & 0.5\\
& $t\bar{t}Z$, leptonic & 2.5 & 1.0 \\\cline{2-4}
& Total & 987.1 & 81.2\\
\hline        
\multicolumn{2}{c|}{Signal} & 763.4 & 112.1 \\
\hline
\multicolumn{2}{c|}{Significance with 20\% systematic uncertainty} & 3.1 & 4.5 \\
\multicolumn{2}{c|}{Significance with 50\% systematic uncertainty} & 1.3 & 1.98 \\
\hline
\end{tabular}
\caption{Number of events from individual background processes and the signal surviving a threshold of 0.9 on the \texttt{XGBOOST} output from two models trained on benchmarks BP2 and BP3 respectively, along with the signal significance for $\mathcal{L}=137$\,fb$^{-1}$.}
\label{tab:significance}
\end{table*}

We quote our results by assuming a 20\%~(50\%) systematic uncertainty, where the signal significance is estimated using the formula in Ref.\,\cite{Adhikary:2020cli}. 
We present our results for $\sqrt{s}=14$\,TeV to make it easier to translate to the case of Run-3 ($\sqrt{s}=13.6$\,TeV) and HL-LHC ($\sqrt{s}=14$\,TeV) as the cross-sections for direct electroweakino production are not expected to change much.
We find that the result sensitively depends on the systematic uncertainty, which can 
have a significant impact for light electroweakinos.
Our result shows that these light Higgsinos are within the reach of LHC and could be probed with upcoming analyses of the Run-2 data or at Run-3 of the LHC, provided the systematic uncertainties can be controlled.

\section{Impact of light staus on the spectrum}
\label{sec:stau}

In our previous scan, we had fixed the soft parameters related to the first and second generation squarks and all the three generations of sleptons. The former are fixed at masses around 5\,TeV and the latter at masses around 2\,TeV, with all the trilinear couplings associated with these squarks and sleptons set to zero. Lighter squarks and sleptons can enter various processes of the neutralino DM and affect its relic density and in turn, impact of DD experimental constraints due to the scaling factor.
For squarks, as discussed previously, the strong limits from the collider searches reduce their effect on these observables. According to Ref.\,\cite{ATLAS:2020syg}, a single non-degenerate squark has to be heavier than $\sim1200$\,GeV for $M_{\tilde{\chi}_1^0}\sim60$\,GeV. We have found that the presence of a 1200\,GeV squark has very little effect on the relic density of DM (less than 2\%) and negligible effect on the DD cross-section.

Light smuons might have important implications for the muon $g-2$ anomaly. The observed discrepancy of the muon $g-2$ measurement with the SM prediction requires an additional contribution of $(24.9\pm4.8)\times10^{-10}$\,\cite{Muong-2:2023cdq,Aoyama:2020ynm} from new physics. In the MSSM, the sign of the contribution to the muon $g-2$ depends on the sign of $\mu$. Therefore, to resolve the muon $g-2$ anomaly within the MSSM, one prefers the positive sign of $\mu$. 
In our analysis, the smuons are fixed to have a high mass around 2\,TeV, therefore their contribution to the muon $g-2$ is negligible. 
For the positive $\mu$ benchmark that we obtain in Sec.\,\ref{sec:collider}, the MSSM contribution comes to be around $\sim 2\times10^{-11}$, which is two orders of magnitude away from the required value. The present limit on the mass of selectrons and smuons from the CMS collaboration is $\gtrsim 700$\,GeV for a 60\,GeV $\tilde{\chi}_1^0$\,\cite{CMS:2020bfa}. When we reduce the smuon mass parameters (both $M_{\tilde{L}_{2L}}$ and $M_{\tilde{e}_{2R}}$) in the positive $\mu$ benchmark from 2\,TeV to 700\,GeV, the MSSM contribution becomes $4.4\times10^{-11}$, which is still not enough to explain the observation.

Among the three generations of sleptons, staus have the weakest limits\,\cite{ATLAS:2019gti}. 
As motivated earlier in Section\,\ref{sec:parameter_space}, for light staus, we are more interested in studying the effect of RH light staus $-$ for this we vary the parameter $M_{\tilde{e}_{3R}}$ from 85\,GeV to 500\,GeV.
We find that the present searches of stau leptons at the LHC, which are already recasted in the \texttt{SModelS} package, do not constrain the scenario of RH staus as the NLSP for the DM mass range under consideration, as shown in Fig.\,\ref{fig:stau_smod} of Appendix\,\ref{app:stau_smod}.

\begin{figure*}[hbt!]
    \centering
    \includegraphics[width=0.5\textwidth]{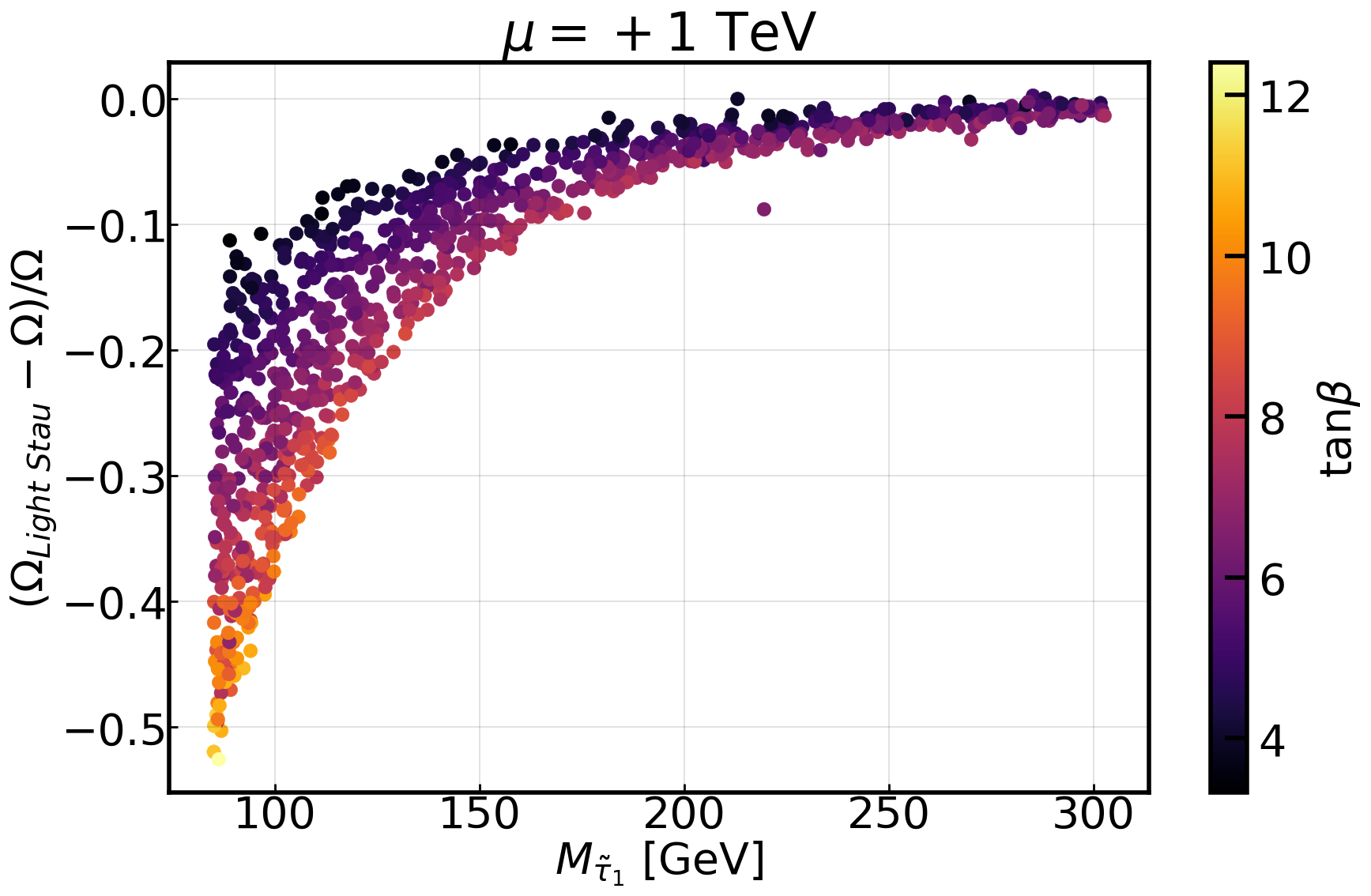}~
    \includegraphics[width=0.5\textwidth]{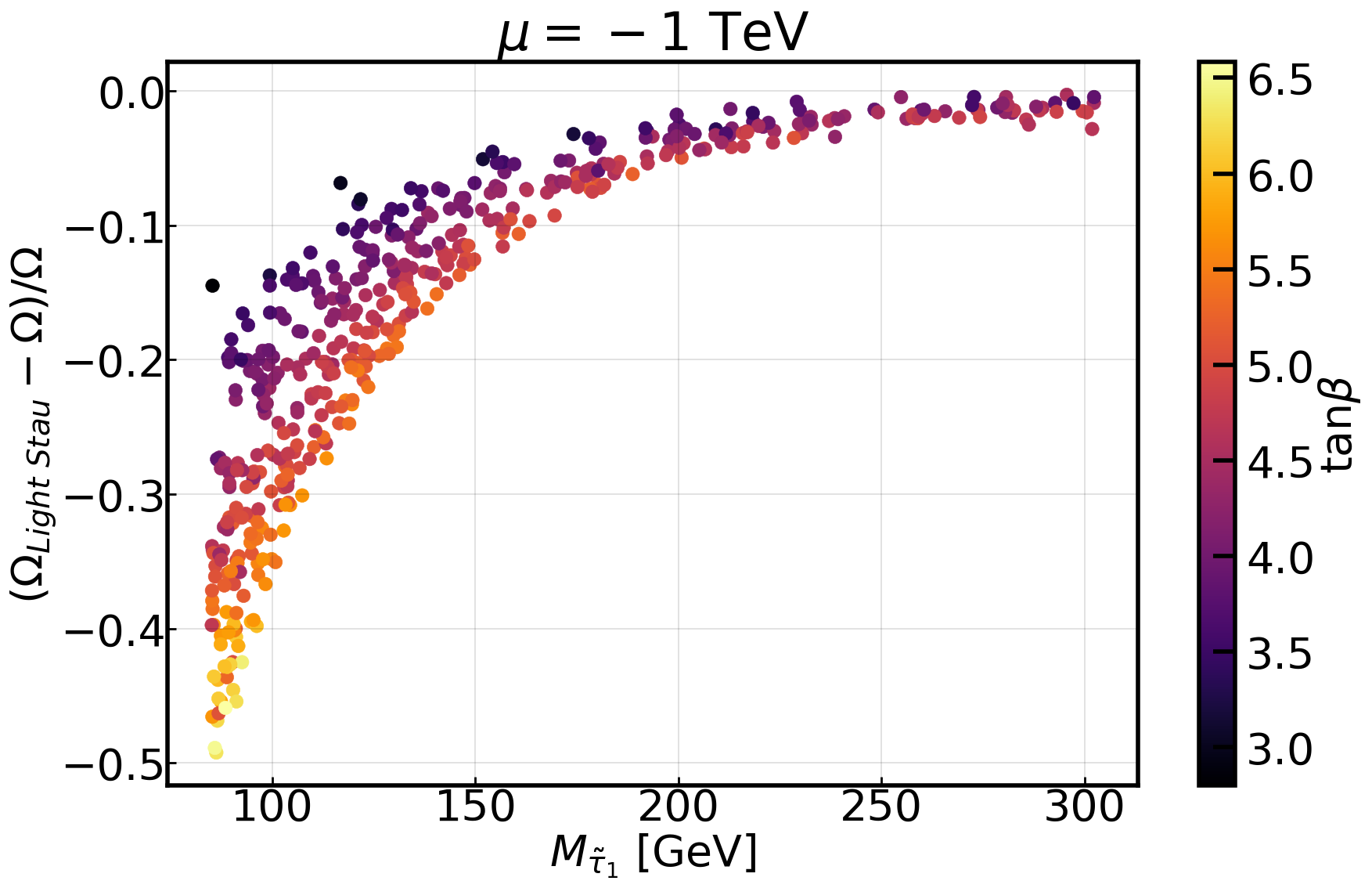}
    \caption{Fractional change in the relic density of the LSP with and without light RH staus as a function of the mass of the lightest stau for $\mu=+1$\,TeV ({\it left}) and $\mu=-1$\,TeV ({\it right}). These parameter points survive all the constraints described earlier and the {\it colorbar} shows the tan\,$\beta$ of these points.}
    \label{fig:stau_fracrelic}
\end{figure*}

\begin{figure*}[hbt!]
    \centering
    \includegraphics[width=0.5\textwidth]{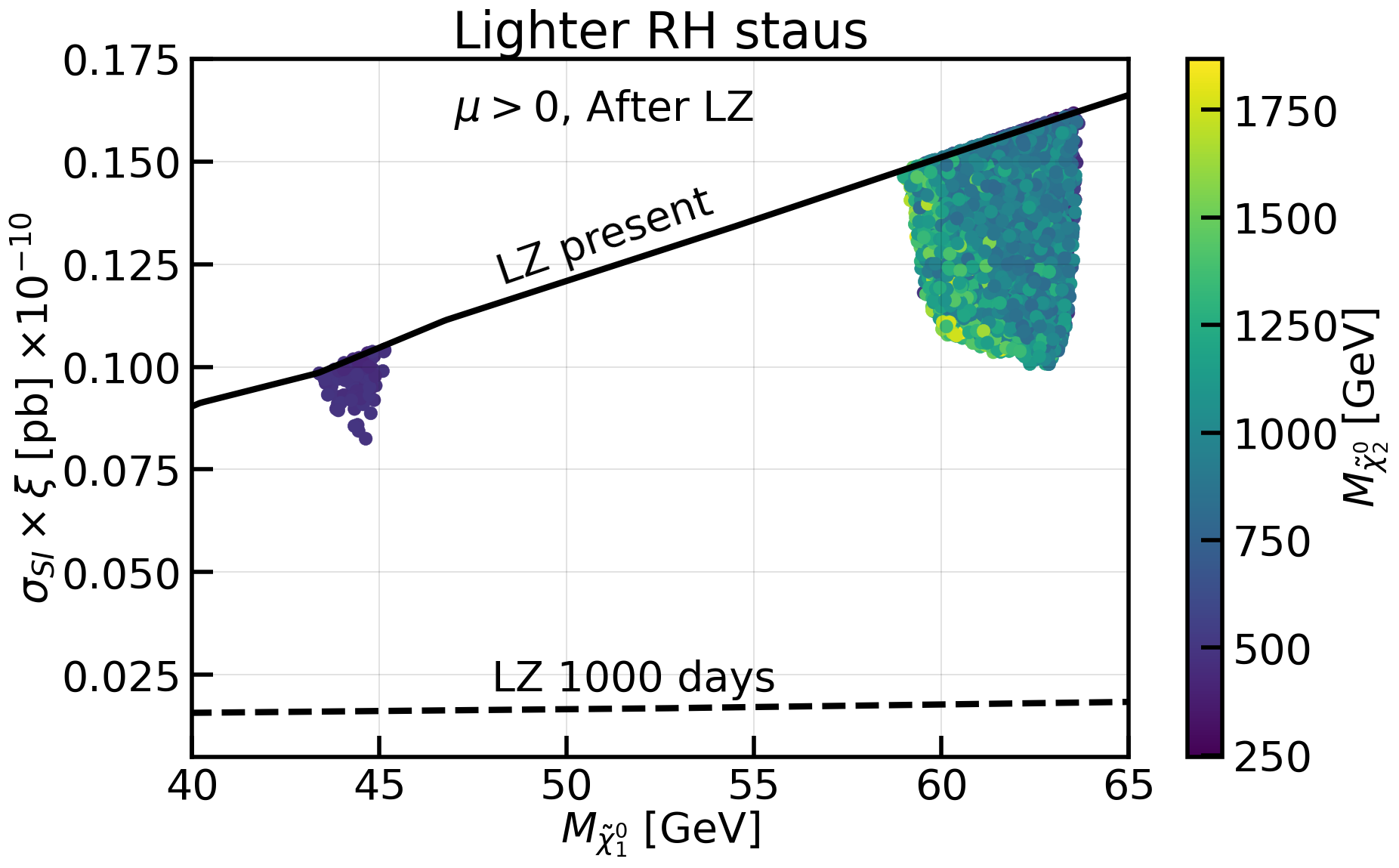}
    \caption{Scaled SI DM-nucleon cross-section ($\sigma_{SI}\times\xi$) for $\mu>0$, with light RH staus ($M_{\tilde{e}_{3R}}=85$\,GeV), as a function of the mass of the LSP neutralino DM in the region of parameter space satisfying LEP, flavor, Higgs constraints, relic density, and DM DD constraints from the XENON-1T, PICO-60, PandaX-4T, and LZ experiments, defined as \textbf{``After LZ''}.}
    \label{fig:stau_DD_result}
\end{figure*}

Since light RH staus are still allowed by the collider constraints, we study their impact on the relic density of the Bino-like LSP that we have studied so far. We observe that for lighter Higgsinos, the couplings $g_{Z\tilde{\chi}_1^0\tilde{\chi}_1^0}$ and $g_{h\tilde{\chi}_1^0\tilde{\chi}_1^0}$ are larger, therefore, we do not expect a large effect from lighter RH staus.
For heavier Higgsinos, say having $|\mu|=1$\,TeV, these coupling have small values due to reduced Higgsino components in the lightest neutralino, and the effect of light RH stau becomes important. To demonstrate this, Fig.\,\ref{fig:stau_fracrelic} shows the fraction of change in the relic density of the LSP DM with and without light RH staus for varying masses of the lightest stau for both positive ({\it left}) and negative ({\it right}) $\mu$, with the Higgsino mass parameter having a value of 1\,TeV in the $h$ funnel. 
We observe that the sign of $\mu$ does not play a significant role, and light staus of 100\,GeV can reduce the relic density by 30-40\%. 

To maximize the effect of a light stau, we fix $M_{\tilde{e}_{3R}}=85$\,GeV corresponding to $M_{\tilde{\tau}_1}=90$-$95$\,GeV, a value above the LEP bound\,\cite{LEP:stau,ALEPH:2001oot,ALEPH:2003acj,DELPHI:2003uqw,L3:2003fyi,OPAL:2003nhx}. We then redo the scan to examine the impact of adding the light stau on the parameter space of the light neutralino thermal dark matter. 
Fig.\,\ref{fig:stau_DD_result} shows the allowed parameter space with the \textbf{``After LZ''} set of cuts in the $\sigma_{SI}\times\xi-M_{\tilde{\chi}_1^0}$ plane for $\mu>0$.
We observe that the light stau reduces the relic density and lowers the scaled SI DD cross-sections, allowing a small region of parameter space in the $Z$ funnel. It also extends the allowed region in the $h$ funnel to lower cross sections. Both the $Z$ and $h$ funnels are within the reach of the LZ projected sensitivity with 1000 days of data.

Our analysis is for a 100\% branching fraction of Higgsinos to the $WZ$ final state. Experimental collaborations also quote their exclusion boundaries assuming 100\% branching to a specific final state and a particular mass hierarchy.
The prospects of the presence of other light SUSY particles might also affect the collider constraints on Higgsinos, if the latter decay into the former with significant branching fractions. One such possibility which we mention in our paper is the presence of light staus having masses between the Bino-like LSP and Higgsinos, which are still allowed by the searches at LHC\,\cite{ATLAS:2019gti} (also see Fig.\,\ref{fig:stau_smod} in Appendix\,\ref{app:stau_smod}). 

\begin{figure*}[hbt!]
    \centering
    \includegraphics[width=0.5\textwidth]{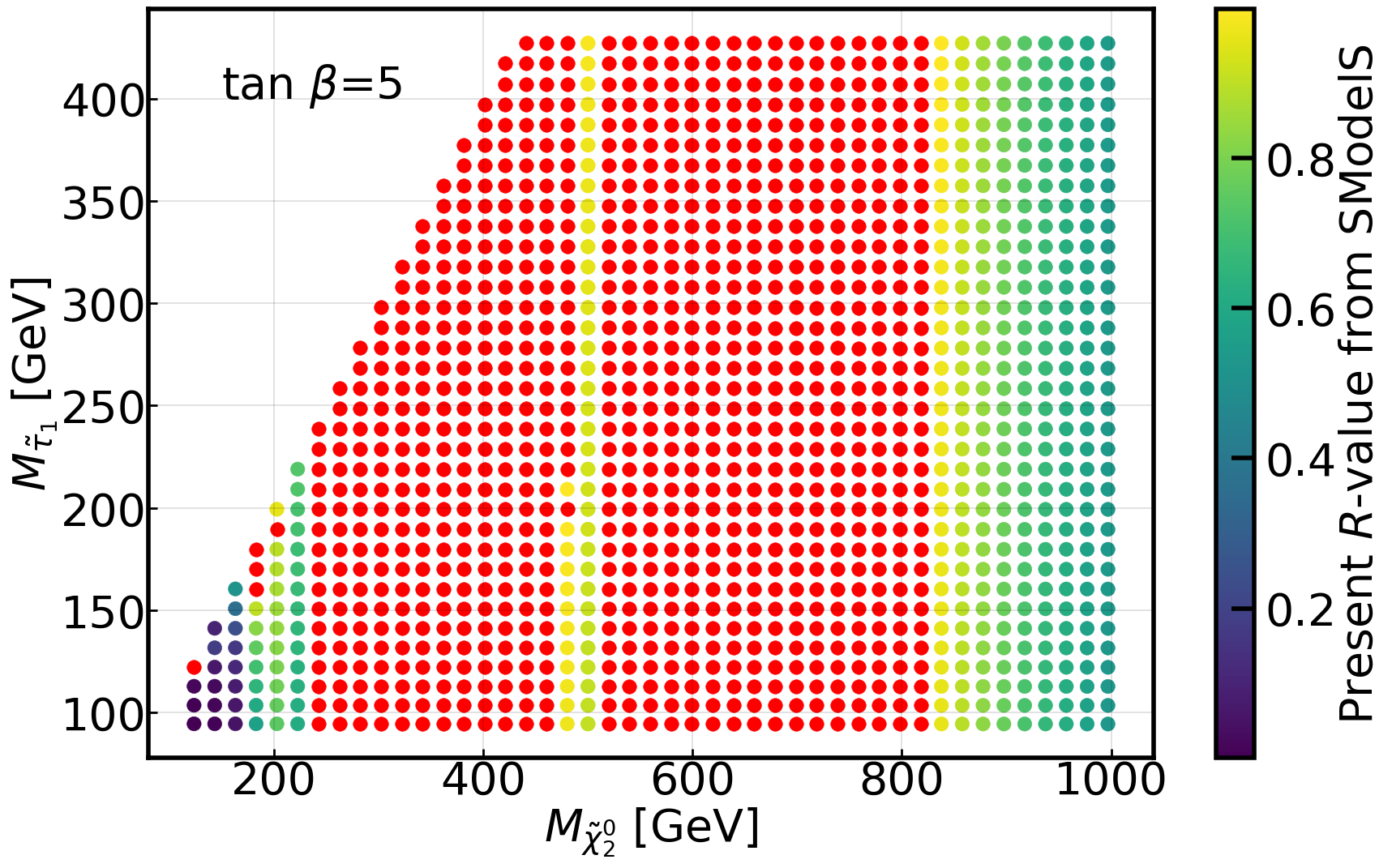}~
    \includegraphics[width=0.5\textwidth]{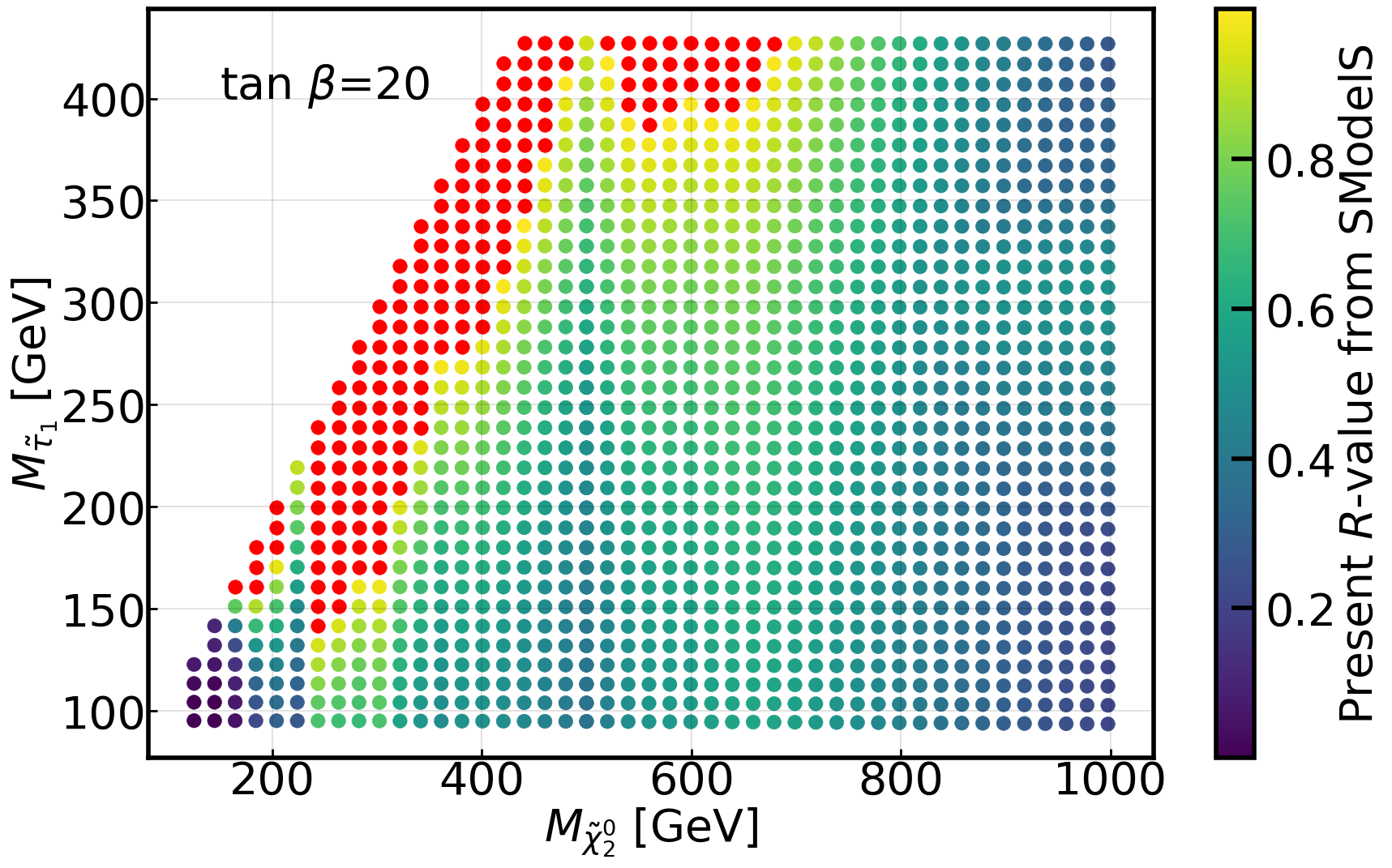}
    \caption{\texttt{SModelS} $R$-values in the plane of mass of the Higgsino-like neutralino and the lightest stau, which is RH, for a range of masses where the Higgsino can kinematically decay to final states involving stau leptons for tan\,$\beta=5$ ({\it left}) and tan\,$\beta=20$ ({\it right}). The points marked in {\it red} are already excluded.}
    \label{fig:higgsino_stau_smod}
\end{figure*}

In order to study the impact of light staus on the exclusion limit of Higgsinos, we perform a scan over the Higgsino-RH stau parameter space, where we fix the Bino mass parameter at 60\,GeV and vary $\mu$ in the range [$100$-$1000$]\,GeV and $M_{\tilde{e}_{3R}}$ in the range [85-500]\,GeV.
Fig.\,\ref{fig:higgsino_stau_smod} shows the $R$-values of the present anaylses as implemented in \texttt{SmodelS} in the plane of mass of the Higgsino-like neutralino and the lightest stau. We present the result for a range of masses where the Higgsino can kinematically decay to final states involving stau leptons for tan\,$\beta=5$ ({\it left}) and tan\,$\beta=20$ ({\it right}). The points marked in {\it red} are already excluded, i.e., have a $R$-value greater than unity\,\footnote{There is a narrow allowed band around Higgsino masses of 500\,GeV in the {\it left panel} of Fig.\,\ref{fig:higgsino_stau_smod}, where all the points have $R$-values greater than 0.9. These are very close to the exclusion limit. This region falls in the gap of two separate analyses $-$ \texttt{CMS-SUS-20-001}\,\cite{CMS:2020bfa} and \texttt{ATLAS-SUSY-2018-41}\,\cite{ATLAS:2021yqv}. 
The first analysis involves leptons in the final state sensitive to smaller Higgsino masses, while the second one
uses boosted hadronically decaying bosons, which is sensitive to higher masses.}. We observe that the presence of such light staus affect the exclusion limit of Higgsino NLSPs. The impact of light RH staus become stronger for higher tan\,$\beta$ values where the branching of Higgsinos to stau leptons increase, thereby, weakening the result. 

\begin{table}[hbt!]
    \resizebox{0.97\textwidth}{!}{
    \centering
    \begin{tabular}{c|c|c|c|c|c}
    \hline
    \multicolumn{4}{c|}{Benchmarks with light staus ({\it mass parameters in} GeV)} & $M_{h} [\Delta_{M_h}^{FH}]~$[GeV] & $\sigma_{SI}\times\xi\times10^{-10}$ [pb] \\\hline
    \multirow{4}{*}{$\mu<0$} & \multirow{2}{*}{$Z$-funnel} & \multirow{2}{*}{BP5} & $M_t=173.21$, $M_1=44$, $M_2=2000$, $\mu=-124$, tan$\beta=5$, $M_A=3000$, & \multirow{2}{*}{125.86[$\pm$0.96]} & \multirow{2}{*}{$7.45\times 10^{-4}$} \\
    &  & & $M_{\tilde{Q}_{3L}}=M_{\tilde{t}_{R}}=M_{\tilde{b}_{R}}=A_t=10000$, $M_3=3000$, $M_{\tilde{e}_{3R}}=85$ & &  \\\cline{2-6}
     & \multirow{2}{*}{$h$-funnel} & \multirow{2}{*}{BP6}
                & $M_t=173.21$, $M_1=68$, $M_2=2000$, $\mu=-150$, tan$\beta=50$, $M_A=3000$, & \multirow{2}{*}{125.65[$\pm$0.63]} & \multirow{2}{*}{0.137} \\
        &       &  & $M_{\tilde{Q}_{3L}}=M_{\tilde{t}_{R}}=M_{\tilde{b}_{R}}=5000$, $A_t=-5000$, $M_3=3000$, $M_{\tilde{e}_{3R}}=85$ &  & \\\hline
     \multirow{4}{*}{$\mu>0$} & \multirow{2}{*}{$Z$-funnel} & \multirow{2}{*}{BP7} & $M_t=173.21$, $M_1=44$, $M_2=2000$, $\mu=500$, tan$\beta=50$, $M_A=6000$, & \multirow{2}{*}{125.11[$\pm$0.99]} & \multirow{2}{*}{0.095} \\
        &    &  & $M_{\tilde{Q}_{3L}}=M_{\tilde{t}_{R}}=M_{\tilde{b}_{R}}=4500$, $A_t=4000$, $M_3=5000$, $M_{\tilde{e}_{3R}}=85$ & & \\\cline{2-6}
        & \multirow{2}{*}{$h$-funnel}  & \multirow{2}{*}{BP8} & $M_t=173.21$, $M_1=62$, $M_2=2000$, $\mu=500$, tan$\beta=20$, $M_A=6000$, & \multirow{2}{*}{124.77[$\pm$0.97]} & \multirow{2}{*}{0.152} \\
        &   &  & $M_{\tilde{Q}_{3L}}=M_{\tilde{t}_{R}}=M_{\tilde{b}_{R}}=4500$, $A_t=4000$, $M_3=5000$, $M_{\tilde{e}_{3R}}=150$ &  & \\\hline
    \end{tabular}}
    \caption{Parameters corresponding to four benchmark points satisfying all the present constraints from the $\mu>0$ and $\mu<0$ scenarios along with their scaled SI DD cross-sections. The mass of the Higgs boson $M_{h}$ and the uncertainty in $M_h$ computed by FeynHiggs~($\Delta_{M_h}^{FH}$) are also shown.}
    \label{tab:benchmarks_stau}
\end{table}

Subsequently, we study the prospect of a \texttt{XGBOOST} based analysis for benchmark points where we have a light stau, which are still allowed by the SUSY searches implemented in \texttt{SModelS-2.3.0}. 
We select four benchmarks $-$ from the $Z$ and $h$ funnel regions of $\mu>0$ and $\mu<0$, each. These are listed in Table\,\ref{tab:benchmarks_stau} with the relevant soft parameters.
BP5 and BP6 correspond to the light Higgsino benchmarks studied in the previous section (BP2 and BP3 in Table\,\ref{tab:benchmarks}) to which we add a light stau with a physical mass around 90\,GeV.
Due to light RH staus, a region of parameter space in the previously excluded $Z$ funnel for positive $\mu$ opens up, and survives both LZ and electroweakino direct search bounds. We select a benchmark from this region, called BP7, with moderate Higgsino mass around 500\,GeV and a very light stau with $M_{\tilde{e}_{3R}}=85$\,GeV. 
For the $h$ funnel of $\mu>0$, Higgsinos up to 850\,GeV masses were excluded for a 100\% branching to the LSP. If RH staus are brought below the Higgsino, then Higgsinos around 500\,GeV masses can still satisfy the collider limits for particular masses of the RH stau, like with $M_{\tilde{e}_{3R}}=150$\,GeV, which we choose as BP8. 

\begin{table*}[!htb]
\centering
\begin{tabular}{c|c|c|c}
\multicolumn{2}{c|}{Number of events for $\mathcal{L}=300$\,fb$^{-1}$} & BP5 & BP6 \\
\hline
\multirow{11}{*}{Backgrounds} & $lll\nu$ & 190.7 & 105.6\\
& $ZZ$, leptonic, $2j$ matched & 39.4 & 26.5\\
& $t\bar{t}$, leptonic & 3500.0 & 1520.5\\
& $VVV$, inclusive & 14.9 & 7.1\\
& $Wh$, inclusive & 61.0 & 27.7\\
& $Zh$, inclusive & 22.8 & 13.5\\
& ggF $h\rightarrow ZZ$, leptonic & 1.2 & 0.5\\
& VBF $h\rightarrow ZZ$, leptonic & 0.2 & 0.05\\
& $t\bar{t}h$, inclusive & 11.0 & 6.4\\
& $t\bar{t}W$, leptonic & 4.7 & 2.1\\
& $t\bar{t}Z$, leptonic & 2.4 & 1.4\\\cline{2-4}
& Total & 3848.3 & 1711.4\\
\hline        
\multicolumn{2}{c|}{Signal} & 5937.9 & 3513.9\\
\hline
\multicolumn{2}{c|}{Significance with 20\% systematic uncertainty} & 5.51 & 6.81\\
\multicolumn{2}{c|}{Significance with 50\% systematic uncertainty} & 2.21 & 2.74\\
\hline
\end{tabular}
\caption{Number of events from individual background processes and the signal surviving a threshold of 0.9 on the \texttt{XGBOOST} output from two models trained on benchmarks BP5 and BP6, respectively, along with the signal significance for $\mathcal{L}=300$\,fb$^{-1}$.}
\label{tab:significance_stau1}
\end{table*}

\begin{table*}[!htb]
\centering
\begin{tabular}{c|c|c|c}
\multicolumn{2}{c|}{Number of events for $\mathcal{L}=300$\,fb$^{-1}$} & BP7 & BP8 \\
\hline
\multirow{11}{*}{Backgrounds} & $lll\nu$ & 40.1 & 35.7\\
& $ZZ$, leptonic, $2j$ matched & 4.9 & 3.43\\
& $t\bar{t}$, leptonic & 1860.1 & 1659.2\\
& $VVV$, inclusive & 16.7 & 16.6\\
& $Wh$, inclusive & 13.8 & 10.5\\
& $Zh$, inclusive & 3.0 & 2.1\\
& ggF $h\rightarrow ZZ$, leptonic & 0.02 & 0.02\\
& VBF $h\rightarrow ZZ$, leptonic & 0.004 & 0.004\\
& $t\bar{t}h$, inclusive & 12.6 & 10.6\\
& $t\bar{t}W$, leptonic & 8.7 & 9.2\\
& $t\bar{t}Z$, leptonic & 2.7 & 2.8\\\cline{2-4}
& Total & 1962.6 & 1750.0\\
\hline        
\multicolumn{2}{c|}{Signal} & 406.8 & 170.2\\
\hline
\multicolumn{2}{c|}{Significance with 5\% systematic uncertainty} & 3.56 & 1.70\\
\hline
\multicolumn{2}{c|}{Significance with 10\% systematic uncertainty} & 1.90 & 0.92\\
\hline
\end{tabular}
\caption{Number of events from individual background processes and the signal surviving a threshold of 0.98 on the \texttt{XGBOOST} output from two models trained on benchmarks BP7 and BP8, respectively, along with the signal significance for $\mathcal{L}=300$\,fb$^{-1}$.}
\label{tab:significance_stau2}
\end{table*}

When the Higgsino decays to staus, we have final states enriched with tau leptons. They in turn decay to electrons, muons, or pions. In our analysis for these benchmarks, we perform a similar analysis like the $3l+$MET, including the hadronic decays of the tau leptons. Table\,\ref{tab:significance_stau1} shows the expected
number of our signal benchmark points BP5 and BP6 along with the background events for a threshold of 0.9 on our \texttt{XGBOOST} output.
We also quote the significance by assuming a 20\% (50\%) systematic uncertainty.
We find that both BP5 and BP6, belonging to the $\mu < 0$ scenario, can be probed with our analysis at the Run-3 of the LHC using 300\,fb$^{-1}$ of data, with a signal significance $\gtrsim 2\sigma$, despite a large systematic uncertainty of $50\%$.
For positive $\mu$, the two benchmarks have higher Higgsino masses (500\,GeV), and therefore, lower production cross-sections. Hence, we put a stronger \texttt{XGBOOST} threshold to reduce the backgrounds further.
Table\,\ref{tab:significance_stau2} shows the expected
number of our signal benchmark points BP7 and BP8 along with the background events for a threshold of 0.98 on our \texttt{XGBOOST} output. For heavier Higgsinos, the systematic uncertainties might be much smaller. We, therefore, quote the significance by assuming a 5\% (10\%) systematic uncertainty. For BP7, we find that if the uncertainty can be brought down to 5\%, we can achieve more than 3$\sigma$ significance, while for BP8, we require the uncertainty to be around 2\% to have 3$\sigma$ significance.

\section{The thermal neutralino in non-standard cosmology}
\label{sec:non_std}

\begin{figure}[hbt!]
    \centering
   \includegraphics[width=0.6\textwidth]{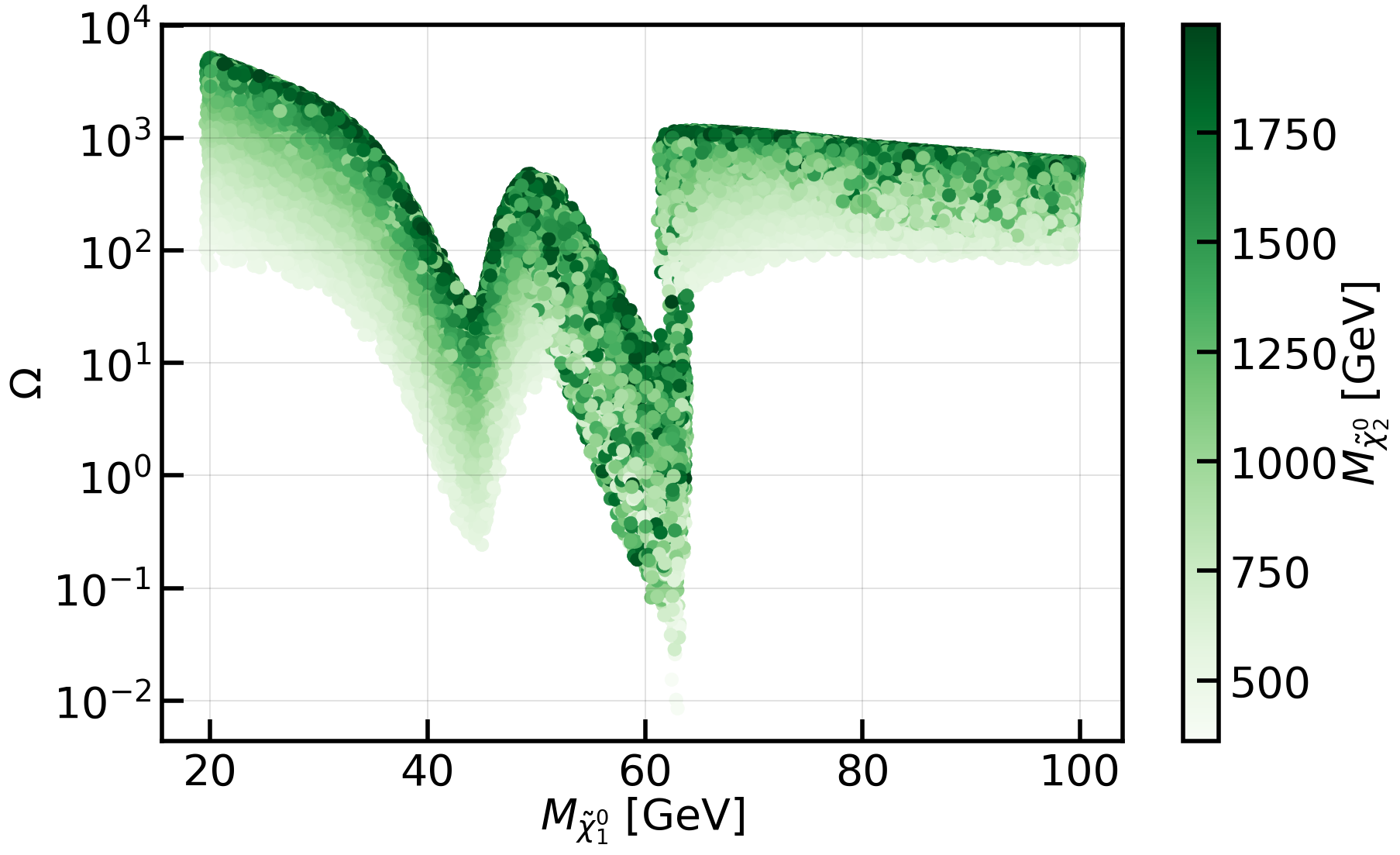}
    \caption{The relic density at freeze-out of the LSP neutralino DM as a function of its mass ($x$-axis) and the mass of the Higgsino-like $\tilde{\chi}_2^0$ ({\it colorbar}), in the region of parameter space satisfying LEP, flavor, Higgs constraints, and DM DD constraints from the XENON-1T, PICO-60, PandaX-4T, and LZ experiments. It is assumed that entropy injection in non-standard cosmology can reduce the relic density of overabundant DM.}
    \label{fig:nonstd}
\end{figure}

Until now, we have worked within the framework of standard cosmological scenario. However, if the neutralino DM is produced thermally in a non-standard cosmology, then the relic density constraint  can be relaxed. This can happen, for example, due to entropy injection in the Universe from the late decay of some particle after the DM freezes out. In this scenario, even if the relic density of the DM at freeze-out is much larger than the present observed relic, it can be diluted due to the increase in the entropy density of the Universe. Ref.\,\cite[Fig.\,8]{Arias:2019uol} shows that for a mass of WIMP DM in the range 40-60\,GeV, the quantity $\langle\sigma v\rangle$, which is the annihilation cross-section multiplied by the DM velocity, can be reduced from the usual value of $10^{-9}$\,GeV$^{-2}$ in standard cosmological scenarios to a value  below $10^{-15}$\,GeV$^{-2}$ in non-standard cosmologies. 
Therefore, in the non-standard cosmology, we can allow for very small DM annihilation cross-sections. The allowed parameter space is not restricted to the funnel regions and the Higgsinos can have masses as large as 2\,TeV, or even heavier in both the $Z$ and $h$ funnels.
Having very small couplings, these points will also satisfy the DD bounds and therefore, will provide interesting benchmarks for probing non-standard cosmologies. 
Fig.\,\ref{fig:nonstd} shows the parameter space surviving the LEP, flavor, Higgs and DM DD constraints in the mass and relic density of the LSP plane, with the colorbar showing the mass of $\tilde{\chi}_2^0$. 

\begin{figure}[hbt!]
    \centering
   \includegraphics[width=0.6\textwidth]{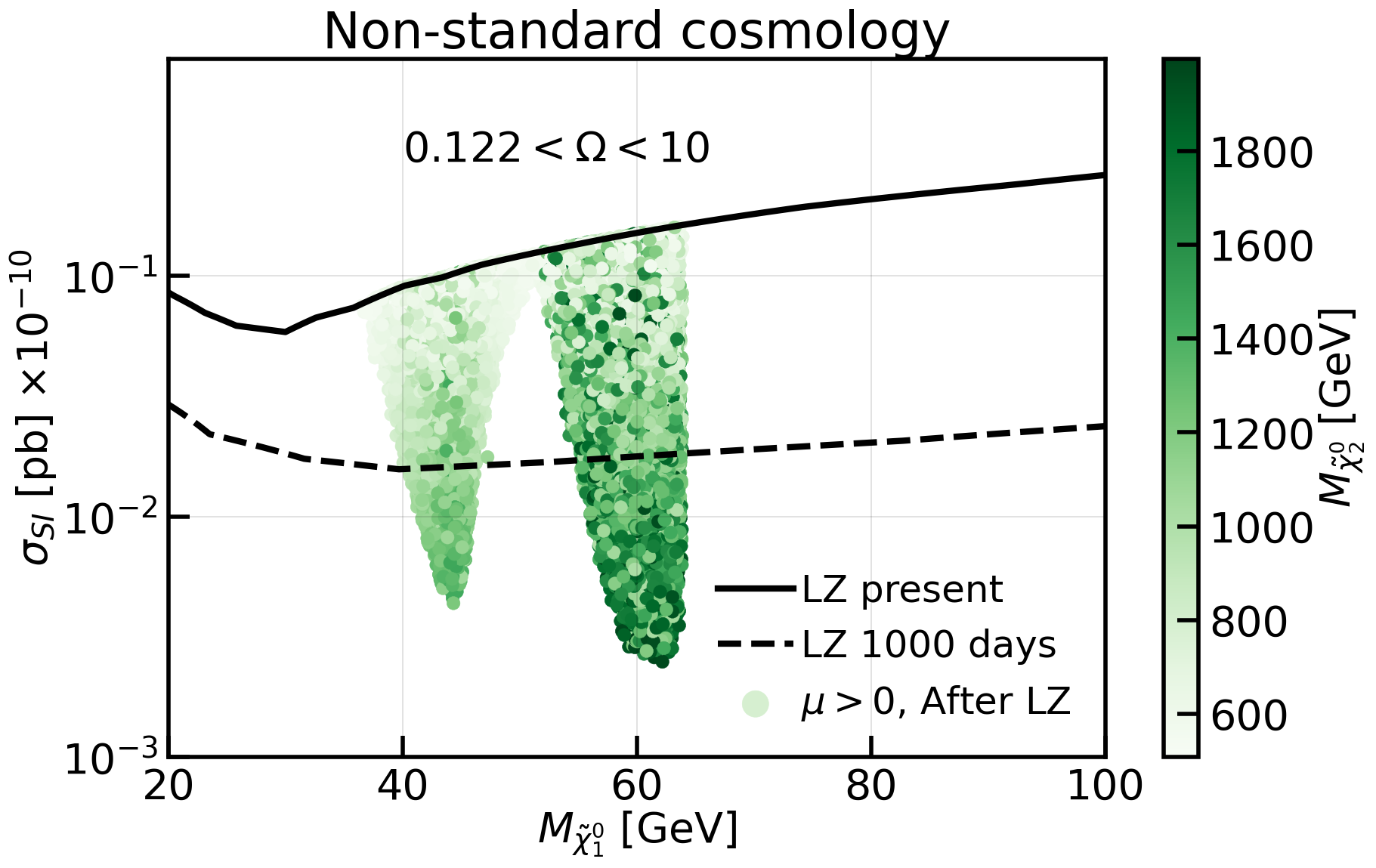}
    \caption{Scaled SI DM-nucleon cross-section ($\sigma_{SI}\times\xi$) for $\mu>0$ in the non-standard cosmological scenario, where the entropy dilution can provide the correct observed DM relic density upto a freeze-out density of around 10, as a function of the mass of the LSP neutralino DM in the region of parameter space satisfying LEP, flavor, Higgs constraints, relic density ($0.122<\Omega<10$), and DM DD constraints from the XENON-1T, PICO-60, PandaX-4T, and LZ experiments.}
    \label{fig:nonstd_DD}
\end{figure}

Fig.\,\ref{fig:nonstd_DD} shows the scaled SI DD cross-section as a function of the DM mass for scenarios where  DM is found to be overabundant in the standard cosmological scenario, but can satisfy the observed relic density due to non-standard cosmology. We apply all the constraints from the set ``After LZ'', where only the relic density constraint is modified to $0.122<\Omega h^2<10$, where we assume that the non-standard cosmology can dilute the DM relic density by a factor of about 100.
We observe that new regions of the parameter space now survive the experimental constraints, in both the $Z$ and $h$ funnel regions. This would allow us to identify scenarios of non-standard cosmology, depending on the nature of the observed signals. 
For instance, if one observes DM in the $Z$-funnel region and simultaneously a LHC signal for Higgsinos heavier than 500\,GeV, it might indicate non-standard cosmology with thermal production of the neutralino DM. Even in the $h$-funnel region, we have an idea of the minimum DD cross-section values that can survive in standard cosmology. Observing a signal with the future LZ data might hint towards a non-standard cosmological picture.  

\begin{figure}[hbt!]
    \centering
    \includegraphics[width=0.47\textwidth]{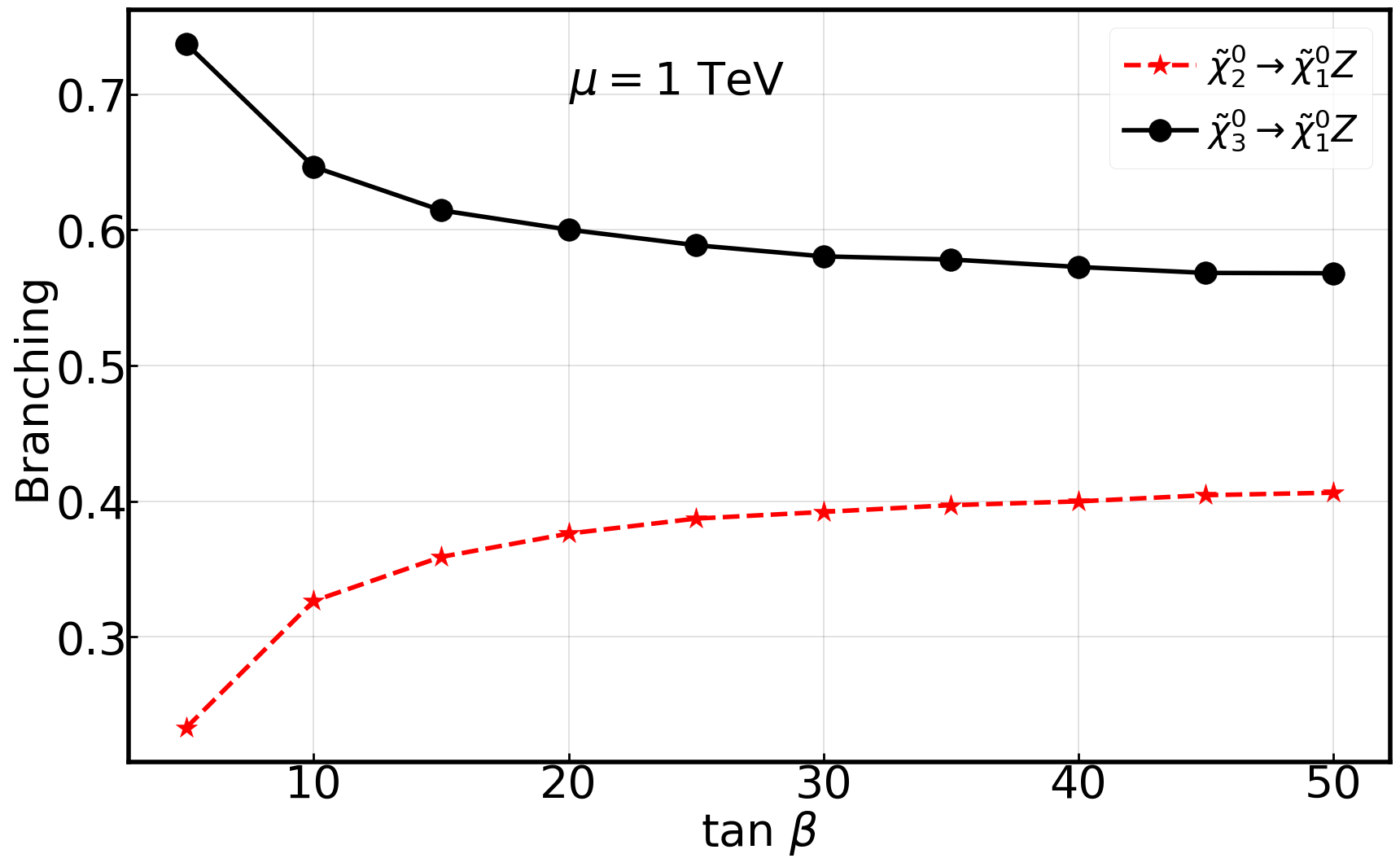}~
    \includegraphics[width=0.47\textwidth]{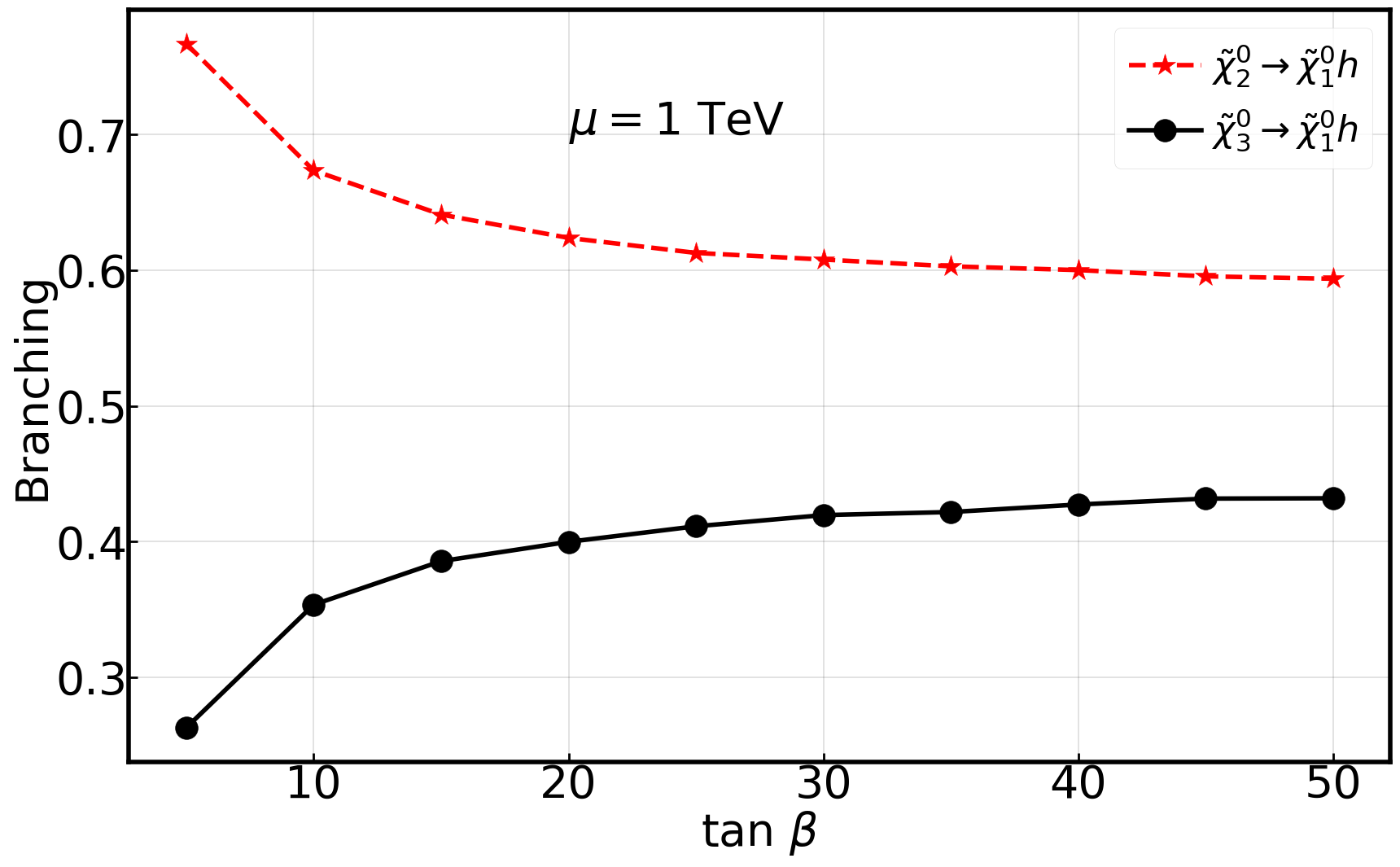}
    \caption{Branching fractions of the two Higgsino-like neutralinos with $\mu=1$\,TeV to $\tilde{\chi}_1^0 Z$ ({\it left}) and $\tilde{\chi}_1^0 h$ ({\it right}) when the staus are heavy.}
    \label{fig:branchings_withoutstau}
\end{figure}

\begin{figure}[hbt!]
    \centering
    \includegraphics[width=0.47\textwidth]{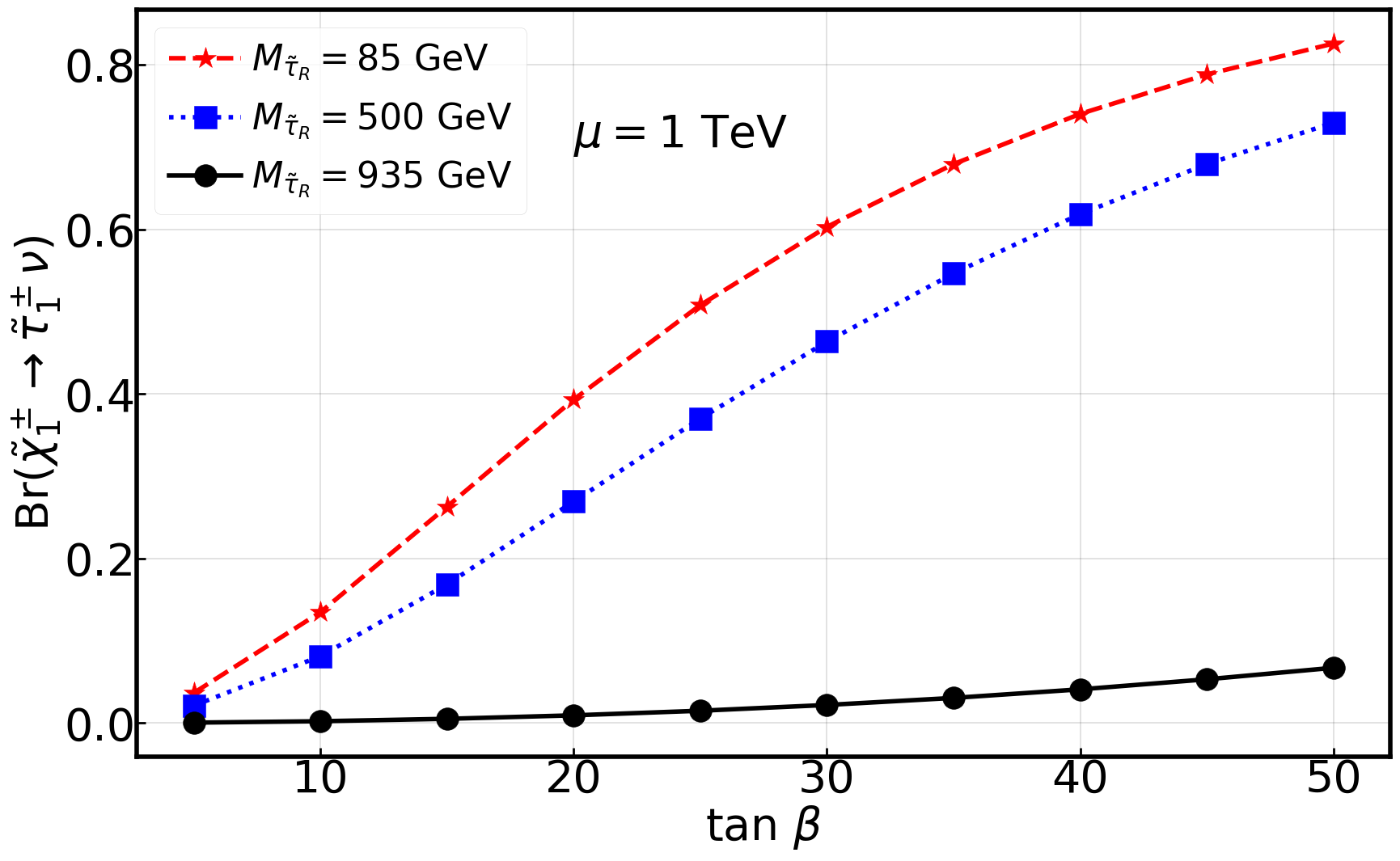}~
    \includegraphics[width=0.47\textwidth]{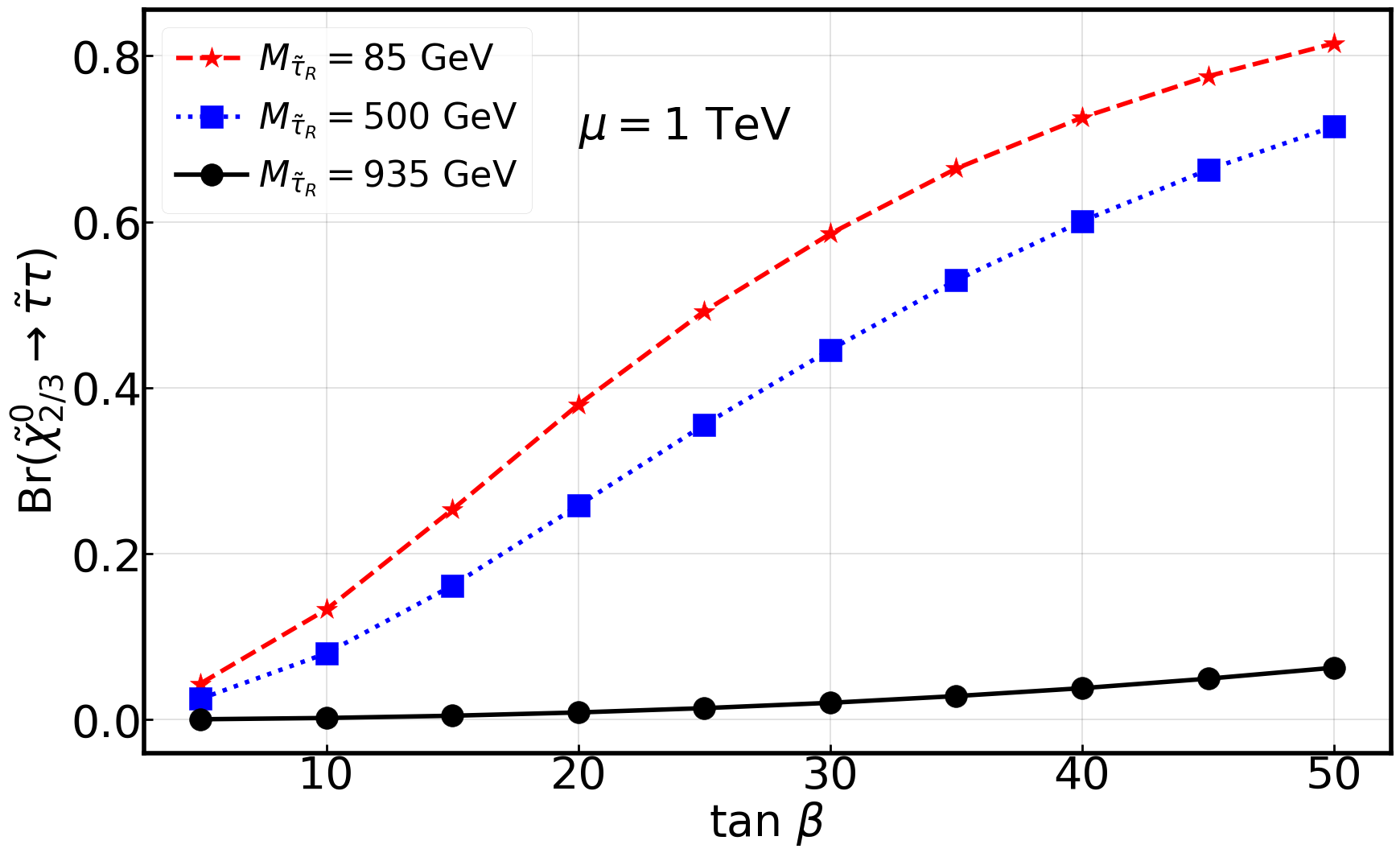}
    \caption{Branching fractions of the Higgsino-like chargino ({\it left}) and the two neutralinos ({\it right}) to final states involving various masses of RH staus.}
    \label{fig:branchings_withstau}
\end{figure}

Additionally, for a DM signal from DD experiments in the $h$ funnel, it is interesting to explore whether collider experiments can provide any hint of non-standard cosmology. If we observe a signal for heavy Higgsinos, having masses around a TeV for high tan\,$\beta$, then it might suggest a non-standard cosmological scenario. To answer the question whether we can get an idea of the tan\,$\beta$ value from the signal, we studied the variation in branching of the Higgsinos to various final states as a function of tan\,$\beta$. 
Fig.\,\ref{fig:branchings_withoutstau} shows the branching of the neutral Higgsinos to the LSP and Z/h. We find that although there is a tan\,$\beta$ dependence, the sum of Br$(\tilde{\chi}_2^0\rightarrow \tilde{\chi}_1^0 Z)+$Br$(\tilde{\chi}_3^0\rightarrow \tilde{\chi}_1^0 Z)$ (or with Higgs boson in the final state), has no variation with tan\,$\beta$. Therefore, it won't be possible to estimate the tan\,$\beta$ value from the branching fraction of Higgsinos to these final states. We then turn to the case of light staus, such that the staus have masses between the Bino and the Higgsinos.
Fig.\,\ref{fig:branchings_withstau} shows the branching of the chargino and the two neutralinos decaying to stau and tau for three different stau masses. We observe a very clear tan\,$\beta$ dependence in this case, especially for lighter staus. If we observe a signal of Higgsinos and could identify the final state with staus, we can get an idea of the tan\,$\beta$ from the branching fraction of Higgsinos decaying to staus. The branching to staus is higher for high tan\,$\beta$, where relic density cannot be satisfied for a TeV scale Higgsino within standard cosmology, and therefore, points toward a non-standard cosmological scenario. 

\section{Conclusion}
\label{sec:conclusion}


In summary, this study shows that the current experiments, especially the recent results from electroweakino searches at the LHC and dark matter DD measurements from LZ, have severely constrained the $\mu>0$ scenario for a light neutralino thermal DM in the pMSSM with 10 free parameters. The DD result of the LZ collaboration is a strong constraint and it affects different regions of the parameter space in the pMSSM depending on the constructive and destructive interferences between the light and the heavy CP-even neutral Higgs bosons.
For heavy staus, the $Z$-funnel is completely excluded, and only heavy Higgsinos ($M_{\tilde{\chi}_1^0}\gtrsim 850$\,GeV) are allowed in the $h$-funnel region. 
In the $\mu<0$ scenario of the same model, the allowed parameter space consists of either Higgsinos heavier than $\sim850$\,GeV in the $h$-funnel or restricted to a narrow region of light Higgsinos having mass of 125-160\,GeV in the $Z$ and $h$-funnels, unlike the $\mu>0$ scenario. For light Higgsinos, there is a constructive interference between the $h$ and $H$ contributions to the DD cross-section for positive $\mu$. Therefore, it is not possible to evade the LZ bounds. There is a destructive interference between the contributions from the two CP-even neutral Higgs bosons for negative $\mu$, thereby, opening up allowed parameter space with light Higgsinos. Our \texttt{XGBOOST} analysis with the $3l+$MET signature shows that these light Higgsinos could be probed with the Run-2 data of 137\,fb$^{-1}$, if the systematic uncertainties lie within 20-30\%. These benchmarks are still allowed by the available recasting frameworks, like \texttt{SModelS} and \texttt{CheckMATE}, which can translate the result of Wino-like NLSPs, provided by experimental collaborations, for Higgsino-like NLSPs. Thus, they form an important target for Run-3 searches.
We also find that the DM bounds on the relic density and the DD SI cross-sections severely constrain any simplified extension of SM with a Majorana fermionic DM coupling only to the discovered Higgs boson.

The situation changes when we have light RH staus as NLSPs. 
They provide an additional annihilation channel for the DM and, therefore, reduce the relic density by a factor depending on the mass of the stau.
Furthermore, they also affect the exclusion limits of Higgsinos in collider searches, when the Higgsino can decay to staus with significant branching fractions. Due to both these effects, we get an allowed region of parameter space even in the $Z$ funnel of $\mu>0$ with lighter staus. In the $h$ funnel of positive $\mu$, the presence of light staus relax the lower limit on Higgsino masses, and 500\,GeV Higgsinos are still allowed in such a scenario. 
Our preliminary analysis of these benchmarks shows they are accessible at Run-3 of LHC if the systematic uncertainties can be controlled. The future direct detection experiments, e.g., 1000 days of LZ, can probe the $Z$ and the $h$ funnels with light staus, as can be seen from Fig.\,\ref{fig:stau_DD_result}.
The status further changes when we go to non-standard cosmological scenarios, where the relic density can be satisfied by late injection of entropy in the Universe. Therefore, a large region of the parameter space becomes available. We discuss some ways to identify a non-standard history of our Universe from the combination of signals in future  DM DD and collider experiments, which could not be realised assuming the standard cosmological model. The future DM DD experiments and the LHC Run-3 have promising prospects in exploring the remaining corners of the pMSSM parameter space with a light neutralino thermal dark matter.

\section*{Acknowledgement}

We thank Sabine Kraml for the useful discussion and help related to the \texttt{SModelS} package. The work of G.B. and R.M.G. was funded in part by the Indo-French Centre for the Promotion of Advanced Research, Grant no: 6304-2. R.M.G. wishes to acknowledge the support of Indian National Science Academy under the award of INSA Senior Scientist Scheme. 
The work of B.B. was supported by the SERB Core Research Grant CRG/2022/001922 and the SERB Matrics Grant MTR/2022/000264.
B.B. and R.S. thank Prabhat Solanki and Camellia Bose for useful discussions. The work of R.K.B. was supported by the World Premier International Research Center Initiative (WPI), MEXT, Japan. R.K.B. thanks the U.S. Department of Energy for the financial support under grant number DE-SC0016013.
R.S. acknowledges the support of the Deutsche Forschungsgemeinschaft (DFG) through the funds provided to the Sino-German
Collaborative Research Center TRR110 “Symmetries and the Emergence of Structure in QCD” (DFG Project-ID 196253076)". R.S. would like to thank the Indian Institute of Science for computational support.

\clearpage

\appendix

\section{Impact of the constraints on the invisible branching of the Higgs boson}
\label{app:h_inv}

\begin{figure}[hbt!]
    \centering
    \includegraphics[width=0.46\textwidth]{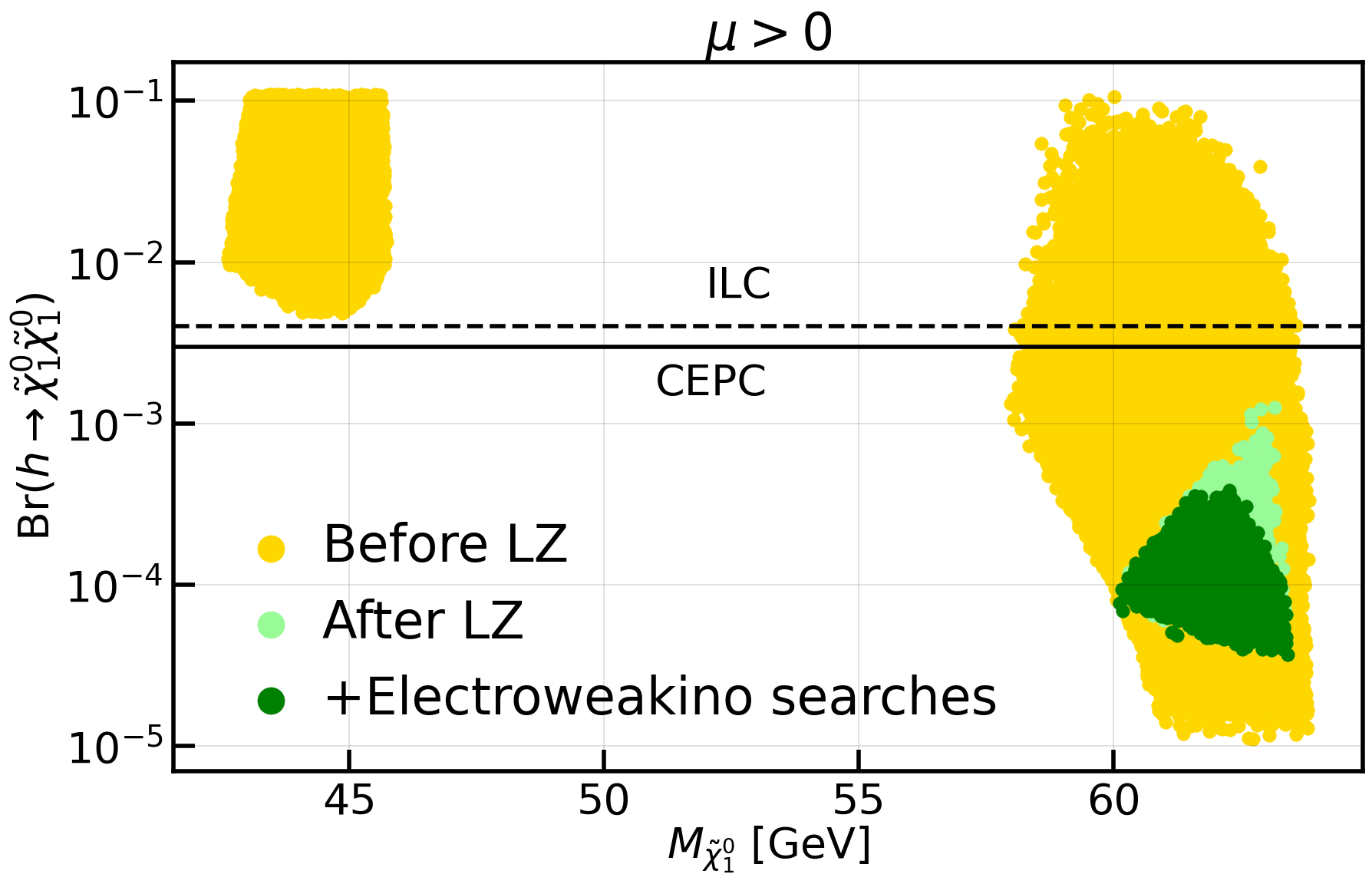}~
    \includegraphics[width=0.46\textwidth]{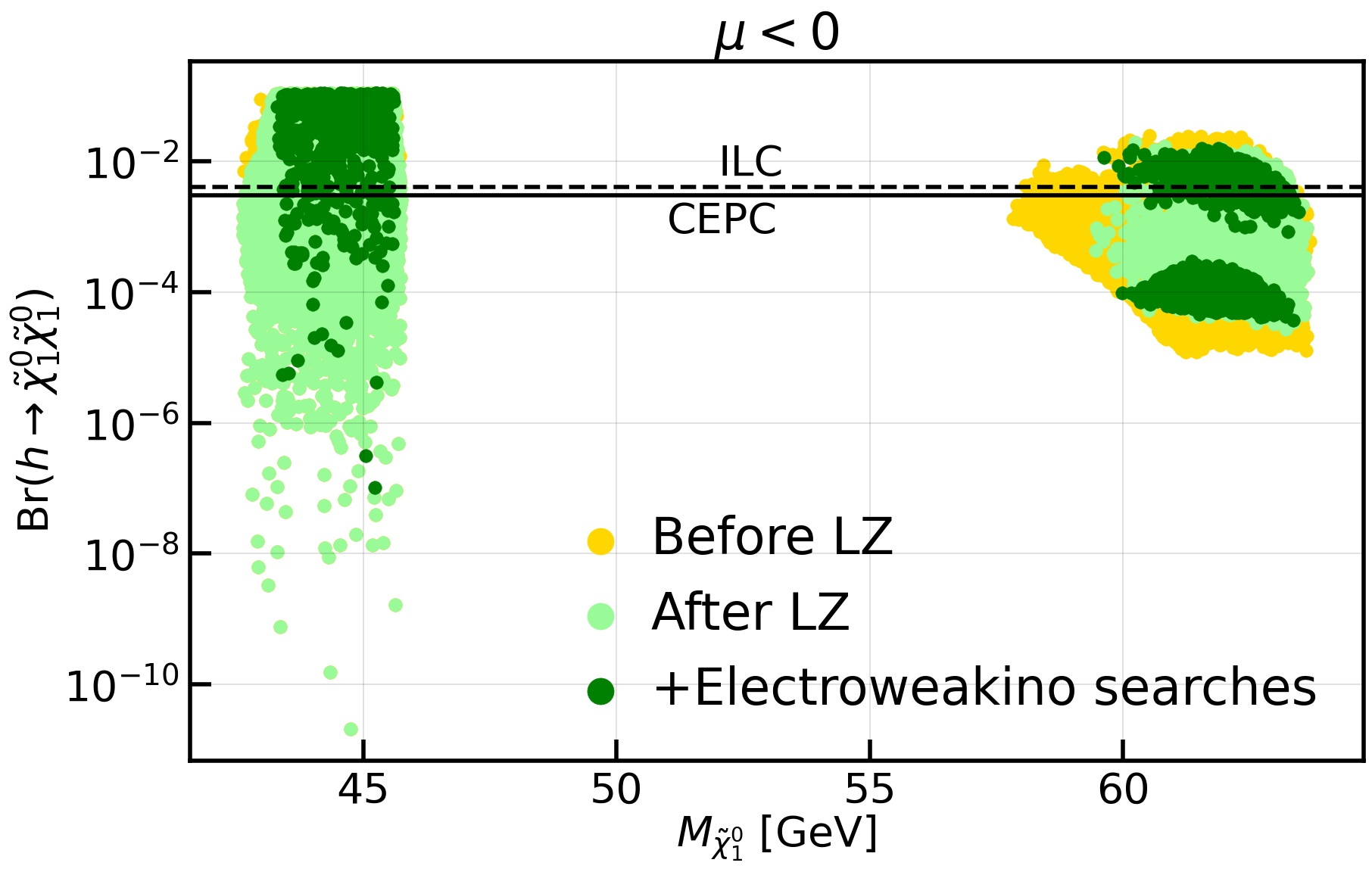}
    \caption{Allowed region of parameter space in the $M_{\tilde{\chi}_1^0}$-Br($h\rightarrow\tilde{\chi}_1^0\tilde{\chi}_1^0$) plane for $\mu>0$ ({\it left}) and $\mu<0$ ({\it right}) with all the constraints 'Before LZ' ({\it yellow}), 'After LZ' ({\it light green}), and from electroweakino searches ({\it dark green}). {\it Dashed} and {\it solid black} lines show the projected sensitivity for Br($h\rightarrow\tilde{\chi}_1^0\tilde{\chi}_1^0$) from the ILC and CEPC experiments, respectively.}
    \label{fig:h_inv}
\end{figure}

\section{Collider limit on right-handed staus}
\label{app:stau_smod}

\begin{figure*}[hbt!]
    \centering
    \includegraphics[width=0.6\textwidth]{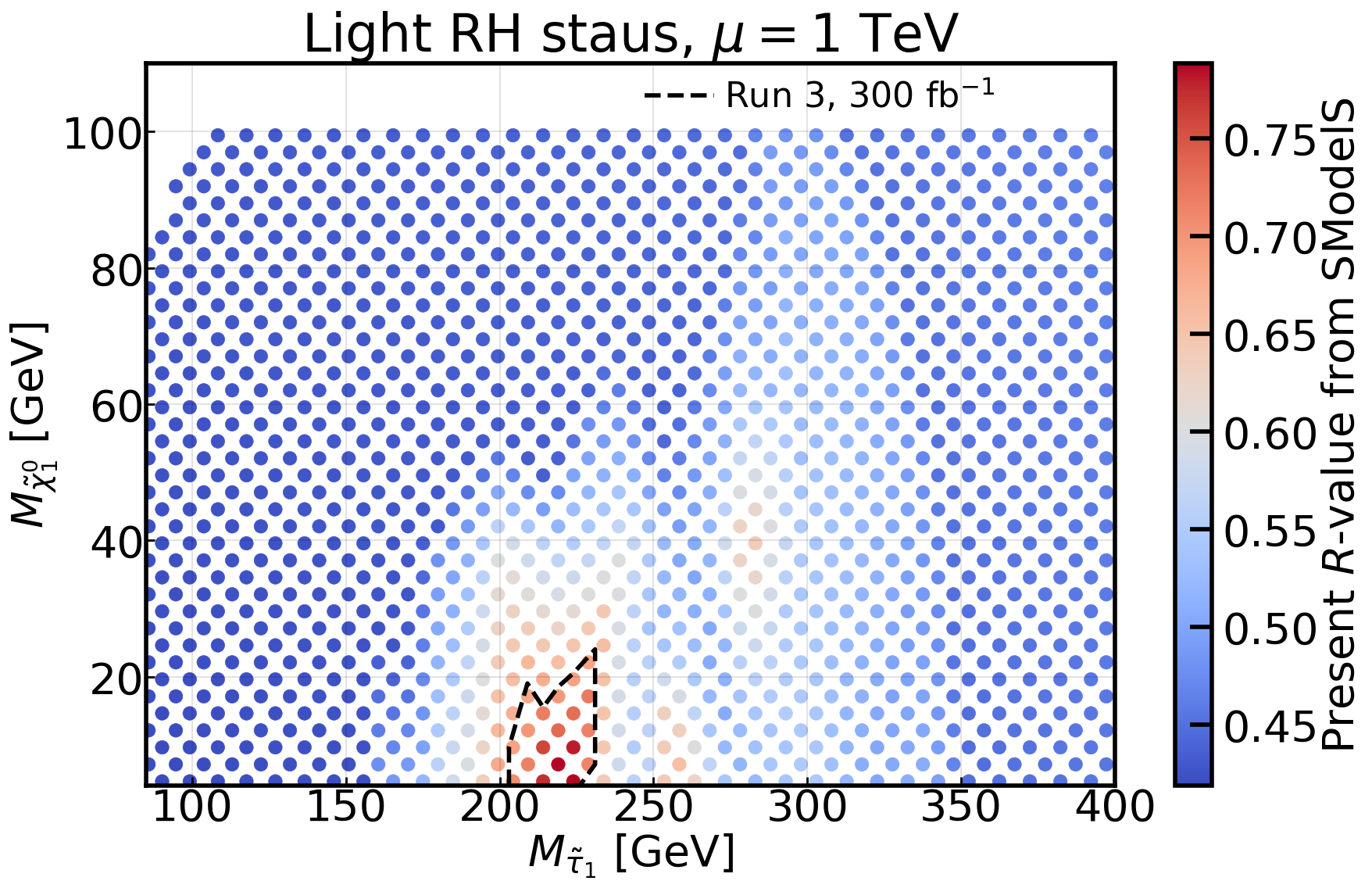}
    \caption{\texttt{SmodelS} $R$-values of present analyses searching for stau leptons at the LHC in the $M_{\tilde{\stau}_1}$-$M_{\tilde{\chi}_1^0}$ plane, where the lightest stau is RH and $M_{\tilde{\chi}_1^0}$ is Bino-like for $\mu=1$\,TeV.}
    \label{fig:stau_smod}
\end{figure*}

Fig.\,\ref{fig:stau_smod} shows the $R$-values of present analyses searching for stau leptons at the LHC, as recasted by the \texttt{SModelS} package in the plane of mass of the lightest stau which is RH, $M_{\tilde{\stau}_1}$ and the mass of the lightest neutralino which is Bino-like, $M_{\tilde{\chi}_1^0}$ for a Higgsino mass parameter of 1\,TeV. We observe that the $R$-values are all less than unity, implying that the current analyses are not sensitive to this region of parameter space. As a result, light staus in the mass range of 90-400 are still allowed, the lower limit coming from the LEP experiment\,\cite{LEP:stau}. We translate these $R$-values with the square root of luminosity for the future runs of LHC. We find that Run-3 with 300\,fb$^{-1}$ integrated luminosity will be sensitive to a small region of parameter space for $M_{\tilde{\chi}_1^0}\lesssim 20$\,GeV and $M_{\tilde{\stau}_1}$ between 200-250\,GeV, and the HL-LHC run with 3\,ab$^{-1}$ of luminosity will be able to probe this whole region of parameter space.

\section{Combination of analyses in \texttt{SModelS}}
\label{app:smodels}

A new important feature in \texttt{SModelS-2.3}\,\cite{MahdiAltakach:2023bdn} is the possibility to combine multiple search analyses. The combination of analyses assumes that these analyses are uncorrelated and therefore, has to be used carefully. One of the difficulties in using this option is that  the names of the analyses that have to be combined should be given as an  input. For a particular point in parameter space, it is not always clear beforehand which analyses are important and whether they can be combined consistently. Another technical challenge is that signal regions within an analysis cannot be combined when one chooses to combine analyses within the current version of SModelS. Nevertheless, it is important to check the effect of combining the most sensitive analyses, at least for the benchmark points, to ascertain whether or not  they are excluded.

Table\,\ref{tab:smodels_combo} shows the comparison of the $R$-values when combining the most sensitive analyses for the  benchmarks points with the ones obtained without combining the analyses. For each case, we also show the names of the most sensitive analyses. 

For BP1 and BP4, which correspond to the heavy higgsinos in the $\mu > 0$ and $\mu < 0$ scenarios, we find that the $R$-value decreases when we combine the most sensitive analyses. This happens in all cases where  the combination of signal regions is more important than combining analyses since both combination cannot be done simultaneously. For BP1 and BP4, the \texttt{ATLAS-SUSY-2018-41} analysis is sensitive to different topologies and gains significantly in sensitivity when combining signal regions. 

For BP2 and BP3, there are very slight changes in the $R$-values when we combine the analyses. For BP5, BP7 and BP8, the $R$-values are very small to begin with, on combining the analyses the $R$-value increases slightly for BP5, but degrades for BP7 and BP8. Lastly, for BP6, which is the stau benchmark in the $\mu<0$ $h$-funnel, we find that combining the analyses increases the $R$-value significantly $-$ from 0.599 to 0.887. We conclude that none of our benchmark points are excluded after we apply the combination of analyses.


\begin{table}[hbt!]
    \centering
    \resizebox{0.7\textwidth}{!}{
    \begin{tabular}{|c|c|c|c|c|}
    \hline
    \multirow{2}{*}{Benchmarks} & \multicolumn{2}{c|}{Without combination of analyses} & \multicolumn{2}{c|}{With combination of analyses} \\
    \cline{2-5}
    & Analysis & $R$-value & Analyses & $R$-value \\
    \hline\hline
    \multirow{5}{*}{BP1} & \multirow{5}{*}{ATLAS-SUSY-2018-41} & \multirow{5}{*}{0.815} & ATLAS-SUSY-2018-05-ewk & \multirow{5}{*}{0.615} \\
        &                    &      & ATLAS-SUSY-2018-41 &      \\
        &                    &      & ATLAS-SUSY-2019-08 &      \\ 
        &                    &      & CMS-SUS-16-039-agg &      \\ 
        &                    &      & CMS-SUS-21-002 &      \\ 
    \hline    
    \multirow{3}{*}{BP2} & \multirow{3}{*}{ATLAS-SUSY-2013-11} & \multirow{3}{*}{0.141} & ATLAS-SUSY-2013-11 & \multirow{3}{*}{0.144}\\
        &                    &       & ATLAS-SUSY-2019-02 &         \\
        &                    &       & CMS-SU-13-012 &         \\
    \hline    
    \multirow{4}{*}{BP3} & \multirow{4}{*}{ATLAS-SUSY-2019-09} & \multirow{4}{*}{0.443} & ATLAS-SUSY-2013-11 & \multirow{4}{*}{0.402} \\
        &                    &       & ATLAS-SUSY-2019-02 &         \\
        &                    &       & ATLAS-SUSY-2019-09 &         \\
        &                    &       & CMS-SUS-13-012 &         \\
    \hline    
    \multirow{5}{*}{BP4} & \multirow{5}{*}{ATLAS-SUSY-2018-41} & \multirow{5}{*}{0.805} & ATLAS-SUSY-2018-05-ewk & \multirow{5}{*}{0.614}\\
        &                    &      & ATLAS-SUSY-2018-41 &      \\
        &                    &      & ATLAS-SUSY-2019-08 &      \\ 
        &                    &      & CMS-SUS-16-039-agg &      \\ 
        &                    &      & CMS-SUS-21-002 &      \\ 
    \hline
    \multirow{6}{*}{BP5} & \multirow{6}{*}{ATLAS-SUSY-2018-05-ewk} & \multirow{6}{*}{0.059} & ATLAS-SUSY-2016-24 & \multirow{6}{*}{0.078} \\
        &                        &       & ATLAS-SUSY-2017-03 &    \\
        &                        &       & ATLAS-SUSY-2018-05-ewk &    \\
        &                        &       & ATLAS-SUSY-2018-41 &    \\
        &                        &       & ATLAS-SUSY-2019-09 &    \\
        &                        &       & CMS-SUS-21-002 &    \\
    \hline
    \multirow{6}{*}{BP6} & \multirow{6}{*}{ATLAS-SUSY-2018-05-ewk} & \multirow{6}{*}{0.599} & ATLAS-SUSY-2016-24 & \multirow{6}{*}{0.887} \\
        &                        &       & ATLAS-SUSY-2017-03 &    \\
        &                        &       & ATLAS-SUSY-2018-05-ewk &    \\
        &                        &       & ATLAS-SUSY-2018-41 &    \\
        &                        &       & ATLAS-SUSY-2019-09 &    \\
        &                        &       & CMS-SUS-21-002 &    \\
    \hline
    \multirow{3}{*}{BP7} & \multirow{3}{*}{ATLAS-SUSY-2013-12} & \multirow{3}{*}{0.061} & ATLAS-SUSY-2013-11 & \multirow{3}{*}{0.023} \\
        &                    &       & ATLAS-SUSY-2019-02 &     \\
        &                    &       & CMS-SUS-13-012 &     \\
    \hline    
    \multirow{3}{*}{BP8} & \multirow{3}{*}{ATLAS-SUSY-2019-09} & \multirow{3}{*}{1.16$\times$10$^{-3}$} & ATLAS-SUSY-2019-02 & \multirow{3}{*}{9.37$\times$10$^{-4}$} \\
        &                    &       & ATLAS-SUSY-2019-09 &      \\
        &                    &       & CMS-SUS-13-012 &      \\
    \hline
    \end{tabular}}
    \caption{Comparison of the $R$-values from the latest version of \texttt{SModelS} (database 2.3.0) on our eight benchmark points without and with the combination of the most sensitive analyses for each point. When not combining analyses, signal regions within an analysis could be combined, which cannot be done when combining analyses.}
    \label{tab:smodels_combo}
\end{table}




\clearpage



\begin{thebibliography}{100}

\bibitem{ATLAS:2024lyh}
{\bfseries ATLAS} Collaboration, G.~Aad {\em et~al.}, ``{Interpretations of the
  ATLAS measurements of Higgs boson production and decay rates and differential
  cross-sections in $pp$ collisions at $\sqrt{s}=13$ TeV},''
  \href{http://arxiv.org/abs/2402.05742}{{\ttfamily arXiv:2402.05742
  [hep-ex]}}.

\bibitem{CMS:2023sdw}
{\bfseries CMS} Collaboration, A.~Tumasyan {\em et~al.}, ``{A search for decays
  of the Higgs boson to invisible particles in events with a top-antitop quark
  pair or a vector boson in proton-proton collisions at $\sqrt{s} = 13\,\text
  {Te}\hspace{-.08em}\text {V} $},''
  \href{http://dx.doi.org/10.1140/epjc/s10052-023-11952-7}{{\em Eur. Phys. J.
  C} {\bfseries 83} no.~10, (2023) 933},
  \href{http://arxiv.org/abs/2303.01214}{{\ttfamily arXiv:2303.01214
  [hep-ex]}}.

\bibitem{Weinberg:1975gm}
S.~Weinberg, ``{Implications of Dynamical Symmetry Breaking},''
  \href{http://dx.doi.org/10.1103/PhysRevD.19.1277}{{\em Phys. Rev. D}
  {\bfseries 13} (1976) 974--996}. [Addendum: Phys.Rev.D 19, 1277--1280
  (1979)].

\bibitem{Gildener:1976ih}
E.~Gildener and S.~Weinberg, ``{Symmetry Breaking and Scalar Bosons},''
  \href{http://dx.doi.org/10.1103/PhysRevD.13.3333}{{\em Phys. Rev. D}
  {\bfseries 13} (1976) 3333}.

\bibitem{PhysRevD.20.2619}
L.~Susskind, ``Dynamics of spontaneous symmetry breaking in the weinberg-salam
  theory,'' \href{http://dx.doi.org/10.1103/PhysRevD.20.2619}{{\em Phys. Rev.
  D} {\bfseries 20} (Nov, 1979) 2619--2625}.
  \url{https://link.aps.org/doi/10.1103/PhysRevD.20.2619}.

\bibitem{L3:1999onh}
{\bfseries L3} Collaboration, M.~Acciarri {\em et~al.}, ``{Search for charginos
  and neutralinos in $e^{+} e^{-}$ collisions at $\sqrt{S}$ = 189-GeV},''
  \href{http://dx.doi.org/10.1016/S0370-2693(99)01388-X}{{\em Phys. Lett. B}
  {\bfseries 472} (2000) 420--433},
  \href{http://arxiv.org/abs/hep-ex/9910007}{{\ttfamily arXiv:hep-ex/9910007}}.

\bibitem{OPAL:2003wxm}
{\bfseries OPAL} Collaboration, G.~Abbiendi {\em et~al.}, ``{Search for
  chargino and neutralino production at s**(1/2) = 192-GeV to 209 GeV at
  LEP},'' \href{http://dx.doi.org/10.1140/epjc/s2004-01758-8}{{\em Eur. Phys.
  J. C} {\bfseries 35} (2004) 1--20},
  \href{http://arxiv.org/abs/hep-ex/0401026}{{\ttfamily arXiv:hep-ex/0401026}}.

\bibitem{Planck:2018vyg}
{\bfseries Planck} Collaboration, N.~Aghanim {\em et~al.}, ``{Planck 2018
  results. VI. Cosmological parameters},''
  \href{http://dx.doi.org/10.1051/0004-6361/201833910}{{\em Astron. Astrophys.}
  {\bfseries 641} (2020) A6}, \href{http://arxiv.org/abs/1807.06209}{{\ttfamily
  arXiv:1807.06209 [astro-ph.CO]}}. [Erratum: Astron.Astrophys. 652, C4
  (2021)].

\bibitem{PhysRevD.37.719}
K.~Griest and H.~E. Haber, ``Invisible decays of higgs bosons in supersymmetric
  models,'' \href{http://dx.doi.org/10.1103/PhysRevD.37.719}{{\em Phys. Rev. D}
  {\bfseries 37} (Feb, 1988) 719--728}.
  \url{https://link.aps.org/doi/10.1103/PhysRevD.37.719}.

\bibitem{Djouadi:1996mj}
A.~Djouadi, P.~Janot, J.~Kalinowski, and P.~M. Zerwas, ``{SUSY decays of Higgs
  particles},'' \href{http://dx.doi.org/10.1016/0370-2693(96)00414-5}{{\em
  Phys. Lett. B} {\bfseries 376} (1996) 220--226},
  \href{http://arxiv.org/abs/hep-ph/9603368}{{\ttfamily arXiv:hep-ph/9603368}}.

\bibitem{Belanger:2000tg}
G.~Belanger, F.~Boudjema, F.~Donato, R.~Godbole, and S.~Rosier-Lees, ``{SUSY
  Higgs at the LHC: Effects of light charginos and neutralinos},''
  \href{http://dx.doi.org/10.1016/S0550-3213(00)00243-1}{{\em Nucl. Phys. B}
  {\bfseries 581} (2000) 3--33},
  \href{http://arxiv.org/abs/hep-ph/0002039}{{\ttfamily arXiv:hep-ph/0002039}}.

\bibitem{Belanger:2001am}
G.~Belanger, F.~Boudjema, A.~Cottrant, R.~M. Godbole, and A.~Semenov, ``{The
  MSSM invisible Higgs in the light of dark matter and g-2},''
  \href{http://dx.doi.org/10.1016/S0370-2693(01)00976-5}{{\em Phys. Lett. B}
  {\bfseries 519} (2001) 93--102},
  \href{http://arxiv.org/abs/hep-ph/0106275}{{\ttfamily arXiv:hep-ph/0106275}}.

\bibitem{Hooper:2002nq}
D.~Hooper and T.~Plehn, ``{Supersymmetric dark matter: How light can the LSP
  be?},'' \href{http://dx.doi.org/10.1016/S0370-2693(03)00548-3}{{\em Phys.
  Lett. B} {\bfseries 562} (2003) 18--27},
  \href{http://arxiv.org/abs/hep-ph/0212226}{{\ttfamily arXiv:hep-ph/0212226}}.

\bibitem{Hisano:2002fk}
J.~Hisano, S.~Matsumoto, and M.~M. Nojiri, ``{Unitarity and higher order
  corrections in neutralino dark matter annihilation into two photons},''
  \href{http://dx.doi.org/10.1103/PhysRevD.67.075014}{{\em Phys. Rev. D}
  {\bfseries 67} (2003) 075014},
  \href{http://arxiv.org/abs/hep-ph/0212022}{{\ttfamily arXiv:hep-ph/0212022}}.

\bibitem{Belanger:2003wb}
G.~Belanger, F.~Boudjema, A.~Cottrant, A.~Pukhov, and S.~Rosier-Lees, ``{Lower
  limit on the neutralino mass in the general MSSM},''
  \href{http://dx.doi.org/10.1088/1126-6708/2004/03/012}{{\em JHEP} {\bfseries
  03} (2004) 012}, \href{http://arxiv.org/abs/hep-ph/0310037}{{\ttfamily
  arXiv:hep-ph/0310037}}.

\bibitem{Dreiner:2009ic}
H.~K. Dreiner, S.~Heinemeyer, O.~Kittel, U.~Langenfeld, A.~M. Weber, and
  G.~Weiglein, ``{Mass Bounds on a Very Light Neutralino},''
  \href{http://dx.doi.org/10.1140/epjc/s10052-009-1042-y}{{\em Eur. Phys. J. C}
  {\bfseries 62} (2009) 547--572},
  \href{http://arxiv.org/abs/0901.3485}{{\ttfamily arXiv:0901.3485 [hep-ph]}}.

\bibitem{AlbornozVasquez:2010nkq}
D.~Albornoz~Vasquez, G.~Belanger, C.~Boehm, A.~Pukhov, and J.~Silk, ``{Can
  neutralinos in the MSSM and NMSSM scenarios still be light?},''
  \href{http://dx.doi.org/10.1103/PhysRevD.82.115027}{{\em Phys. Rev. D}
  {\bfseries 82} (2010) 115027},
  \href{http://arxiv.org/abs/1009.4380}{{\ttfamily arXiv:1009.4380 [hep-ph]}}.

\bibitem{Calibbi:2011ug}
L.~Calibbi, T.~Ota, and Y.~Takanishi, ``{Light Neutralino in the MSSM: a
  playground for dark matter, flavor physics and collider experiments},''
  \href{http://dx.doi.org/10.1007/JHEP07(2011)013}{{\em JHEP} {\bfseries 07}
  (2011) 013}, \href{http://arxiv.org/abs/1104.1134}{{\ttfamily arXiv:1104.1134
  [hep-ph]}}.

\bibitem{Choudhury:2012tc}
A.~Choudhury and A.~Datta, ``{Many faces of low mass neutralino dark matter in
  the unconstrained MSSM, LHC data and new signals},''
  \href{http://dx.doi.org/10.1007/JHEP06(2012)006}{{\em JHEP} {\bfseries 06}
  (2012) 006}, \href{http://arxiv.org/abs/1203.4106}{{\ttfamily arXiv:1203.4106
  [hep-ph]}}.

\bibitem{Arbey:2012na}
A.~Arbey, M.~Battaglia, and F.~Mahmoudi, ``{Light Neutralino Dark Matter in the
  pMSSM: Implications of LEP, LHC and Dark Matter Searches on SUSY Particle
  Spectra},'' \href{http://dx.doi.org/10.1140/epjc/s10052-012-2169-9}{{\em Eur.
  Phys. J. C} {\bfseries 72} (2012) 2169},
  \href{http://arxiv.org/abs/1205.2557}{{\ttfamily arXiv:1205.2557 [hep-ph]}}.

\bibitem{Dreiner:2012ex}
H.~K. Dreiner, J.~S. Kim, and O.~Lebedev, ``{First LHC Constraints on
  Neutralinos},'' \href{http://dx.doi.org/10.1016/j.physletb.2012.07.058}{{\em
  Phys. Lett. B} {\bfseries 715} (2012) 199--202},
  \href{http://arxiv.org/abs/1206.3096}{{\ttfamily arXiv:1206.3096 [hep-ph]}}.

\bibitem{Boehm:2013qva}
C.~Boehm, P.~S.~B. Dev, A.~Mazumdar, and E.~Pukartas, ``{Naturalness of Light
  Neutralino Dark Matter in pMSSM after LHC, XENON100 and Planck Data},''
  \href{http://dx.doi.org/10.1007/JHEP06(2013)113}{{\em JHEP} {\bfseries 06}
  (2013) 113}, \href{http://arxiv.org/abs/1303.5386}{{\ttfamily arXiv:1303.5386
  [hep-ph]}}.

\bibitem{Choudhury:2013jpa}
A.~Choudhury and A.~Datta, ``{Neutralino dark matter confronted by the LHC
  constraints on Electroweak SUSY signals},''
  \href{http://dx.doi.org/10.1007/JHEP09(2013)119}{{\em JHEP} {\bfseries 09}
  (2013) 119}, \href{http://arxiv.org/abs/1305.0928}{{\ttfamily arXiv:1305.0928
  [hep-ph]}}.

\bibitem{Ananthanarayan:2013fga}
B.~Ananthanarayan, J.~Lahiri, P.~N. Pandita, and M.~Patra, ``{Invisible decays
  of the lightest Higgs boson in supersymmetric models},''
  \href{http://dx.doi.org/10.1103/PhysRevD.87.115021}{{\em Phys. Rev. D}
  {\bfseries 87} no.~11, (2013) 115021},
  \href{http://arxiv.org/abs/1306.1291}{{\ttfamily arXiv:1306.1291 [hep-ph]}}.

\bibitem{Calibbi:2013poa}
L.~Calibbi, J.~M. Lindert, T.~Ota, and Y.~Takanishi, ``{Cornering light
  Neutralino Dark Matter at the LHC},''
  \href{http://dx.doi.org/10.1007/JHEP10(2013)132}{{\em JHEP} {\bfseries 10}
  (2013) 132}, \href{http://arxiv.org/abs/1307.4119}{{\ttfamily arXiv:1307.4119
  [hep-ph]}}.

\bibitem{Belanger:2013pna}
G.~B\'elanger, G.~Drieu La~Rochelle, B.~Dumont, R.~M. Godbole, S.~Kraml, and
  S.~Kulkarni, ``{LHC constraints on light neutralino dark matter in the
  MSSM},'' \href{http://dx.doi.org/10.1016/j.physletb.2013.09.059}{{\em Phys.
  Lett. B} {\bfseries 726} (2013) 773--780},
  \href{http://arxiv.org/abs/1308.3735}{{\ttfamily arXiv:1308.3735 [hep-ph]}}.

\bibitem{Chakraborti:2014gea}
M.~Chakraborti, U.~Chattopadhyay, A.~Choudhury, A.~Datta, and S.~Poddar, ``{The
  Electroweak Sector of the pMSSM in the Light of LHC - 8 TeV and Other
  Data},'' \href{http://dx.doi.org/10.1007/JHEP07(2014)019}{{\em JHEP}
  {\bfseries 07} (2014) 019}, \href{http://arxiv.org/abs/1404.4841}{{\ttfamily
  arXiv:1404.4841 [hep-ph]}}.

\bibitem{Han:2014nba}
T.~Han, Z.~Liu, and S.~Su, ``{Light Neutralino Dark Matter: Direct/Indirect
  Detection and Collider Searches},''
  \href{http://dx.doi.org/10.1007/JHEP08(2014)093}{{\em JHEP} {\bfseries 08}
  (2014) 093}, \href{http://arxiv.org/abs/1406.1181}{{\ttfamily arXiv:1406.1181
  [hep-ph]}}.

\bibitem{Cahill-Rowley:2014boa}
M.~Cahill-Rowley, R.~Cotta, A.~Drlica-Wagner, S.~Funk, J.~Hewett, A.~Ismail,
  T.~Rizzo, and M.~Wood, ``{Complementarity of dark matter searches in the
  phenomenological MSSM},''
  \href{http://dx.doi.org/10.1103/PhysRevD.91.055011}{{\em Phys. Rev. D}
  {\bfseries 91} no.~5, (2015) 055011},
  \href{http://arxiv.org/abs/1405.6716}{{\ttfamily arXiv:1405.6716 [hep-ph]}}.

\bibitem{Bhattacherjee:2015sga}
B.~Bhattacherjee, A.~Chakraborty, and A.~Choudhury, ``{Status of the MSSM Higgs
  sector using global analysis and direct search bounds, and future prospects
  at the High Luminosity LHC},''
  \href{http://dx.doi.org/10.1103/PhysRevD.92.093007}{{\em Phys. Rev. D}
  {\bfseries 92} no.~9, (2015) 093007},
  \href{http://arxiv.org/abs/1504.04308}{{\ttfamily arXiv:1504.04308
  [hep-ph]}}.

\bibitem{Belanger:2015vwa}
G.~Belanger, D.~Ghosh, R.~Godbole, and S.~Kulkarni, ``{Light stop in the MSSM
  after LHC Run 1},'' \href{http://dx.doi.org/10.1007/JHEP09(2015)214}{{\em
  JHEP} {\bfseries 09} (2015) 214},
  \href{http://arxiv.org/abs/1506.00665}{{\ttfamily arXiv:1506.00665
  [hep-ph]}}.

\bibitem{Chakraborti:2015mra}
M.~Chakraborti, U.~Chattopadhyay, A.~Choudhury, A.~Datta, and S.~Poddar,
  ``{Reduced LHC constraints for higgsino-like heavier electroweakinos},''
  \href{http://dx.doi.org/10.1007/JHEP11(2015)050}{{\em JHEP} {\bfseries 11}
  (2015) 050}, \href{http://arxiv.org/abs/1507.01395}{{\ttfamily
  arXiv:1507.01395 [hep-ph]}}.

\bibitem{Hamaguchi:2015rxa}
K.~Hamaguchi and K.~Ishikawa, ``{Prospects for Higgs- and Z-resonant Neutralino
  Dark Matter},'' \href{http://dx.doi.org/10.1103/PhysRevD.93.055009}{{\em
  Phys. Rev. D} {\bfseries 93} no.~5, (2016) 055009},
  \href{http://arxiv.org/abs/1510.05378}{{\ttfamily arXiv:1510.05378
  [hep-ph]}}.

\bibitem{Cao:2015efs}
J.~Cao, Y.~He, L.~Shang, W.~Su, and Y.~Zhang, ``{Testing the light dark matter
  scenario of the MSSM at the LHC},''
  \href{http://dx.doi.org/10.1007/JHEP03(2016)207}{{\em JHEP} {\bfseries 03}
  (2016) 207}, \href{http://arxiv.org/abs/1511.05386}{{\ttfamily
  arXiv:1511.05386 [hep-ph]}}.

\bibitem{ATLAS:2015wrn}
{\bfseries ATLAS} Collaboration, G.~Aad {\em et~al.}, ``{Summary of the ATLAS
  experiment\textquoteright{}s sensitivity to supersymmetry after LHC Run 1
  \textemdash{} interpreted in the phenomenological MSSM},''
  \href{http://dx.doi.org/10.1007/JHEP10(2015)134}{{\em JHEP} {\bfseries 10}
  (2015) 134}, \href{http://arxiv.org/abs/1508.06608}{{\ttfamily
  arXiv:1508.06608 [hep-ex]}}.

\bibitem{Bagnaschi:2015eha}
E.~A. Bagnaschi {\em et~al.}, ``{Supersymmetric Dark Matter after LHC Run 1},''
  \href{http://dx.doi.org/10.1140/epjc/s10052-015-3718-9}{{\em Eur. Phys. J. C}
  {\bfseries 75} (2015) 500}, \href{http://arxiv.org/abs/1508.01173}{{\ttfamily
  arXiv:1508.01173 [hep-ph]}}.

\bibitem{Choudhury:2016lku}
A.~Choudhury and S.~Mondal, ``{Revisiting the Exclusion Limits from Direct
  Chargino-Neutralino Production at the LHC},''
  \href{http://dx.doi.org/10.1103/PhysRevD.94.055024}{{\em Phys. Rev. D}
  {\bfseries 94} no.~5, (2016) 055024},
  \href{http://arxiv.org/abs/1603.05502}{{\ttfamily arXiv:1603.05502
  [hep-ph]}}.

\bibitem{Barman:2016jov}
R.~K. Barman, B.~Bhattacherjee, A.~Choudhury, D.~Chowdhury, J.~Lahiri, and
  S.~Ray, ``{Current status of MSSM Higgs sector with LHC 13 TeV data},''
  \href{http://dx.doi.org/10.1140/epjp/i2019-12566-5}{{\em Eur. Phys. J. Plus}
  {\bfseries 134} no.~4, (2019) 150},
  \href{http://arxiv.org/abs/1608.02573}{{\ttfamily arXiv:1608.02573
  [hep-ph]}}.

\bibitem{Barman:2017swy}
R.~K. Barman, G.~Belanger, B.~Bhattacherjee, R.~Godbole, G.~Mendiratta, and
  D.~Sengupta, ``{Invisible decay of the Higgs boson in the context of a
  thermal and nonthermal relic in MSSM},''
  \href{http://dx.doi.org/10.1103/PhysRevD.95.095018}{{\em Phys. Rev.}
  {\bfseries D95} no.~9, (2017) 095018},
\href{http://arxiv.org/abs/1703.03838}{{\ttfamily arXiv:1703.03838 [hep-ph]}}.

\bibitem{Abdughani:2017dqs}
M.~Abdughani, L.~Wu, and J.~M. Yang, ``{Status and prospects of light
  bino\textendash{}higgsino dark matter in natural SUSY},''
  \href{http://dx.doi.org/10.1140/epjc/s10052-017-5485-2}{{\em Eur. Phys. J. C}
  {\bfseries 78} no.~1, (2018) 4},
  \href{http://arxiv.org/abs/1705.09164}{{\ttfamily arXiv:1705.09164
  [hep-ph]}}.

\bibitem{Roszkowski:2017nbc}
L.~Roszkowski, E.~M. Sessolo, and S.~Trojanowski, ``{WIMP dark matter
  candidates and searches\textemdash{}current status and future prospects},''
  \href{http://dx.doi.org/10.1088/1361-6633/aab913}{{\em Rept. Prog. Phys.}
  {\bfseries 81} no.~6, (2018) 066201},
  \href{http://arxiv.org/abs/1707.06277}{{\ttfamily arXiv:1707.06277
  [hep-ph]}}.

\bibitem{Pozzo:2018anw}
G.~Pozzo and Y.~Zhang, ``{Constraining resonant dark matter with combined LHC
  electroweakino searches},''
  \href{http://dx.doi.org/10.1016/j.physletb.2018.12.062}{{\em Phys. Lett. B}
  {\bfseries 789} (2019) 582--591},
  \href{http://arxiv.org/abs/1807.01476}{{\ttfamily arXiv:1807.01476
  [hep-ph]}}.

\bibitem{GAMBIT:2018gjo}
{\bfseries GAMBIT} Collaboration, P.~Athron {\em et~al.}, ``{Combined collider
  constraints on neutralinos and charginos},''
  \href{http://dx.doi.org/10.1140/epjc/s10052-019-6837-x}{{\em Eur. Phys. J. C}
  {\bfseries 79} no.~5, (2019) 395},
  \href{http://arxiv.org/abs/1809.02097}{{\ttfamily arXiv:1809.02097
  [hep-ph]}}.

\bibitem{Wang:2020dtb}
K.~Wang and J.~Zhu, ``{Funnel annihilations of light dark matter and the
  invisible decay of the Higgs boson},''
  \href{http://dx.doi.org/10.1103/PhysRevD.101.095028}{{\em Phys. Rev. D}
  {\bfseries 101} no.~9, (2020) 095028},
  \href{http://arxiv.org/abs/2003.01662}{{\ttfamily arXiv:2003.01662
  [hep-ph]}}.

\bibitem{KumarBarman:2020ylm}
R.~Kumar~Barman, G.~Belanger, and R.~M. Godbole, ``{Status of low mass LSP in
  SUSY},'' \href{http://dx.doi.org/10.1140/epjst/e2020-000198-1}{{\em Eur.
  Phys. J. ST} {\bfseries 229} no.~21, (2020) 3159--3185},
  \href{http://arxiv.org/abs/2010.11674}{{\ttfamily arXiv:2010.11674
  [hep-ph]}}.

\bibitem{VanBeekveld:2021tgn}
M.~Van~Beekveld, W.~Beenakker, M.~Schutten, and J.~De~Wit, ``{Dark matter,
  fine-tuning and $(g-2)_{\mu}$ in the pMSSM},''
  \href{http://dx.doi.org/10.21468/SciPostPhys.11.3.049}{{\em SciPost Phys.}
  {\bfseries 11} no.~3, (2021) 049},
  \href{http://arxiv.org/abs/2104.03245}{{\ttfamily arXiv:2104.03245
  [hep-ph]}}.

\bibitem{Barman:2022jdg}
R.~K. Barman, G.~B\'elanger, B.~Bhattacherjee, R.~M. Godbole, and R.~Sengupta,
  ``{Is Light Neutralino Thermal Dark Matter in the Phenomenological Minimal
  Supersymmetric Standard Model Ruled Out?},''
  \href{http://dx.doi.org/10.1103/PhysRevLett.131.011802}{{\em Phys. Rev.
  Lett.} {\bfseries 131} no.~1, (2023) 011802},
  \href{http://arxiv.org/abs/2207.06238}{{\ttfamily arXiv:2207.06238
  [hep-ph]}}.

\bibitem{ATLAS:2020zms}
{\bfseries ATLAS} Collaboration, G.~Aad {\em et~al.}, ``{Search for heavy Higgs
  bosons decaying into two tau leptons with the ATLAS detector using $pp$
  collisions at $\sqrt{s}=13$ TeV},''
  \href{http://dx.doi.org/10.1103/PhysRevLett.125.051801}{{\em Phys. Rev.
  Lett.} {\bfseries 125} no.~5, (2020) 051801},
  \href{http://arxiv.org/abs/2002.12223}{{\ttfamily arXiv:2002.12223
  [hep-ex]}}.

\bibitem{CMS:2020bfa}
{\bfseries CMS} Collaboration, A.~M. Sirunyan {\em et~al.}, ``{Search for
  supersymmetry in final states with two oppositely charged same-flavor leptons
  and missing transverse momentum in proton-proton collisions at $\sqrt{s} =$
  13 TeV},'' \href{http://dx.doi.org/10.1007/JHEP04(2021)123}{{\em JHEP}
  {\bfseries 04} (2021) 123}, \href{http://arxiv.org/abs/2012.08600}{{\ttfamily
  arXiv:2012.08600 [hep-ex]}}.

\bibitem{ATLAS:2021moa}
{\bfseries ATLAS} Collaboration, G.~Aad {\em et~al.}, ``{Search for
  chargino\textendash{}neutralino pair production in final states with three
  leptons and missing transverse momentum in $\sqrt{s} = 13$~TeV pp collisions
  with the ATLAS detector},''
  \href{http://dx.doi.org/10.1140/epjc/s10052-021-09749-7}{{\em Eur. Phys. J.
  C} {\bfseries 81} no.~12, (2021) 1118},
  \href{http://arxiv.org/abs/2106.01676}{{\ttfamily arXiv:2106.01676
  [hep-ex]}}.

\bibitem{ATLAS:2021yqv}
{\bfseries ATLAS} Collaboration, G.~Aad {\em et~al.}, ``{Search for charginos
  and neutralinos in final states with two boosted hadronically decaying bosons
  and missing transverse momentum in $pp$ collisions at $\sqrt {s}$ = 13\,\,TeV
  with the ATLAS detector},''
  \href{http://dx.doi.org/10.1103/PhysRevD.104.112010}{{\em Phys. Rev. D}
  {\bfseries 104} no.~11, (2021) 112010},
  \href{http://arxiv.org/abs/2108.07586}{{\ttfamily arXiv:2108.07586
  [hep-ex]}}.

\bibitem{CMS:2022sfi}
{\bfseries CMS} Collaboration, ``{Search for electroweak production of
  charginos and neutralinos at $\sqrt{s}$ =13 TeV in final states containing
  hadronic decays of WW, WZ, or WH and missing transverse momentum},''
  \href{http://arxiv.org/abs/2205.09597}{{\ttfamily arXiv:2205.09597
  [hep-ex]}}.

\bibitem{ATLAS:2022yvh}
{\bfseries ATLAS} Collaboration, G.~Aad {\em et~al.}, ``{Search for invisible
  Higgs-boson decays in events with vector-boson fusion signatures using 139
  $\text{fb}^{-1}$ of proton-proton data recorded by the ATLAS experiment},''
  \href{http://arxiv.org/abs/2202.07953}{{\ttfamily arXiv:2202.07953
  [hep-ex]}}.

\bibitem{XENON:2018voc}
{\bfseries XENON} Collaboration, E.~Aprile {\em et~al.}, ``{Dark Matter Search
  Results from a One Ton-Year Exposure of XENON1T},''
  \href{http://dx.doi.org/10.1103/PhysRevLett.121.111302}{{\em Phys. Rev.
  Lett.} {\bfseries 121} no.~11, (2018) 111302},
  \href{http://arxiv.org/abs/1805.12562}{{\ttfamily arXiv:1805.12562
  [astro-ph.CO]}}.

\bibitem{XENON:2019rxp}
{\bfseries XENON} Collaboration, E.~Aprile {\em et~al.}, ``{Constraining the
  spin-dependent WIMP-nucleon cross sections with XENON1T},''
  \href{http://dx.doi.org/10.1103/PhysRevLett.122.141301}{{\em Phys. Rev.
  Lett.} {\bfseries 122} no.~14, (2019) 141301},
  \href{http://arxiv.org/abs/1902.03234}{{\ttfamily arXiv:1902.03234
  [astro-ph.CO]}}.

\bibitem{PICO:2019vsc}
{\bfseries PICO} Collaboration, C.~Amole {\em et~al.}, ``{Dark Matter Search
  Results from the Complete Exposure of the PICO-60 C$_3$F$_8$ Bubble
  Chamber},'' \href{http://dx.doi.org/10.1103/PhysRevD.100.022001}{{\em Phys.
  Rev. D} {\bfseries 100} no.~2, (2019) 022001},
  \href{http://arxiv.org/abs/1902.04031}{{\ttfamily arXiv:1902.04031
  [astro-ph.CO]}}.

\bibitem{PandaX-4T:2021bab}
{\bfseries PandaX-4T} Collaboration, Y.~Meng {\em et~al.}, ``{Dark Matter
  Search Results from the PandaX-4T Commissioning Run},''
  \href{http://dx.doi.org/10.1103/PhysRevLett.127.261802}{{\em Phys. Rev.
  Lett.} {\bfseries 127} no.~26, (2021) 261802},
  \href{http://arxiv.org/abs/2107.13438}{{\ttfamily arXiv:2107.13438
  [hep-ex]}}.

\bibitem{Aalbers:2022fxq}
J.~Aalbers {\em et~al.}, ``{First Dark Matter Search Results from the
  LUX-ZEPLIN (LZ) Experiment},''
  \href{http://arxiv.org/abs/2207.03764}{{\ttfamily arXiv:2207.03764
  [hep-ex]}}.

\bibitem{PandaX:2022xas}
{\bfseries PandaX} Collaboration, Z.~Huang {\em et~al.}, ``{Constraints on the
  axial-vector and pseudo-scalar mediated WIMP-nucleus interactions from
  PandaX-4T experiment},''
  \href{http://dx.doi.org/10.1016/j.physletb.2022.137487}{{\em Phys. Lett. B}
  {\bfseries 834} (2022) 137487},
  \href{http://arxiv.org/abs/2208.03626}{{\ttfamily arXiv:2208.03626
  [hep-ex]}}.

\bibitem{XENON:2023cxc}
{\bfseries XENON} Collaboration, E.~Aprile {\em et~al.}, ``{First Dark Matter
  Search with Nuclear Recoils from the XENONnT Experiment},''
  \href{http://dx.doi.org/10.1103/PhysRevLett.131.041003}{{\em Phys. Rev.
  Lett.} {\bfseries 131} no.~4, (2023) 041003},
  \href{http://arxiv.org/abs/2303.14729}{{\ttfamily arXiv:2303.14729
  [hep-ex]}}.

\bibitem{Djouadi:2005gj}
A.~Djouadi, ``{The Anatomy of electro-weak symmetry breaking. II. The Higgs
  bosons in the minimal supersymmetric model},''
  \href{http://dx.doi.org/10.1016/j.physrep.2007.10.005}{{\em Phys. Rept.}
  {\bfseries 459} (2008) 1--241},
  \href{http://arxiv.org/abs/hep-ph/0503173}{{\ttfamily arXiv:hep-ph/0503173}}.

\bibitem{ATLAS:2020syg}
{\bfseries ATLAS} Collaboration, G.~Aad {\em et~al.}, ``{Search for squarks and
  gluinos in final states with jets and missing transverse momentum using 139
  fb$^{-1}$ of $\sqrt{s}$ =13 TeV $pp$ collision data with the ATLAS
  detector},'' \href{http://dx.doi.org/10.1007/JHEP02(2021)143}{{\em JHEP}
  {\bfseries 02} (2021) 143}, \href{http://arxiv.org/abs/2010.14293}{{\ttfamily
  arXiv:2010.14293 [hep-ex]}}.

\bibitem{Carena:2018nlf}
M.~Carena, J.~Osborne, N.~R. Shah, and C.~E.~M. Wagner, ``{Supersymmetry and
  LHC Missing Energy Signals},''
  \href{http://dx.doi.org/10.1103/PhysRevD.98.115010}{{\em Phys. Rev. D}
  {\bfseries 98} no.~11, (2018) 115010},
  \href{http://arxiv.org/abs/1809.11082}{{\ttfamily arXiv:1809.11082
  [hep-ph]}}.

\bibitem{Olive_2014}
K.~Olive, ``Review of particle physics,''
  \href{http://dx.doi.org/10.1088/1674-1137/38/9/090001}{{\em Chinese Physics
  C} {\bfseries 38} no.~9, (Aug, 2014) 090001}.
  \url{https://dx.doi.org/10.1088/1674-1137/38/9/090001}.

\bibitem{Workman:2022ynf}
{\bfseries Particle Data Group} Collaboration, R.~L. Workman and Others,
  ``{Review of Particle Physics},''
  \href{http://dx.doi.org/10.1093/ptep/ptac097}{{\em PTEP} {\bfseries 2022}
  (2022) 083C01}.

\bibitem{ATLAS:2019gti}
{\bfseries ATLAS} Collaboration, G.~Aad {\em et~al.}, ``{Search for direct stau
  production in events with two hadronic $\tau$-leptons in $\sqrt{s} = 13$ TeV
  $pp$ collisions with the ATLAS detector},''
  \href{http://dx.doi.org/10.1103/PhysRevD.101.032009}{{\em Phys. Rev. D}
  {\bfseries 101} no.~3, (2020) 032009},
  \href{http://arxiv.org/abs/1911.06660}{{\ttfamily arXiv:1911.06660
  [hep-ex]}}.

\bibitem{Heinemeyer:1998yj}
S.~Heinemeyer, W.~Hollik, and G.~Weiglein, ``{FeynHiggs: A Program for the
  calculation of the masses of the neutral CP even Higgs bosons in the MSSM},''
  \href{http://dx.doi.org/10.1016/S0010-4655(99)00364-1}{{\em Comput. Phys.
  Commun.} {\bfseries 124} (2000) 76--89},
  \href{http://arxiv.org/abs/hep-ph/9812320}{{\ttfamily arXiv:hep-ph/9812320}}.

\bibitem{Heinemeyer:1998np}
S.~Heinemeyer, W.~Hollik, and G.~Weiglein, ``{The Masses of the neutral CP -
  even Higgs bosons in the MSSM: Accurate analysis at the two loop level},''
  \href{http://dx.doi.org/10.1007/s100529900006}{{\em Eur. Phys. J. C}
  {\bfseries 9} (1999) 343--366},
  \href{http://arxiv.org/abs/hep-ph/9812472}{{\ttfamily arXiv:hep-ph/9812472}}.

\bibitem{Degrassi:2002fi}
G.~Degrassi, S.~Heinemeyer, W.~Hollik, P.~Slavich, and G.~Weiglein, ``{Towards
  high precision predictions for the MSSM Higgs sector},''
  \href{http://dx.doi.org/10.1140/epjc/s2003-01152-2}{{\em Eur. Phys. J. C}
  {\bfseries 28} (2003) 133--143},
  \href{http://arxiv.org/abs/hep-ph/0212020}{{\ttfamily arXiv:hep-ph/0212020}}.

\bibitem{Frank:2006yh}
M.~Frank, T.~Hahn, S.~Heinemeyer, W.~Hollik, H.~Rzehak, and G.~Weiglein, ``{The
  Higgs Boson Masses and Mixings of the Complex MSSM in the
  Feynman-Diagrammatic Approach},''
  \href{http://dx.doi.org/10.1088/1126-6708/2007/02/047}{{\em JHEP} {\bfseries
  02} (2007) 047}, \href{http://arxiv.org/abs/hep-ph/0611326}{{\ttfamily
  arXiv:hep-ph/0611326}}.

\bibitem{Hahn:2013ria}
T.~Hahn, S.~Heinemeyer, W.~Hollik, H.~Rzehak, and G.~Weiglein,
  ``{High-Precision Predictions for the Light CP -Even Higgs Boson Mass of the
  Minimal Supersymmetric Standard Model},''
  \href{http://dx.doi.org/10.1103/PhysRevLett.112.141801}{{\em Phys. Rev.
  Lett.} {\bfseries 112} no.~14, (2014) 141801},
  \href{http://arxiv.org/abs/1312.4937}{{\ttfamily arXiv:1312.4937 [hep-ph]}}.

\bibitem{Bahl:2016brp}
H.~Bahl and W.~Hollik, ``{Precise prediction for the light MSSM Higgs boson
  mass combining effective field theory and fixed-order calculations},''
  \href{http://dx.doi.org/10.1140/epjc/s10052-016-4354-8}{{\em Eur. Phys. J. C}
  {\bfseries 76} no.~9, (2016) 499},
  \href{http://arxiv.org/abs/1608.01880}{{\ttfamily arXiv:1608.01880
  [hep-ph]}}.

\bibitem{Bahl:2017aev}
H.~Bahl, S.~Heinemeyer, W.~Hollik, and G.~Weiglein, ``{Reconciling EFT and
  hybrid calculations of the light MSSM Higgs-boson mass},''
  \href{http://dx.doi.org/10.1140/epjc/s10052-018-5544-3}{{\em Eur. Phys. J. C}
  {\bfseries 78} no.~1, (2018) 57},
  \href{http://arxiv.org/abs/1706.00346}{{\ttfamily arXiv:1706.00346
  [hep-ph]}}.

\bibitem{Bahl:2018qog}
H.~Bahl, T.~Hahn, S.~Heinemeyer, W.~Hollik, S.~Pa\ss{}ehr, H.~Rzehak, and
  G.~Weiglein, ``{Precision calculations in the MSSM Higgs-boson sector with
  FeynHiggs 2.14},'' \href{http://dx.doi.org/10.1016/j.cpc.2019.107099}{{\em
  Comput. Phys. Commun.} {\bfseries 249} (2020) 107099},
  \href{http://arxiv.org/abs/1811.09073}{{\ttfamily arXiv:1811.09073
  [hep-ph]}}.

\bibitem{Belanger:2004yn}
G.~Belanger, F.~Boudjema, A.~Pukhov, and A.~Semenov, ``{micrOMEGAs: Version
  1.3},'' \href{http://dx.doi.org/10.1016/j.cpc.2005.12.005}{{\em Comput. Phys.
  Commun.} {\bfseries 174} (2006) 577--604},
  \href{http://arxiv.org/abs/hep-ph/0405253}{{\ttfamily arXiv:hep-ph/0405253}}.

\bibitem{Belanger:2006is}
G.~Belanger, F.~Boudjema, A.~Pukhov, and A.~Semenov, ``{MicrOMEGAs 2.0: A
  Program to calculate the relic density of dark matter in a generic model},''
  \href{http://dx.doi.org/10.1016/j.cpc.2006.11.008}{{\em Comput. Phys.
  Commun.} {\bfseries 176} (2007) 367--382},
  \href{http://arxiv.org/abs/hep-ph/0607059}{{\ttfamily arXiv:hep-ph/0607059}}.

\bibitem{Belanger:2008sj}
G.~Belanger, F.~Boudjema, A.~Pukhov, and A.~Semenov, ``{Dark matter direct
  detection rate in a generic model with micrOMEGAs 2.2},''
  \href{http://dx.doi.org/10.1016/j.cpc.2008.11.019}{{\em Comput. Phys.
  Commun.} {\bfseries 180} (2009) 747--767},
  \href{http://arxiv.org/abs/0803.2360}{{\ttfamily arXiv:0803.2360 [hep-ph]}}.

\bibitem{Belanger:2010gh}
G.~Belanger, F.~Boudjema, P.~Brun, A.~Pukhov, S.~Rosier-Lees, P.~Salati, and
  A.~Semenov, ``{Indirect search for dark matter with micrOMEGAs2.4},''
  \href{http://dx.doi.org/10.1016/j.cpc.2010.11.033}{{\em Comput. Phys.
  Commun.} {\bfseries 182} (2011) 842--856},
  \href{http://arxiv.org/abs/1004.1092}{{\ttfamily arXiv:1004.1092 [hep-ph]}}.

\bibitem{Belanger:2013oya}
G.~Belanger, F.~Boudjema, A.~Pukhov, and A.~Semenov, ``{micrOMEGAs$\_$3: A
  program for calculating dark matter observables},''
  \href{http://dx.doi.org/10.1016/j.cpc.2013.10.016}{{\em Comput. Phys.
  Commun.} {\bfseries 185} (2014) 960--985},
  \href{http://arxiv.org/abs/1305.0237}{{\ttfamily arXiv:1305.0237 [hep-ph]}}.

\bibitem{Belanger:2020gnr}
G.~Belanger, A.~Mjallal, and A.~Pukhov, ``{Recasting direct detection limits
  within micrOMEGAs and implication for non-standard Dark Matter scenarios},''
  \href{http://dx.doi.org/10.1140/epjc/s10052-021-09012-z}{{\em Eur. Phys. J.
  C} {\bfseries 81} no.~3, (2021) 239},
  \href{http://arxiv.org/abs/2003.08621}{{\ttfamily arXiv:2003.08621
  [hep-ph]}}.

\bibitem{CMS:2020xrn}
{\bfseries CMS} Collaboration, A.~M. Sirunyan {\em et~al.}, ``{A measurement of
  the Higgs boson mass in the diphoton decay channel},''
  \href{http://dx.doi.org/10.1016/j.physletb.2020.135425}{{\em Phys. Lett. B}
  {\bfseries 805} (2020) 135425},
  \href{http://arxiv.org/abs/2002.06398}{{\ttfamily arXiv:2002.06398
  [hep-ex]}}.

\bibitem{Allanach:2004rh}
B.~C. Allanach, A.~Djouadi, J.~L. Kneur, W.~Porod, and P.~Slavich, ``{Precise
  determination of the neutral Higgs boson masses in the MSSM},''
  \href{http://dx.doi.org/10.1088/1126-6708/2004/09/044}{{\em JHEP} {\bfseries
  09} (2004) 044}, \href{http://arxiv.org/abs/hep-ph/0406166}{{\ttfamily
  arXiv:hep-ph/0406166}}.

\bibitem{Heinemeyer:2007aq}
S.~Heinemeyer, W.~Hollik, H.~Rzehak, and G.~Weiglein, ``{The Higgs sector of
  the complex MSSM at two-loop order: QCD contributions},''
  \href{http://dx.doi.org/10.1016/j.physletb.2007.07.030}{{\em Phys. Lett. B}
  {\bfseries 652} (2007) 300--309},
  \href{http://arxiv.org/abs/0705.0746}{{\ttfamily arXiv:0705.0746 [hep-ph]}}.

\bibitem{Borowka:2015ura}
S.~Borowka, T.~Hahn, S.~Heinemeyer, G.~Heinrich, and W.~Hollik,
  ``{Renormalization scheme dependence of the two-loop QCD corrections to the
  neutral Higgs-boson masses in the MSSM},''
  \href{http://dx.doi.org/10.1140/epjc/s10052-015-3648-6}{{\em Eur. Phys. J. C}
  {\bfseries 75} no.~9, (2015) 424},
  \href{http://arxiv.org/abs/1505.03133}{{\ttfamily arXiv:1505.03133
  [hep-ph]}}.

\bibitem{Camargo-Molina:2013sta}
J.~E. Camargo-Molina, B.~O'Leary, W.~Porod, and F.~Staub, ``{Stability of the
  CMSSM against sfermion VEVs},''
  \href{http://dx.doi.org/10.1007/JHEP12(2013)103}{{\em JHEP} {\bfseries 12}
  (2013) 103}, \href{http://arxiv.org/abs/1309.7212}{{\ttfamily arXiv:1309.7212
  [hep-ph]}}.

\bibitem{Chowdhury:2013dka}
D.~Chowdhury, R.~M. Godbole, K.~A. Mohan, and S.~K. Vempati, ``{Charge and
  Color Breaking Constraints in MSSM after the Higgs Discovery at LHC},''
  \href{http://dx.doi.org/10.1007/JHEP02(2014)110}{{\em JHEP} {\bfseries 02}
  (2014) 110}, \href{http://arxiv.org/abs/1310.1932}{{\ttfamily arXiv:1310.1932
  [hep-ph]}}. [Erratum: JHEP 03, 149 (2018)].

\bibitem{Blinov:2013fta}
N.~Blinov and D.~E. Morrissey, ``{Vacuum Stability and the MSSM Higgs Mass},''
  \href{http://dx.doi.org/10.1007/JHEP03(2014)106}{{\em JHEP} {\bfseries 03}
  (2014) 106}, \href{http://arxiv.org/abs/1310.4174}{{\ttfamily arXiv:1310.4174
  [hep-ph]}}.

\bibitem{ALEPH:2005ab}
{\bfseries ALEPH, DELPHI, L3, OPAL, SLD, LEP Electroweak Working Group, SLD
  Electroweak Group, SLD Heavy Flavour Group} Collaboration, S.~Schael {\em
  et~al.}, ``{Precision electroweak measurements on the $Z$ resonance},''
  \href{http://dx.doi.org/10.1016/j.physrep.2005.12.006}{{\em Phys. Rept.}
  {\bfseries 427} (2006) 257--454},
  \href{http://arxiv.org/abs/hep-ex/0509008}{{\ttfamily arXiv:hep-ex/0509008}}.

\bibitem{HFLAV:2016hnz}
{\bfseries HFLAV} Collaboration, Y.~Amhis {\em et~al.}, ``{Averages of
  $b$-hadron, $c$-hadron, and $\tau$-lepton properties as of summer 2016},''
  \href{http://dx.doi.org/10.1140/epjc/s10052-017-5058-4}{{\em Eur. Phys. J. C}
  {\bfseries 77} no.~12, (2017) 895},
  \href{http://arxiv.org/abs/1612.07233}{{\ttfamily arXiv:1612.07233
  [hep-ex]}}.

\bibitem{CMS:2014xfa}
{\bfseries CMS, LHCb} Collaboration, V.~Khachatryan {\em et~al.},
  ``{Observation of the rare $B^0_s\to\mu^+\mu^-$ decay from the combined
  analysis of CMS and LHCb data},''
  \href{http://dx.doi.org/10.1038/nature14474}{{\em Nature} {\bfseries 522}
  (2015) 68--72}, \href{http://arxiv.org/abs/1411.4413}{{\ttfamily
  arXiv:1411.4413 [hep-ex]}}.

\bibitem{Belle:2010xzn}
{\bfseries Belle} Collaboration, K.~Hara {\em et~al.}, ``{Evidence for $B^- ->
  \tau^- \bar{\nu}$ with a Semileptonic Tagging Method},''
  \href{http://dx.doi.org/10.1103/PhysRevD.82.071101}{{\em Phys. Rev. D}
  {\bfseries 82} (2010) 071101},
  \href{http://arxiv.org/abs/1006.4201}{{\ttfamily arXiv:1006.4201 [hep-ex]}}.

\bibitem{Bechtle:2013xfa}
P.~Bechtle, S.~Heinemeyer, O.~St\r{a}l, T.~Stefaniak, and G.~Weiglein,
  ``{$HiggsSignals$: Confronting arbitrary Higgs sectors with measurements at
  the Tevatron and the LHC},''
  \href{http://dx.doi.org/10.1140/epjc/s10052-013-2711-4}{{\em Eur. Phys. J. C}
  {\bfseries 74} no.~2, (2014) 2711},
  \href{http://arxiv.org/abs/1305.1933}{{\ttfamily arXiv:1305.1933 [hep-ph]}}.

\bibitem{Stal:2013hwa}
O.~St\r{a}l and T.~Stefaniak, ``{Constraining extended Higgs sectors with
  HiggsSignals},'' \href{http://dx.doi.org/10.22323/1.180.0314}{{\em PoS}
  {\bfseries EPS-HEP2013} (2013) 314},
  \href{http://arxiv.org/abs/1310.4039}{{\ttfamily arXiv:1310.4039 [hep-ph]}}.

\bibitem{Bechtle:2014ewa}
P.~Bechtle, S.~Heinemeyer, O.~St\r{a}l, T.~Stefaniak, and G.~Weiglein,
  ``{Probing the Standard Model with Higgs signal rates from the Tevatron, the
  LHC and a future ILC},''
  \href{http://dx.doi.org/10.1007/JHEP11(2014)039}{{\em JHEP} {\bfseries 11}
  (2014) 039}, \href{http://arxiv.org/abs/1403.1582}{{\ttfamily arXiv:1403.1582
  [hep-ph]}}.

\bibitem{Bechtle:2008jh}
P.~Bechtle, O.~Brein, S.~Heinemeyer, G.~Weiglein, and K.~E. Williams,
  ``{HiggsBounds: Confronting Arbitrary Higgs Sectors with Exclusion Bounds
  from LEP and the Tevatron},''
  \href{http://dx.doi.org/10.1016/j.cpc.2009.09.003}{{\em Comput. Phys.
  Commun.} {\bfseries 181} (2010) 138--167},
  \href{http://arxiv.org/abs/0811.4169}{{\ttfamily arXiv:0811.4169 [hep-ph]}}.

\bibitem{Bechtle:2011sb}
P.~Bechtle, O.~Brein, S.~Heinemeyer, G.~Weiglein, and K.~E. Williams,
  ``{HiggsBounds 2.0.0: Confronting Neutral and Charged Higgs Sector
  Predictions with Exclusion Bounds from LEP and the Tevatron},''
  \href{http://dx.doi.org/10.1016/j.cpc.2011.07.015}{{\em Comput. Phys.
  Commun.} {\bfseries 182} (2011) 2605--2631},
  \href{http://arxiv.org/abs/1102.1898}{{\ttfamily arXiv:1102.1898 [hep-ph]}}.

\bibitem{Bechtle:2012lvg}
P.~Bechtle, O.~Brein, S.~Heinemeyer, O.~Stal, T.~Stefaniak, G.~Weiglein, and
  K.~Williams, ``{Recent Developments in HiggsBounds and a Preview of
  HiggsSignals},'' \href{http://dx.doi.org/10.22323/1.156.0024}{{\em PoS}
  {\bfseries CHARGED2012} (2012) 024},
  \href{http://arxiv.org/abs/1301.2345}{{\ttfamily arXiv:1301.2345 [hep-ph]}}.

\bibitem{Bechtle:2013wla}
P.~Bechtle, O.~Brein, S.~Heinemeyer, O.~St\r{a}l, T.~Stefaniak, G.~Weiglein,
  and K.~E. Williams, ``{$\mathsf{HiggsBounds}-4$: Improved Tests of Extended
  Higgs Sectors against Exclusion Bounds from LEP, the Tevatron and the LHC},''
  \href{http://dx.doi.org/10.1140/epjc/s10052-013-2693-2}{{\em Eur. Phys. J. C}
  {\bfseries 74} no.~3, (2014) 2693},
  \href{http://arxiv.org/abs/1311.0055}{{\ttfamily arXiv:1311.0055 [hep-ph]}}.

\bibitem{Bechtle:2015pma}
P.~Bechtle, S.~Heinemeyer, O.~Stal, T.~Stefaniak, and G.~Weiglein, ``{Applying
  Exclusion Likelihoods from LHC Searches to Extended Higgs Sectors},''
  \href{http://dx.doi.org/10.1140/epjc/s10052-015-3650-z}{{\em Eur. Phys. J. C}
  {\bfseries 75} no.~9, (2015) 421},
  \href{http://arxiv.org/abs/1507.06706}{{\ttfamily arXiv:1507.06706
  [hep-ph]}}.

\bibitem{Aghanim:2018eyx}
{\bfseries Planck} Collaboration, N.~Aghanim {\em et~al.}, ``{Planck 2018
  results. VI. Cosmological parameters},''
\href{http://arxiv.org/abs/1807.06209}{{\ttfamily arXiv:1807.06209
  [astro-ph.CO]}}.

\bibitem{Belanger:2018ccd}
G.~B\'elanger, F.~Boudjema, A.~Goudelis, A.~Pukhov, and B.~Zaldivar,
  ``{micrOMEGAs5.0 : Freeze-in},''
  \href{http://dx.doi.org/10.1016/j.cpc.2018.04.027}{{\em Comput. Phys.
  Commun.} {\bfseries 231} (2018) 173--186},
  \href{http://arxiv.org/abs/1801.03509}{{\ttfamily arXiv:1801.03509
  [hep-ph]}}.

\bibitem{Carena:1999py}
M.~Carena, D.~Garcia, U.~Nierste, and C.~E.~M. Wagner, ``{Effective Lagrangian
  for the $\bar{t} b H^{+}$ interaction in the MSSM and charged Higgs
  phenomenology},'' \href{http://dx.doi.org/10.1016/S0550-3213(00)00146-2}{{\em
  Nucl. Phys. B} {\bfseries 577} (2000) 88--120},
  \href{http://arxiv.org/abs/hep-ph/9912516}{{\ttfamily arXiv:hep-ph/9912516}}.

\bibitem{Hall:1993gn}
L.~J. Hall, R.~Rattazzi, and U.~Sarid, ``{The Top quark mass in supersymmetric
  SO(10) unification},'' \href{http://dx.doi.org/10.1103/PhysRevD.50.7048}{{\em
  Phys. Rev. D} {\bfseries 50} (1994) 7048--7065},
  \href{http://arxiv.org/abs/hep-ph/9306309}{{\ttfamily arXiv:hep-ph/9306309}}.

\bibitem{Guasch:2003cv}
J.~Guasch, P.~Hafliger, and M.~Spira, ``{MSSM Higgs decays to bottom quark
  pairs revisited},'' \href{http://dx.doi.org/10.1103/PhysRevD.68.115001}{{\em
  Phys. Rev. D} {\bfseries 68} (2003) 115001},
  \href{http://arxiv.org/abs/hep-ph/0305101}{{\ttfamily arXiv:hep-ph/0305101}}.

\bibitem{Dawson:2011pe}
S.~Dawson, C.~B. Jackson, and P.~Jaiswal, ``{SUSY QCD Corrections to Higgs-b
  Production : Is the $\Delta_b$ Approximation Accurate?},''
  \href{http://dx.doi.org/10.1103/PhysRevD.83.115007}{{\em Phys. Rev. D}
  {\bfseries 83} (2011) 115007},
  \href{http://arxiv.org/abs/1104.1631}{{\ttfamily arXiv:1104.1631 [hep-ph]}}.

\bibitem{Kraml:2013mwa}
S.~Kraml, S.~Kulkarni, U.~Laa, A.~Lessa, W.~Magerl, D.~Proschofsky-Spindler,
  and W.~Waltenberger, ``{SModelS: a tool for interpreting simplified-model
  results from the LHC and its application to supersymmetry},''
  \href{http://dx.doi.org/10.1140/epjc/s10052-014-2868-5}{{\em Eur. Phys. J. C}
  {\bfseries 74} (2014) 2868}, \href{http://arxiv.org/abs/1312.4175}{{\ttfamily
  arXiv:1312.4175 [hep-ph]}}.

\bibitem{Ambrogi:2017neo}
F.~Ambrogi, S.~Kraml, S.~Kulkarni, U.~Laa, A.~Lessa, V.~Magerl, J.~Sonneveld,
  M.~Traub, and W.~Waltenberger, ``{SModelS v1.1 user manual: Improving
  simplified model constraints with efficiency maps},''
  \href{http://dx.doi.org/10.1016/j.cpc.2018.02.007}{{\em Comput. Phys.
  Commun.} {\bfseries 227} (2018) 72--98},
  \href{http://arxiv.org/abs/1701.06586}{{\ttfamily arXiv:1701.06586
  [hep-ph]}}.

\bibitem{Dutta:2018ioj}
J.~Dutta, S.~Kraml, A.~Lessa, and W.~Waltenberger, ``{SModelS extension with
  the CMS supersymmetry search results from Run 2},''
  \href{http://dx.doi.org/10.31526/LHEP.1.2018.02}{{\em LHEP} {\bfseries 1}
  no.~1, (2018) 5--12}, \href{http://arxiv.org/abs/1803.02204}{{\ttfamily
  arXiv:1803.02204 [hep-ph]}}.

\bibitem{Heisig:2018kfq}
J.~Heisig, S.~Kraml, and A.~Lessa, ``{Constraining new physics with searches
  for long-lived particles: Implementation into SModelS},''
  \href{http://dx.doi.org/10.1016/j.physletb.2018.10.049}{{\em Phys. Lett. B}
  {\bfseries 788} (2019) 87--95},
  \href{http://arxiv.org/abs/1808.05229}{{\ttfamily arXiv:1808.05229
  [hep-ph]}}.

\bibitem{Ambrogi:2018ujg}
F.~Ambrogi {\em et~al.}, ``{SModelS v1.2: long-lived particles, combination of
  signal regions, and other novelties},''
  \href{http://dx.doi.org/10.1016/j.cpc.2019.07.013}{{\em Comput. Phys.
  Commun.} {\bfseries 251} (2020) 106848},
  \href{http://arxiv.org/abs/1811.10624}{{\ttfamily arXiv:1811.10624
  [hep-ph]}}.

\bibitem{Khosa:2020zar}
C.~K. Khosa, S.~Kraml, A.~Lessa, P.~Neuhuber, and W.~Waltenberger, ``{SModelS
  Database Update v1.2.3},''
  \href{http://dx.doi.org/10.31526/lhep.2020.158}{{\em LHEP} {\bfseries 2020}
  (2020) 158}, \href{http://arxiv.org/abs/2005.00555}{{\ttfamily
  arXiv:2005.00555 [hep-ph]}}.

\bibitem{Alguero:2020grj}
G.~Alguero, S.~Kraml, and W.~Waltenberger, ``{A SModelS interface for pyhf
  likelihoods},'' \href{http://dx.doi.org/10.1016/j.cpc.2021.107909}{{\em
  Comput. Phys. Commun.} {\bfseries 264} (2021) 107909},
  \href{http://arxiv.org/abs/2009.01809}{{\ttfamily arXiv:2009.01809
  [hep-ph]}}.

\bibitem{Alguero:2021dig}
G.~Alguero, J.~Heisig, C.~Khosa, S.~Kraml, S.~Kulkarni, A.~Lessa,
  H.~Reyes-Gonz\'alez, W.~Waltenberger, and A.~Wongel, ``{Constraining new
  physics with SModelS version 2},''
  \href{http://arxiv.org/abs/2112.00769}{{\ttfamily arXiv:2112.00769
  [hep-ph]}}.

\bibitem{Asner:2013psa}
D.~M. Asner {\em et~al.}, ``{ILC Higgs White Paper},'' in {\em {Community
  Summer Study 2013}: {Snowmass on the Mississippi}}.
\newblock 10, 2013.
\newblock \href{http://arxiv.org/abs/1310.0763}{{\ttfamily arXiv:1310.0763
  [hep-ph]}}.

\bibitem{An:2018dwb}
F.~An {\em et~al.}, ``{Precision Higgs physics at the CEPC},''
  \href{http://dx.doi.org/10.1088/1674-1137/43/4/043002}{{\em Chin. Phys. C}
  {\bfseries 43} no.~4, (2019) 043002},
  \href{http://arxiv.org/abs/1810.09037}{{\ttfamily arXiv:1810.09037
  [hep-ex]}}.

\bibitem{Carena:2003aj}
M.~Carena, A.~de~Gouvea, A.~Freitas, and M.~Schmitt, ``{Invisible Z boson
  decays at e+ e- colliders},''
  \href{http://dx.doi.org/10.1103/PhysRevD.68.113007}{{\em Phys. Rev. D}
  {\bfseries 68} (2003) 113007},
  \href{http://arxiv.org/abs/hep-ph/0308053}{{\ttfamily arXiv:hep-ph/0308053}}.

\bibitem{Baro:2007em}
N.~Baro, F.~Boudjema, and A.~Semenov, ``{Full one-loop corrections to the relic
  density in the MSSM: A Few examples},''
  \href{http://dx.doi.org/10.1016/j.physletb.2008.01.031}{{\em Phys. Lett. B}
  {\bfseries 660} (2008) 550--560},
  \href{http://arxiv.org/abs/0710.1821}{{\ttfamily arXiv:0710.1821 [hep-ph]}}.

\bibitem{Baro:2009na}
N.~Baro, F.~Boudjema, G.~Chalons, and S.~Hao, ``{Relic density at one-loop with
  gauge boson pair production},''
  \href{http://dx.doi.org/10.1103/PhysRevD.81.015005}{{\em Phys. Rev. D}
  {\bfseries 81} (2010) 015005},
  \href{http://arxiv.org/abs/0910.3293}{{\ttfamily arXiv:0910.3293 [hep-ph]}}.

\bibitem{Banerjee:2021hal}
S.~Banerjee, F.~Boudjema, N.~Chakrabarty, and H.~Sun, ``{Relic density of dark
  matter in the inert doublet model beyond leading order for the low mass
  region: 4. The Higgs resonance region},''
  \href{http://dx.doi.org/10.1103/PhysRevD.104.075005}{{\em Phys. Rev. D}
  {\bfseries 104} (2021) 075005},
  \href{http://arxiv.org/abs/2101.02170}{{\ttfamily arXiv:2101.02170
  [hep-ph]}}.

\bibitem{MahdiAltakach:2023bdn}
M.~Mahdi~Altakach, S.~Kraml, A.~Lessa, S.~Narasimha, T.~Pascal, and
  W.~Waltenberger, ``{SModelS v2.3: Enabling global likelihood analyses},''
  \href{http://dx.doi.org/10.21468/SciPostPhys.15.5.185}{{\em SciPost Phys.}
  {\bfseries 15} no.~5, (2023) 185},
  \href{http://arxiv.org/abs/2306.17676}{{\ttfamily arXiv:2306.17676
  [hep-ph]}}.

\bibitem{Dercks:2016npn}
D.~Dercks, N.~Desai, J.~S. Kim, K.~Rolbiecki, J.~Tattersall, and T.~Weber,
  ``{CheckMATE 2: From the model to the limit},''
  \href{http://dx.doi.org/10.1016/j.cpc.2017.08.021}{{\em Comput. Phys.
  Commun.} {\bfseries 221} (2017) 383--418},
  \href{http://arxiv.org/abs/1611.09856}{{\ttfamily arXiv:1611.09856
  [hep-ph]}}.

\bibitem{Mario_P_Giordani_2006}
M.~P. Giordani, ``Beyond the standard model physics at the tevatron,''
  \href{http://dx.doi.org/10.1088/1742-6596/53/1/021}{{\em Journal of Physics:
  Conference Series} {\bfseries 53} no.~1, (Nov, 2006) 329}.
  \url{https://dx.doi.org/10.1088/1742-6596/53/1/021}.

\bibitem{xgboost}
``{XGBOOST Documentation}.''
\newblock \url{https://xgboost.readthedocs.io/en/stable/}.

\bibitem{higgsxs}
``{Higgs cross sections for HL-LHC and HE-LHC}.''
\newblock
  \url{https://twiki.cern.ch/twiki/bin/view/LHCPhysics/HiggsEuropeanStrategy/}.

\bibitem{Adhikary:2020cli}
A.~Adhikary, N.~Chakrabarty, I.~Chakraborty, and J.~Lahiri, ``{Probing the
  $H^\pm W^{\mp } Z$ interaction at the high energy upgrade of the LHC},''
  \href{http://dx.doi.org/10.1140/epjc/s10052-021-09335-x}{{\em Eur. Phys. J.
  C} {\bfseries 81} no.~6, (2021) 554},
  \href{http://arxiv.org/abs/2010.14547}{{\ttfamily arXiv:2010.14547
  [hep-ph]}}.

\bibitem{Muong-2:2023cdq}
{\bfseries Muon g-2} Collaboration, D.~P. Aguillard {\em et~al.},
  ``{Measurement of the Positive Muon Anomalous Magnetic Moment to 0.20~ppm},''
  \href{http://dx.doi.org/10.1103/PhysRevLett.131.161802}{{\em Phys. Rev.
  Lett.} {\bfseries 131} no.~16, (2023) 161802},
  \href{http://arxiv.org/abs/2308.06230}{{\ttfamily arXiv:2308.06230
  [hep-ex]}}.

\bibitem{Aoyama:2020ynm}
T.~Aoyama {\em et~al.}, ``{The anomalous magnetic moment of the muon in the
  Standard Model},''
  \href{http://dx.doi.org/10.1016/j.physrep.2020.07.006}{{\em Phys. Rept.}
  {\bfseries 887} (2020) 1--166},
  \href{http://arxiv.org/abs/2006.04822}{{\ttfamily arXiv:2006.04822
  [hep-ph]}}.

\bibitem{LEP:stau}
``The lep susy working group and the aleph, delphi, l3 and opal experiments,''.
  \url{http://lepsusy.web.cern.ch/lepsusy/Welcome.html}.

\bibitem{ALEPH:2001oot}
{\bfseries ALEPH} Collaboration, A.~Heister {\em et~al.}, ``{Search for scalar
  leptons in e+ e- collisions at center-of-mass energies up to 209-GeV},''
  \href{http://dx.doi.org/10.1016/S0370-2693(01)01494-0}{{\em Phys. Lett. B}
  {\bfseries 526} (2002) 206--220},
  \href{http://arxiv.org/abs/hep-ex/0112011}{{\ttfamily arXiv:hep-ex/0112011}}.

\bibitem{ALEPH:2003acj}
{\bfseries ALEPH} Collaboration, A.~Heister {\em et~al.}, ``{Absolute mass
  lower limit for the lightest neutralino of the MSSM from e+ e- data at
  s**(1/2) up to 209-GeV},''
  \href{http://dx.doi.org/10.1016/j.physletb.2003.12.066}{{\em Phys. Lett. B}
  {\bfseries 583} (2004) 247--263}.

\bibitem{DELPHI:2003uqw}
{\bfseries DELPHI} Collaboration, J.~Abdallah {\em et~al.}, ``{Searches for
  supersymmetric particles in e+ e- collisions up to 208-GeV and interpretation
  of the results within the MSSM},''
  \href{http://dx.doi.org/10.1140/epjc/s2003-01355-5}{{\em Eur. Phys. J. C}
  {\bfseries 31} (2003) 421--479},
  \href{http://arxiv.org/abs/hep-ex/0311019}{{\ttfamily arXiv:hep-ex/0311019}}.

\bibitem{L3:2003fyi}
{\bfseries L3} Collaboration, P.~Achard {\em et~al.}, ``{Search for scalar
  leptons and scalar quarks at LEP},''
  \href{http://dx.doi.org/10.1016/j.physletb.2003.10.010}{{\em Phys. Lett. B}
  {\bfseries 580} (2004) 37--49},
  \href{http://arxiv.org/abs/hep-ex/0310007}{{\ttfamily arXiv:hep-ex/0310007}}.

\bibitem{OPAL:2003nhx}
{\bfseries OPAL} Collaboration, G.~Abbiendi {\em et~al.}, ``{Search for
  anomalous production of dilepton events with missing transverse momentum in
  e+ e- collisions at s**(1/2) = 183-Gev to 209-GeV},''
  \href{http://dx.doi.org/10.1140/epjc/s2003-01466-y}{{\em Eur. Phys. J. C}
  {\bfseries 32} (2004) 453--473},
  \href{http://arxiv.org/abs/hep-ex/0309014}{{\ttfamily arXiv:hep-ex/0309014}}.

\bibitem{Arias:2019uol}
P.~Arias, N.~Bernal, A.~Herrera, and C.~Maldonado, ``{Reconstructing
  Non-standard Cosmologies with Dark Matter},''
  \href{http://dx.doi.org/10.1088/1475-7516/2019/10/047}{{\em JCAP} {\bfseries
  10} (2019) 047}, \href{http://arxiv.org/abs/1906.04183}{{\ttfamily
  arXiv:1906.04183 [hep-ph]}}.

\end{thebibliography}


\providecommand{\href}[2]{#2}\begingroup\raggedright\endgroup

\end{document}